\documentclass[twocolumn,aps,prb,showpacs]{revtex4-1}
\usepackage{amsmath} 
\usepackage{amssymb} 
\usepackage{graphicx} 
\usepackage{epsfig}
\usepackage{epstopdf}
\usepackage{color}
\usepackage{rotating}
 
\begin{document} 
\title{Pseudogap and singlet formation in cuprate and organic superconductors} 
\author{J. Merino$^1$ and O. Gunnarsson$^2$} 
\affiliation{$^1$ Departamento de F\'isica Te\'orica de la Materia Condensada, Condensed
Matter Physics Center (IFIMAC) and Instituto Nicol\'as Cabrera, Universidad 
Aut\'onoma de Madrid, Madrid 28049, Spain \\
$^2$ Max-Planck-Institut f\"ur Festk\"orperforschung, Heisenbergstrasse 1, D-70569 Stuttgart, Germany} 
 
\begin{abstract}
The pseudogap phase occurring in cuprate and organic superconductors is analyzed based on 
the dynamical cluster approximation (DCA) approach to the Hubbard model. In this method a 
cluster embedded in a self-consistent bath is studied. With increasing Coulomb repulsion, 
$U$, the antinodal point [${\bf k}=(\pi,0)$] displays a gradual suppression of spectral 
density of states around the Fermi energy which is not observed at the nodal point 
[${\bf k}=(\pi/2,\pi/2)$]. The opening of the antinodal pseudogap is found to be related  
to the internal structure of the cluster and the much weaker bath-cluster couplings
at the antinodal than nodal point. The role played by internal cluster correlations is 
elucidated from a simple four-level model. For small $U$, the cluster levels form Kondo 
singlets with their baths leading to a peak in the spectral density. As $U$ is increased 
a localized state is formed  localizing the electrons in the cluster. If this cluster localized
state is non-degenerate, the Kondo effect is destroyed and a pseudogap opens up in the spectra
at the anti-nodal point. The pseudogap can be understood in terms of destructive 
interference between different paths for electrons hopping between the cluster and the bath.
However, electrons at the nodal points remain in Kondo states up to larger $U$ since they are 
more strongly coupled to the bath.  The strong correlation between the $(\pi,0)$ and $(0,\pi)$ 
cluster levels in the localized state leads to a large correlation energy gain which is important
for localizing electrons and opening up a pseudogap at the anti-nodal point. Such scenario is 
in contrast with two independent Mott transitions found in two-band systems with different 
bandwidths in which the localized cluster electron does not correlate strongly with any 
other cluster electron for intermediate $U$. The important intracluster sector correlations 
are associated with the resonating valence bond (RVB) character of the cluster ground state 
containing $d$-wave singlet pairs. The low energy excitations determining the pseudogap have suppressed d-wave 
pairing indicating that the pseudogap can be related to breaking very short-range $d$-wave 
pairs. Geometrical frustration on the anisotropic triangular lattice relevant to $\kappa$-(BEDT-TTF)$_2$X      
leads to a switch in the character of the ground state of the cluster at intermediate hopping 
ratios $t'/t\sim 0.7$.  Electron doping of the frustrated square lattice destroys the pseudogap 
in agreement with photoemission experiments on cuprates, due to a larger Schrieffer-Wolff 
exchange coupling, $J_K$, and a stronger cluster-bath coupling for the antinodal point. 
\end{abstract}
\date{\today} 
\pacs{71.30.+h; 71.27.+a; 71.10.Fd}
\maketitle 
\section{Introduction}\label{sec:1}

Cuprate and organic superconductors show a pseudogap in their phase diagram  over 
a substantial temperature and doping range. The pseudogap shows 
up as a suppression of the many-body density of states at the Fermi energy in 
a number of experiments. For hole-doped cuprates this shows up as a dip in the 
photoemission spectrum in the ${\bf k}=(\pi,0)$ direction, while there is a peak 
in the ${\bf k}=(\pi/2,\pi/2)$ direction.\cite{photoemission} The precise origin of the pseudogap
is still a topic of debate.  The understanding of the pseudogap is believed to be
important for the electronic properties of cuprates, in particular, the mechanism of
high-T$_c$ superconductivity.

Both cuprates and the organic layered superconductors $\kappa$-(BEDT-TTF)$_2$X are strongly
correlated two-dimensional systems, which are Mott\cite{Mott,Imada} insulators or close to
Mott insulators. For the cuprates a square lattice is used and for the organics an anisotropic triangular lattice
which are described by nearest neighbor and next-nearest-neighbor hopping. Undoped cuprates are
antiferromagnetic insulators, driven by a large Coulomb interaction $U$. Under doping the
cuprates become metallic and the pseudogap is observed at small dopings in the underdoped
range. On the other hand, the organics are half-filled Mott insulators at ambient
pressure. These systems, which have weaker Coulomb repulsion than the cuprates,
become metallic under hydrostatic pressure or by substitution of the anions $X$.  Thereby the lattice parameter
in the planes and the geometrical frustration can be varied. Hence, the cuprates are benchmark
systems to analyze the doping-driven Mott transition whereas the organics
are ideal examples of the Coulomb-driven Mott transition.

Early observation of the pseudogap in cuprates were made in, e.g., spin-lattice relaxation
times,\cite{spinlattice} Knight shifts,\cite{Knight} resistivity,\cite{resistivity}, photoemission
spectra,\cite{photoemission} Raman scattering,\cite{Raman} tunneling\cite{tunneling}
and scanning tunneling microscopy data.\cite{STM} For a review see Timusk and Statt.\cite{Timusk}
Evidence of a pseudogap in metallic
$\kappa$-(BEDT-TTF)$_2$X salts close to an antiferromagnetic Mott insulator comes from
the $T$-dependence of magnetic susceptibility, $\chi(T)$, NMR relaxation rate,
$1 \over T_1T$ and Knight shift, $K(T)$ experiments.\cite{Kanoda06}  A significant
decrease in $\chi(T)$ is observed below 50 K becoming steeper as the Mott transition
is approached.\cite{Kawamoto97} For instance, $^{13}$C NMR $1 \over T_1T$ and $K(T)$
experiments \cite{Mayaffre94} on $\kappa$-(ET)$_2$Cu[N(CN)$_2$]Br display a suppression
also below about 50 K becoming more pronounced in deuterated samples which are effectively
closer to the Mott transition. Under hydrostatic pressures of 4 Kbar, $1 \over T_1T$ becomes nearly constant reestablishing
conventional metallic behavior and no pseudogap.  Metallic states obtained from the spin liquid\cite{Shimizu03}
Mott insulator,$\kappa$-(ET)$_2$Cu$_2$(CN)$_3$ under pressures\cite{Kurosaki05} above 0.4 GPa
do not show signs of a pseudogap in $1 \over T_1T$ experiments.

Despite their similarities \cite{McKenzie} there are also differences between
cuprates and organics, doping being the crucial tuning parameter for cuprates
and frustration and lattice parameter for the organics. Nevertheless, the pseudogap
phenomenon appears similar in both types of systems, and closely related
models are used to describe them.  We will therefore address both classes
of systems here.

There have been a large number of theories addressing the pseudogap in cuprates.
Calculations based on the $t$-$J$ model, the Dynamical Mean-Field Theory (DMFT) and
the non-crossing approximation obtained a pseudogap in the local spectral function and related it to antiferromagnetic
fluctuations.\cite{Woelfle}  Calculations using embedded cluster methods,\cite{Maier} e.g.,
the dynamical cluster approximation (DCA), have reproduced a ${\bf k}$-dependent pseudogap for
the Hubbard model of cuprates.\cite{Civelli,Macridin,Kyung,Ferrero09,Millis,Liebscha,Sordi}
The pseudogap was interpreted in terms of antiferromagnetic correlations,\cite{Macridin,Kyung}
a momentum selective metal insulator transition\cite{Ferrero09,Millis} or a low-frequency
collective mode.\cite{Liebscha} There has been much work relating the pseudogap to preformed
superconducting pairs,\cite{Emery,Xu,Kohsaka,Wang} which have not reached phase
coherence and superconductivity at the temperature $T$ and doping studied. On the other
hand it has been argued that the pseudogap and superconductivity phases compete.\cite{Gull}

In the context of the $\kappa$-(BEDT-TTF)$_2$X organics, there have been Path Integral
Renormalization Group (PIRG),\cite{Morita,Yoshioka}  variational cluster perturbation
theory, \cite{Sahebsara} cluster DMFT,\cite{Parcollet,Ohashi,Liebsch09} finite-$T$ Lanczos
diagonalization \cite{Jure} and cluster perturbation
theory calculations\cite{Kang} on the Hubbard model on an anisotropic triangular lattice.
DMFT \cite{DMFT,Georges,KotliarDMFT} has been very successful in describing the crossover from
a Fermi liquid to a 'bad' metal at low temperature, $T^*$, observed in resistivity,\cite{Merino1,Limelette}
optical conductivity,\cite{Merino2} ultrasonic attenuation\cite{attenuation} and
thermopower\cite{thermopower,Merino1} of $\kappa$-(BEDT-TTF)$_2$X.  A pseudogap was
found,\cite{Parcollet,Kang} from CDMFT on four site clusters and assigned to antiferromagnetic
fluctuations.\cite{Kang} The relevance of short range spin fluctuations to the pseudogap phase
in organics has been discussed.\cite{Powell09} Cluster DMFT calculations show a reentrant
behavior\cite{Ohashi,Park08,Liebsch09} for the metal-insulator transition in agreement with
experiment.\cite{Kagawa} Several\cite{Civelli,Kyung,Leo04,Liebscha,Sordi,Parcollet,Ohashi,Liebsch08,Liebsch09}
of the embedded cluster calculations have used clusters with four or fewer sites, which complicates
the discussion of the large differences between ${\bf k}=(\pi,0)$ and $(\pi/2,\pi/2)$,
since the latter ${\bf k}$-vector is not present for most of these small clusters 
and interpolation procedures have been used.

The dynamical cluster approximation (DCA) treats a cluster of $N_c$ atoms embedded 
in a self-consistent bath. In much of the discussions, we describe the cluster in 
terms of the $N_c$ one-particle levels with different ${\bf K}$ vectors, where each state 
couples to its own bath.  We perform numerical experiments to identify important 
factors for the pseudogap.  We show that for, e.g., a $N_c=8$ cluster, a pseudogap 
is formed also for a non self-consistent (metallic) bath. In contrast a non self-consistent 
DMFT ($N_c=1$) calculation always gives a Kondo peak 
for a metallic bath. This shows the importance of the internal structure of the cluster. 
In a second numerical experiment, we exchange the baths for the $(\pi,0)$ and $(0,\pi)$ 
cluster levels  with the $(\pm \pi/2,\pm \pi/2)$ levels. The pseudogap then appears at $(\pi/2,\pi/2)$ 
instead of $(\pi,0)$. This shows that the bath also plays a crucial role. The essential 
aspect is that the coupling to the ${\bf K}=(\pi,0)$ and $(0,\pi)$ levels is much weaker 
than the coupling to the $(\pm \pi/2,\pm \pi/2)$ levels. The coupling to the bath of a 
${\bf K}$-point is related to the second moment of the one-particle eigenvalue in a patch 
around the ${\bf K}$-point.  This is much larger for the  $(\pm \pi/2,\pm\pi/2)$ levels 
than for the $(\pi,0)$ and $(0,\pi)$ levels.  

Guided by a small ($N_c=4$) cluster, giving a pseudogap, we construct a very simple
four-level model, with two cluster levels each coupling to one bath level. The two
cluster levels represent the $(\pi,0)$ and $(0,\pi)$ ${\bf K}$-levels on the cluster.
For a small Coulomb interaction $U$, the cluster levels form Kondo-like states with the
bath. As $U$ is increased, a localized state is formed on the cluster. If this localized state
is nondegenerate, the Kondo-like states are destroyed and a pseudogap is formed.
If the localized cluster state instead is a triplet, a new Kondo state is formed with
the bath as $U$ is increased and there is no pseudogap for moderate values of $U$.
We analyze the formation of a pseudogap or a Kondo-like peak in terms of interference effects.
By comparing correlation functions, we find that a $N_c=8$ DCA calculation
behaves in a similar way as the case with a localized nondegenerate cluster state. We
emphasize the importance of the internal structure of the cluster. This leads to a competition 
between Kondo-like effects and the formation of a localized state on the cluster. Let us consider 
a $N_c=8$ calculation with both $(\pi,0)$, $(0,\pi)$ and $(\pm \pi/2,\pm \pi/2)$ cluster 
levels. For a small value of $U$, electrons in these levels form Kondo-like states with their baths. 
Electrons in the different cluster levels are only moderately correlated. As $U$ is increased, 
the energy gain from the Kondo-like coupling is reduced and it becomes favorable to localize 
electrons in the $(\pi,0)$ and $(0,\pi)$ cluster levels. This process, however, differs in essential
 ways from two independent Mott transitions in the two channels, as described in the dynamical mean-field
theory (DMFT). Electrons in the $(\pi,0)$ and $(0,\pi)$ levels become strongly correlated, in a similar
way as for an isolated four site cluster. The corresponding energy gain is important for
driving the transition. The coupling to the bath is much larger for the  $(\pm \pi/2,\pm\pi/2)$ 
than for the $(\pi,0)$ and $(0,\pi)$ levels. The $(\pm \pi/2,\pm\pi/2)$ electrons 
therefore remain in Kondo-like states up to larger values of $U$. As $U$ is increased 
further, it becomes favorable to also localize the $(\pm \pi/2,\pm\pi/2)$ electrons.
All the electrons in the different cluster levels then become strongly correlated, leading to a large contribution to 
the energy. We also show why the pseudogap is lost for a frustrated electron-doped cuprate.

Very similar work has been done by Ferrero {\it et al.},\cite{Ferrero07} De Leo {\it et al.},\cite{Leo04}
and Capone {\it et al.},\cite{Capone} who considered two- and three-orbital Anderson
models and a four-impurity model. This work emphasized the interorbital interaction
and the cross over from a Kondo screened state to an unscreened state. The work mainly
addressed A$_3$C$_{60}$ (A= K, Rb) and did not address the ${\bf k}$ dependence of the
pseudogap for the cuprates.

The work presented here is an in-depth extension of our previous short paper\cite{MerinoGunnarsson}
and is organized as follows. In Sec. \ref{sec:2} we introduce the Hubbard model relevant to the
cuprates and organics and summarize key aspects of the DCA method. In Sec. \ref{sec:3} we show
general aspects of the model obtained within DCA such as spin and pairing correlations, double
occupancies and phase diagram.  We discuss how the pseudogap observed close to the Mott transition is
already present in the quantum impurity problem even without the self-consistency condition.
This allow us to concentrate on first iteration results.  Some general features of the spectra
are found already in a two-level model presented in Sec.~\ref{sec:4}. To obtain a pseudogap,
however, we have to introduce a four-level model in Sec.~\ref{sec:5}. We present an explanation
of the pseudogap in terms of destructive interference in Sec.~\ref{sec:6}. This analysis is
extended in Sec. \ref{sec:7}  to understanding the pseudogap in $N_c=8$ clusters which contain both
$(\pi,0)$ and $(\pi/2,\pi/2)$ sectors. We compare correlation functions to show that the $N_c=8$
DCA calculation behaves in a similar way as the four-level model. We also analyze the evolution
of the pseudogap with geometrical frustration and doping relevant to the cuprates. The character
and implications of the pseudogap found within quantum cluster theories are discussed in Secs.
\ref{sec:8} and \ref{sec:9}.

\section{Model and method}\label{sec:2}

\begin{figure}
{\rotatebox{0}{\resizebox{3.0cm}{!}{\includegraphics {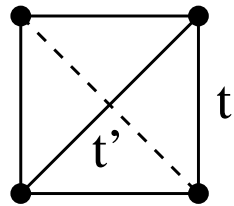}}}}                                        
\caption{\label{fig:2.1}
Hopping integrals for the Hubbard model. For cuprates the hopping integral $t'$
along both diagonals (full and dashed lines) are included (square lattice), while for the organics 
only the hopping along one diagonal (full line) is included (triangular lattice).}
\end{figure}
A minimal model for understanding the properties of layered organic compounds or 
 cuprates is a Hubbard model on a two-dimensional lattice
\begin{equation}\label{eq:2.1}
H =t \sum_{\langle ij \rangle,\sigma} (c^\dagger_{i \sigma}
c^{\phantom \dagger}_{j \sigma} + c^\dagger_{j \sigma} c^{\phantom \dagger}_{i \sigma})
+t'\sum_{\langle\langle ij \rangle\rangle,\sigma} (c^\dagger_{i
\sigma} c^{\phantom \dagger}_{j \sigma} + c^\dagger_{j \sigma} c^{\phantom \dagger}_{i \sigma})
\nonumber
\end{equation}\vspace*{-5mm}
\begin{equation}
+ U \sum_{i} n_{i\uparrow} n_{i\downarrow} -\mu \sum_{i \sigma}
c^{\dagger}_{i\sigma} c^{\phantom \dagger}_{i \sigma} \quad, 
\label{eq:Hubbard}
\end{equation}
where $\sigma$ is the spin index, $U$ is an on-site Coulomb interaction,
$t$ describes the nearest neighbor ($\langle ij \rangle$)
hopping and $t'$ the second nearest neighbor hopping ($\langle\langle ij \rangle\rangle$).
These hoppings are shown in Fig.~\ref{fig:2.1}. For the cuprates we use 
hopping along both diagonals with $t'/t\le 0$ ("square lattice") and for organics hopping
along one diagonal with $t'/t\ge 0$ ("triangular lattice"). The hole creation and destruction 
operators are given by $c^\dagger_{i \sigma}$
and $c_{i \sigma}$. We fix $t=-0.04$ eV which is the appropriate value for 
describing the lowest energy band arising from the antibonding orbitals of  
the dimerized molecules in organic materials.\cite{yOrganics} For the cuprates $t=-0.4$ 
eV is more appropriate.\cite{Prelovsek} We vary $U$ and the $t'/t $ ratio.
All units are given in eV unless explictly stated.

The Coulomb part of the Hamiltonian is rewritten as 
\begin{eqnarray}\label{eq:2.1a}
&&U\sum_{i} n_{i\uparrow} n_{i\downarrow}=U\sum_{i} (n_{i\uparrow}-n_0)( n_{i\downarrow}-n_0)\\
&&+Un_0\sum_{i\sigma}n_{i\sigma}-U\sum_i n_0^2, \nonumber
\end{eqnarray}
where $n_0$ may be chosen as $n/2$, where $n$ is the average occupancy. 
The first term in Eq.~(\ref{eq:2.1a}) is the
new many-body term, while the second term is a one-body term. In the limit of a small
$U$, the Hartree-Fock self-energy, $\Sigma_{i\sigma}^{\rm HF}=U\langle n_{i-\sigma}\rangle$, is a 
good approximation to the self-energy $\Sigma_{i\sigma}$. $Un_0$, with $n_0=n/2$,
is then an approximation to the self-energy, and the remaining part of the self-energy, $\Sigma_c\equiv \Sigma-\Sigma^{HF}$
below, is reduced. With this choice $\Sigma_c$ goes to zero as the frequency goes 
to infinity.

The model above is solved using the dynamical cluster approximation (DCA).\cite{Maier}                     
Here we summarize the main steps in the theory. The Brillouin zone is divided into 
$N_c$ patches denoted by the momenta ${\bf K}$, at each patch 
center. The ${\bf k}$ vectors of a patch are given by ${\bf K}+\tilde {\bf k}$,
where $\tilde{\bf k}$ can run over a finite but large number of points. The problem is reduced to solving
a cluster embedded in an effective bath. The bath is described by the cluster excluded Green's function
$G_0(z)$, where $G_0$ is a $N_c \times N_c$ matrix. Solving the cluster in
this bath gives a cluster Green's function $G_c$. From this a self-energy is extracted 
\begin{equation}\label{eq:2.2}
\Sigma_c(z)=G_0^{-1}(z)-G_c^{-1}(z).
\end{equation}
A coarse-grained Green's function is calculated
\begin{equation}\label{eq:2.3}
\bar G(z)={N_c\over N}\sum_{\tilde {\bf k}}\lbrace [G_0^0(\tilde {\bf k},z)]^{-1}-\Sigma_c(z)\rbrace^{-1},
\end{equation}
where $[G_0^0(\tilde {\bf k},z)]_{{\bf K},{\bf K}}=1/(z-\Delta-\varepsilon_{{\bf K}+\tilde {\bf k}})$ is a free-electron
Green's function,  
\begin{equation}\label{eq:2.3a}
\Delta=n_0 U-\mu
\end{equation}
and $N$ is the total number of ${\bf K+\tilde k}$-points.
A new bath Green's function is extracted
\begin{equation}\label{eq:2.4}
G_0(z)=[\bar G^{-1}(z)+\Sigma_c(z)]^{-1},
\end{equation}
and the approach is iterated to self-consistency.
Here we mainly study $N_c=4$, 8 and 16. We use the clusters of Betts {\it et al.}.\cite{Betts}
For instance we use clusters 8A and 16B in their notations. Here we use a spin independent $G_0$.

\begin{figure}
{\rotatebox{-90}{\resizebox{6.0cm}{!}{\includegraphics {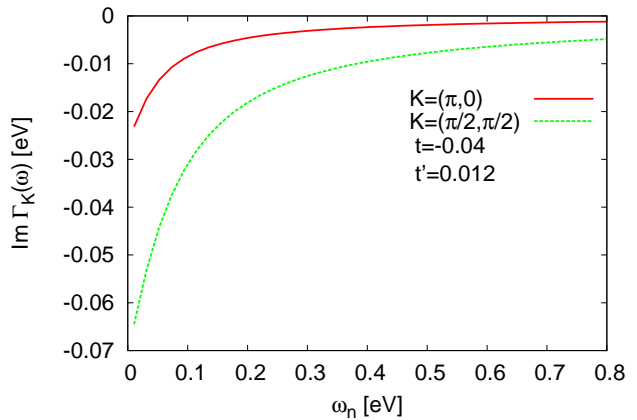}}}}                                        
\caption{\label{fig:2.2}
Coupling $\Gamma_K({\rm i}\omega_n)$ for ${\bf K}=(\pi,0)$ and $(\pi/2,\pi/2)$
for $\Sigma_c(z)\equiv 0$.
The parameters are $N_c=8$, $t=-0.04$ eV, $t'=0.012$ eV, $U=0$ eV, $\mu=0$ eV and $T=38.4$ K. The figure
illustrates how the coupling is much stronger for ${\bf K}=(\pi/2,\pi/2)$ than
$(\pi,0)$. 
}
\end{figure}

The bath Green's function can also be written as
\begin{equation}\label{eq:2.4a}
G_0(z)\equiv [z-\Delta-\bar \varepsilon_{\bf K}-\Gamma_{\bf K}(z)]^{-1},
\end{equation}
where $\bar \varepsilon_{\bf K}=(N_c/N)\sum_{\tilde {\bf k}} \varepsilon_{\bf K+\tilde k}$.
For the case of $\Sigma_c(z)\equiv 0$ we then obtain
\begin{equation}\label{eq:2.5}
{N_c\over N}\sum_{\tilde {\bf k}}{1\over z-\Delta-\varepsilon_{\bf K+\tilde k}}={1\over z-\Delta-\bar \varepsilon_{\bf K}-\Gamma_{\bf K}(z)}.
\end{equation}
We can obtain some additional understanding by expanding the left and right hand sides of Eq.~(\ref{eq:2.5}) 
to second order in $1/z$. This leads to the sum rule\cite{ErikDCA} 
\begin{equation}\label{eq:2.6}
{1\over \pi}\int {\rm Im} \Gamma_{\bf K}(\varepsilon)d\varepsilon ={N_c\over N} 
\sum_{\tilde {\bf k}}(\varepsilon_{\bf K+\tilde k}-\bar \varepsilon_{\bf K})^2,
\end{equation}
where $\Gamma_{\bf K}(i\omega_n)$ has been analytically continued to real $\varepsilon$.
If the bath is described by a finite number of levels with the energies $\varepsilon_i$
and couplings $V_{{\bf K}i}$ to the cluster state ${\bf K}$, $\Gamma_{\bf K}(\varepsilon)=
\sum_i |V_{{\bf K}i}|^2/(\varepsilon-\varepsilon_i)$ and the sum rule above is related
to $\sum_i|V_{{\bf K}i}|^2$.
This result shows that the second moment of the $\varepsilon_{\bf k}$ inside a patch is 
a  measure of the coupling to the bath. Due to the weak dispersion around 
${\bf K=}(\pi,0)$, $\Gamma_{\bf K}(z)$ is much smaller than around ${\bf K=}(\pi/2,\pi/2)$.
For $N_c=8$ the difference is roughly a factor of three to four. This is illustrated in Fig.~\ref{fig:2.2}. 
This result is crucial for the following discussion.
As is discussed in Sec.~\ref{sec:7ab} the coupling can also depend 
on the details of the density of states and the position of the chemical potential $\mu$
in a rather important way.  

In the DCA procedure, the hopping integrals used in the embedded cluster differ from the ones
in the Hubbard model in Eq.~(\ref{eq:2.1}). For instance for the cluster 8A the nearest and
second nearest hopping integrals are $t_{\rm cluster}=0.81t$ and $t'_{\rm cluster}=1.27t'$,
respectively. For $N_c=8$ second nearest neighbor sites are connected by a direct hopping
as well as a hopping due to periodic boundary conditions. These two integrals are added 
in the DCA, which leads to the large prefactor 1.27. 
In the following, we discuss results in the DCA by performing calculations
for isolated clusters. For these clusters we then use the rescaled values of the hopping
parameters to be able to compare as closely as possible with the DCA results.

The Green's function can be written as 
\begin{equation}\label{eq:2.7}
G({\bf K},\omega_n)={1 \over i\omega_n+\mu-\varepsilon_{\bf K}-n_0U-\Sigma_c({\bf K},\omega_n)}.
\end{equation}
This form can be analytically continued to real frequencies using Maximum Entropy.\cite{Jarrell}
This provides a good approximation for the spectrum of the wave vector ${\bf K}$ for the homogeneous
system. In Sec.~\ref{sec:9} we deal with an inhomogeneous system. Then we Fourier transform
the real space Green's function of the cluster to reciprocal space and obtain spectra of the
cluster. For a homogeneous system this would correspond to averaging the spectrum over
the patch surrounding a ${\bf K}$-point, which leads to a broader spectrum than Eq.~(\ref{eq:2.7}).  

For a small range of $U$and $T$ values, both a metallic and insulating state can be obtained,
depending on the state of the first iteration. In this case we have studied the metallic phase,
unless explicitly stating the opposite. For large values of $U$ the system supports an antiferromagnetic 
solution with a polarized bath. In these cases we have nevertheless used a paramagnetic bath.

\section{Results}\label{sec:3}

In this section we present results of the properties of the Hubbard model within the DCA. In 
the following sections we will analyze and discuss the origin of the pseudogap in quantum 
cluster approaches.
 
\subsection{Spin and pairing correlations}\label{sec:3s}

\begin{figure}
{\rotatebox{-90}{\resizebox{5.0cm}{!}{\includegraphics {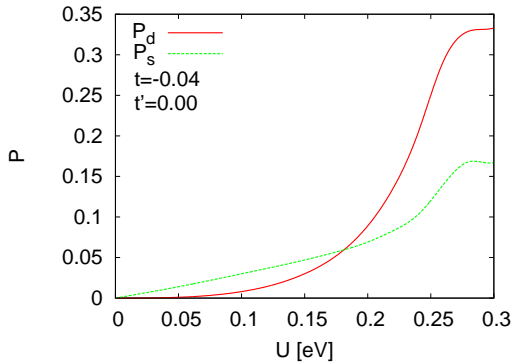}}}}                                        
\caption{(Color online) Dependence of pairing amplitudes $P$ on  the
Coulomb repulsion, $U$, in the half-filled Hubbard model on a square lattice. 
Extended $s$-wave (dotted curve) and $d_{x^2-y^2}$-wave (full curve) pairing 
channels are considered. The parameters are $t=-0.04$ eV, $t'=0$ and $T=38.4$ K,
and the calculations are self-consistent.
}\label{fig:pairing}
\end{figure}

We introduce a pairing operator
\begin{equation}\label{eq:7.3}
\Delta_i^{\dagger}={1\over 2}\sum_{\delta}f(\delta)c^{\dagger}_{i\uparrow}c^{\dagger}_{i+\delta \downarrow},
\end{equation}
where the sum over $\delta$ runs over the nearest neighbor lattice sites. If $i+\delta$ falls outside the cluster,
it is periodically continued inside the cluster. The function $f(\delta)$ describes the symmetry 
of the pairing. We use $f(\pm 1,0)=1$ and $f(0,\pm 1)=1$ for extended $s$-wave symmetry,    
$f(\pm 1,0)=1$ and $f(0,\pm 1)=-1$ for $d_{x^2-y^2}$ symmetry 
 and $f(1,1)=f(-1,-1)=1$ and f(1,-1)=f(-1,1)=-1 for $d_{xy}$ symmetry.
We then calculate 
\begin{eqnarray}\label{eq:7.4}
&&P={1\over N_c}\large(\sum_{ij}\langle \Delta_i^{\dagger}\Delta_j^{\phantom \dagger}\rangle \\
&&-{1\over 4}\sum_{ij \delta \gamma}f(\delta)f(\gamma)\langle c^{\dagger}_{i \uparrow}c^{\phantom \dagger}_{j \uparrow} \rangle
\langle c^{\dagger}_{i+\delta \downarrow}c^{\phantom \dagger}_{j+\gamma \downarrow} \rangle\large), \nonumber
\end{eqnarray}
where we have subtracted a term so that $P=0$ for noninteracting electrons. 
Without the subtraction term and for small cluster, $P$ can sometimes be 
large because of a large contribution for $i=j$. Since a tendency to superconductivity
is indicated by substantial contributions from $i$ and $j$ far apart, it is 
desirable to reduce the importance of the contributions for $i=j$. 

Fig.~\ref{fig:pairing} shows a rapid increase in the $d_{x^2-y^2}$ channel as $U$ is increased
and the system approaches a Mott transition. This pairing channel wins
over the extended $s$-wave channel for large values of $U$.

\begin{figure}
\epsfig{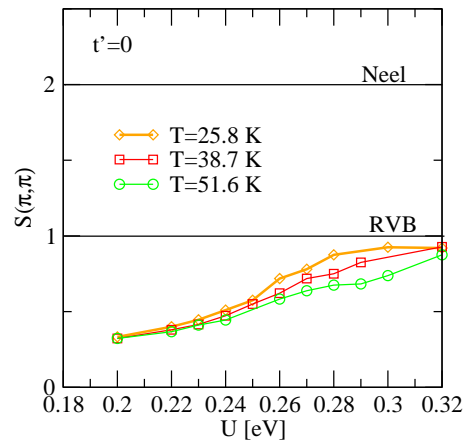}
\caption{(Color online) Magnetic correlations across the metal insulator transition from self-consistent DCA for $N_c=8$.
The dependence of the spin structure factor at $(\pi,\pi)$ on $U$ for $t'=0$ is compared to a full
antiferromagnetic N\'eel state and pure RVB states (marked with the
horizontal lines). Self-consistent DCA calculations are shown. The figure 
shows how spin correlations obtained from DCA saturate at the
value expected for the RVB state in the Mott insulating phase.   
}
\label{fig:spinRVB}
\end{figure}

In Fig. \ref{fig:spinRVB} we show the longitudinal spin-spin correlation function of the electrons:
\begin{equation}
S({\bf q})={1 \over N_c} \sum_{ij}e^{i{\bf q} {\bf R}_{ij} }\langle S^z_i S^z_j   \rangle,
\end{equation}
where $S^z_i=(n_{i\uparrow}-n_{i\downarrow})/2$ is the $z$-component of ${\bf S}_i$, which
is evaluated at the antiferromagnetic ordering vector ${\bf Q}=(\pi,\pi)$.

We compare the results of the average value for $S({\bf Q})$ obtained from DCA with the
case in which we have perfect N\'eel order and the case of the pure
RVB state which is the exact ground state of the Heisenberg model on the $N_c=8$ cluster 
(see discussion in Sec. \ref{sec:3r}). 
The spin correlations are found to saturate to the RVB value as the Mott transition is approached. This value is 
$S(\pi, \pi)=1$, which is half of the value that would be found for pure N\'eel states. 
At moderate $U$ values before the Mott transition occurs the spin correlations are not quite 
saturated to the pure RVB state. For example, for $U=0.25$ eV, within the pseudogap phase 
we find that $S(\pi,\pi)=0.6$, still far from the pure spin liquid state value.
In all cases we have considered a paramagnetic solution.

\subsection{Sector correlation functions and RVB}\label{sec:3r}

We now analyze in detail the ground state properties of isolated clusters
 in order to gain insight in the predictions of DCA embedded cluster calculations.
 We perform a detailed comparison of the exact cluster properties with
 predictions from RVB theory.
 
\subsubsection{Cluster correlations of the isolated cluster: unfrustrated case}

The ground state can be characterized through the correlation functions between
nodal $(\pm \pi/2, \pm \pi/2)$  and antinodal $(\pi,0), (0,\pi)$
cluster ${\bf K}$-sectors
\begin{equation}\label{eq:3.1a}
 C_{ {\bf K } \sigma, {\bf K'} \sigma'}=\langle n_{{\bf K} \sigma} n_{{\bf K'} \sigma'} \rangle 
- \langle n_{{\bf K}\sigma}  \rangle \langle n_{{\bf K'} \sigma'} \rangle.
\end{equation}
As we will see in Secs.~\ref{sec:5} and \ref{sec:7},
a positive correlation of  $C_{ {\bf K } \uparrow , {\bf K } \downarrow}>0$ for ${\bf K}=(\pi/2,\pi/2)$ 
and $(\pi,0)$ sectors is typically found at large $U$ using the DCA for $t'=0$.  This corresponds to 
a localized state in the cluster which correlates well with the opening of a gap in the ${\bf K}$-sector 
spectra at the Fermi surface.  Increasing the cluster size, $N_c$, leads to a reduction of the sector 
correlations.  In order to understand this behavior, we compare the exact ground state of  the cluster 
with Anderson's  RVB wavefunction. We show that the ground state of the $N_c=4$ and  8 clusters is 
exactly described by a nearest-neighbor RVB (NN-RVB). This is also a very good approximation of the ground 
state of the $N_c=16$ cluster.

Using the ED formulation we obtain the isolated cluster correlations between different sectors. The 
results for different cluster sizes are displayed in Tables \ref{table1} and \ref{table2}.
\begin{table}
\caption{Correlations between ${\bf K}=(\pi,0)$ and ${\bf K'}= (0,\pi)$ sectors obtained with ED on the isolated clusters of size N$_c$ in the limit $U>>|t|$ and $t'=0$.}
\label{table1}
\begin{tabular}{llll}
 N$_c$ &  C$_{ {\bf K } \uparrow, {\bf K} \downarrow}$ &   C$_{ {\bf K } \uparrow, {\bf K'} \uparrow}$ &    C$_{ {\bf K } \uparrow, {\bf K'} \downarrow}$\\
\hline
4 & 0.125 & -0.166    &   -0.208     \\
8 & 0.063   &  -0.125     &  -0.188   \\
16 & 0.031  &  -0.093   &    -0.153    \\
\hline
\end{tabular}
\end{table}
\begin{table}
\caption{Correlations between ${\bf K}=(\pi/2,\pi/2)$ and  ${\bf K'}=(-\pi/2,-\pi/2)$ sectors obtained with ED on the isolated clusters of size N$_c$ in the large $U>>|t|$ limit and $t'=0$. }
\label{table2}
\begin{tabular}{llll}
 N$_c$ & C$_{ {\bf K } \uparrow, {\bf K} \downarrow}$ &   C$_{ {\bf K } \uparrow, {\bf K'} \uparrow}$ &    C$_{ {\bf K } \uparrow, {\bf K'} \downarrow}$  \\
\hline
8 &    0.063  & -0.125   & -0.188  \\
16 &   0.031  &    -0.093  & -0.153 \\
\hline
\end{tabular}
\end{table}

These correlations may be compared with results for the correlations using the 
NN-RVB wavefunction as introduced by Liang, Doucot and Anderson \cite{Liang88}
for bipartite lattices. Such a wavefunction was introduced in the context of the $S=1/2$
Heisenberg antiferromagnet on the square lattice ($t'=0$) to compare to the exact singlet ground 
state. The NN-RVB is a superposition of singlets between neighboring sites taken from A to B 
sublattice with equal positive bond amplitudes:
\begin{equation}
| \Psi_0 \rangle =\sum_{i_\alpha,j_\beta} (i_1j_1)(i_2j_2)...(i_nj_n)
\label{RVB}
\end{equation}
 where $i_\alpha$ ($j_\beta$) denote neighbor sites in the A-sublattice (B-sublattice) and $(i_\alpha$ $j_\beta$)
denotes a singlet. Using this construction, the Marshall 
 sign convention\cite{Marshall} for the amplitude of ground state configurations in bipartite lattices is automatically satisfied.  
We restrict our analysis to the NN-RVB state avoiding the arbitrariness in the decay of singlet bond 
amplitudes with the relative distance.

The exact ground state of the Hubbard model on the $N_c=4$ cluster for $U>>t$  coincides 
with the\cite{Fazekas} NN-RVB (Eq. \ref{RVB}):
\begin{eqnarray}\label{eq:RVB4}
&&|\Psi_0 \rangle = (12)(34)+(14)(32) \nonumber  \\
&&={1 \over \sqrt 3 } \left( |1\uparrow,2\downarrow,3\uparrow,4\downarrow \rangle + |1\downarrow,2\uparrow,3\downarrow,4\uparrow \rangle \right)
\nonumber \\
&&-{1 \over 2 \sqrt 3} (  |1\uparrow,2\downarrow,3\downarrow,4\uparrow \rangle + |1\downarrow,2\uparrow,3\uparrow,4\downarrow \rangle \\
&&+ |1\uparrow,2\uparrow,3\downarrow,4\downarrow \rangle + |1\downarrow,2\downarrow,3\uparrow,4\uparrow \rangle  ), \nonumber 
\end{eqnarray}
 where $(12)(34)$  and $(14)(32)$ describe the four site plaquettes factorized
 into horizontal and vertical singlets, respectively (with 1 and 3 being on a diagonal).  In $N_c=8$ clusters, all 
 singlet bonds $(ij)$ that can be formed between A and B sublattices 
 are nearest neighbors so that the NN-RVB is also the exact ground state of this cluster. 
 
 \begin{table}
\caption{Correlations between  ${\bf K}=(\pi,0)$ and ${\bf K'}= (0,\pi)$  sectors obtained with the NN-RVB wavefunction on  isolated clusters of size N$_c$.}
\label{table3}
\begin{tabular}{llll}
N$_c$ &  C$_{ {\bf K } \uparrow, {\bf K} \downarrow}$ &   C$_{ {\bf K } \uparrow, {\bf K'} \uparrow}$ &    C$_{ {\bf K } \uparrow, {\bf K'} \downarrow}$\\
\hline
4 & 0.125 & -0.166   &   -0.208      \\
8 & 0.063   &  -0.125     &  -0.188  \\
16 & 0.031    &   -0.071    &  -0.111  \\
\hline
\end{tabular}
\end{table}

\begin{table}
\caption{Correlations between ${\bf K}=(\pi/2,\pi/2)$ and  ${\bf K'}=(-\pi/2,-\pi/2)$ sectors obtained with the NN-RVB on the isolated clusters of size N$_c$. }
\label{table4}
\begin{tabular}{llll}
 N$_c$ &  C$_{ {\bf K } \uparrow, {\bf K} \downarrow}$ &   C$_{ {\bf K } \uparrow, {\bf K'} \uparrow}$ &    C$_{ {\bf K } \uparrow, {\bf K'} \downarrow}$   \\
\hline
8 &    0.063   & -0.125   & -0.188  \\
16 &   0.031   &   -0.071   &   -0.111  \\
\hline
\end{tabular}
\end{table}

The sector correlations obtained exactly on the different 
clusters at large $U$ are shown in Tables~\ref{table1} and \ref{table2} 
and the ones obtained using the NN-RVB are shown in Tables~\ref{table3} 
and \ref{table4}.  The overlap between the NN-RVB, $|RVB \rangle$ and the 
ground state of the cluster is:  $| \langle RVB|\Psi_0 \rangle| =1$ for $N_c=4, 8$, and  
$| \langle RVB|\Psi_0 \rangle|=0.93 $ for  $N_c=16$.  This indicates that the ground states of the $N_c=4$ and 
$N_c=8$ clusters are pure short range RVB spin liquid states.\cite{Tang} The correlations between different sectors of 
the ground state of the $N_c=16$ cluster deviate somewhat from the pure NN-RVB wavefunction whereas the intrasector 
correlations for $N_c=16$ agree with the NN-RVB state.
The ground state of the $N_c=16$ cluster is very close but does not  exactly coincide with the NN-RVB spin liquid state.
The differences may be attributed to neglecting more distant bonds in the bond factorization. 

\subsubsection{Ground state cluster properties in the thermodynamic limit for t'=0}

\begin{figure}
\epsfig{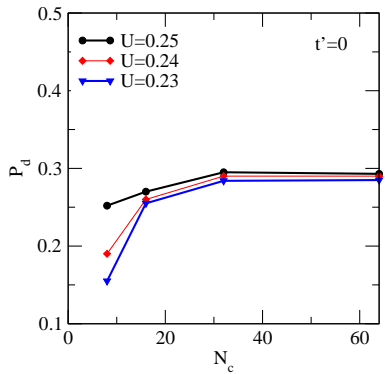}
\caption{(Color online)  $d$-wave pairing correlations for different clusters 
from self-consistent DCA for $t'=0$. The correlation length at $T=38.4$ K can be obtained 
from the saturation of P$_d$. 
}
\label{fig:Pd}
\end{figure}

In order to gain insight on the ground state properties of the cluster in the thermodynamic limit we have explored 
the pairing correlations on clusters of up to $N_c=64$ sites.  The dependence of $P_d$ with $N_c$ 
as obtained from DCA is shown in Fig. \ref{fig:Pd} at fixed temperature $T=38.4$ K and different  $U$. 
$P_d$ saturates with cluster size, meaning that there is only very short range correlation. 

\begin{figure}
\epsfig{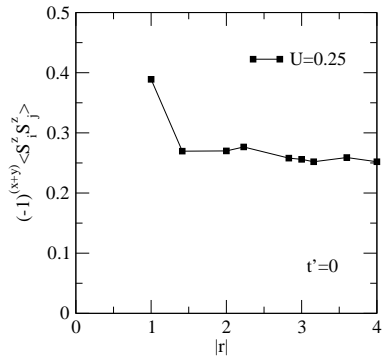}
\caption{(Color online)  Decay of spin correlations with relative distance, $|{ \bf r}|$, between
sites $i$ and $j$ inside the $N_c=64$ cluster for $t=-0.04$ eV and $t'=0$ at $T=38.4$ K in a
self-consistent calculation.
}
\label{fig:szsz}
\end{figure}

The behavior of AF spin correlations in large systems can be obtained from
 DCA calculations on the largest $Nc=64$ site cluster we have considered. In Fig. \ref{fig:szsz} we show the dependence of
of $(-1)^{(r_x+r_y)}\langle S^z_i S^z_j \rangle $ between sites $i$ and $j$ at a relative distance:
${\bf r}={\bf R}_i-{\bf R}_j$ inside the $N_c=64$ cluster. The spin correlations are found to
decay very slowly with the relative distance between sites, $|{\bf r}|=\sqrt{r_x^2+r_y^2}$. Since
the calculation is performed at low but finite temperatures Mermin-Wagner theorem predicts
no long range order in the two-dimensional lattice. On the other hand, the slow decay observed
is consistent with the long range order expected at $T=0$ in the Heisenberg model on the
square lattice in the thermodynamic limit.\cite{Oitmaa78}  This is also consistent with the
need of including further distant singlet bonds between sites when describing the exact
ground state correlations of clusters with $N_c>8$ using RVB states  discussed
previously.\cite{Liang88}

\subsubsection{Cluster correlations on the isolated cluster: geometrical frustration effects}

We now turn on the geometrical frustration, $t'/t$, which is present both for 
cuprate and organic superconductors. Here we focus on the organic superconductors with a triangular lattice.
Below we perform calculations for isolated clusters to understand the results better.
As discussed in Sec.~\ref{sec:2}, in DCA the hopping integrals of the embedded cluster are
rescaled from the values in the Hubbard model. Below we give the hoppings of the Hubbard model,
$t$ and $t'$. The corresponding hopping parameters for the cluster are then ($N_c=8$ and cluster 8A
according to Ref. \onlinecite{Betts}) $t_{\rm cluster}=0.81t$ and $t'_{\rm cluster}=1.27t'$.
These parameters are then also used for the isolated cluster.
We show in tables  \ref{table:4} and \ref{table:5} results for varying degree of geometrical 
frustration, $t'/t$.  Apart from overall quantitative changes we find a qualitative change in
C$_{(\pi/2,\pi/2) \uparrow, (-\pi/2,-\pi/2) \downarrow}$  which displays a positive correlation
for $t'/t>0.7$, instead of the negative correlation found for the unfrustrated case: $t'=0$.
A similar change is seen for $C_{(\pi,0)\uparrow, (0,\pi)\downarrow}$.

In order to understand the origin of this sudden change of correlation
functions we have explored pairing correlations. In table \ref{table:6} we show
the dependence of pairing correlations on $t'/t$ for a $N_c=8$ cluster.
The figure shows how the d$_{x^2-y^2}$ and extended-$s$ pairing are
strongly suppressed and the $d_{xy}$ becomes positive and large at $t'\ge 0.7t$.
The fact that at the same time C$_{(\pi/2,\pi/2) \uparrow,(-\pi/2,-\pi/2) \downarrow}$
becomes positive is a consequence of $d_{xy}$ pairing since
C$_{(\pi/2,\pi/2)  \uparrow, (-\pi/2,-\pi/2)\downarrow}$ is related to the strength
of pairing correlations at the nodal points. This sudden switch in the pairing
and correlation functions occurs at a crossing of cluster states. A similar
change from d$_{x^2-y^2}$ to d$_{xy}$ has been found previously\cite{Leo04} in
smaller $N_c=4$ clusters on a frustrated square lattice.

\begin{table}
\caption{\label{table:4} Dependence of correlations between ${\bf K}=(\pi,0)$ and ${\bf K'}= (0,\pi)$ for different $t'/t$ from ED on N$_c=8$ isolated clusters.
The hopping $t_{\rm cluster}=-0.0324$ ($t=-0.04$) eV and $U=0.3$ eV are used in a triangular lattice model for organics.}
\begin{tabular}{llll}
t'/t &  C$_{ {\bf K } \uparrow, {\bf K} \downarrow}$ &   C$_{ {\bf K } \uparrow, {\bf K'} \uparrow}$ &    C$_{ {\bf K } \uparrow, {\bf K'} \downarrow}$\\
\hline
0  &   0.072  & -0.131 & -0.191  \\
0.5  &  0.072    & -0.131 &  -0.191   \\
0.7  &   0.072   & -0.131 & -0.191  \\
0.8 &   0.063 & -0.012 & 0.032    \\
1. &     0.057  & -0.0092 &  0.031  \\
\hline
\end{tabular}
\end{table}

\begin{table}
\caption{\label{table:5} Dependence of correlations between ${\bf K}=(\pi/2,\pi/2)$ and ${\bf K'}= (-\pi/2,-\pi/2)$ for different $t'/t$ from ED on N$_c=8$ isolated clusters.
The hopping $t_{\rm cluster}=-0.0324$ ($t=-0.04$) eV and $U=0.3$ eV are used in a triangular lattice model for organics.}
\begin{tabular}{llll}
t'/t &  C$_{ {\bf K } \uparrow, {\bf K} \downarrow}$ &   C$_{ {\bf K } \uparrow, {\bf K'} \uparrow}$ &    C$_{ {\bf K } \uparrow, {\bf K'} \downarrow}$\\
\hline
0  &    0.072   & -0.131 & -0.191 \\
0.5  &  0.072   & -0.131 & -0.191  \\
0.7  &  0.072    & -0.131 & -0.191 \\
0.8 &   0.048   & -0.012  & 0.048  \\
1. &    0.044  &  -0.0092  &0.044  \\
\hline
\end{tabular}
\end{table}

\begin{figure}
{\rotatebox{-90}{\resizebox{6.0cm}{!}{\includegraphics {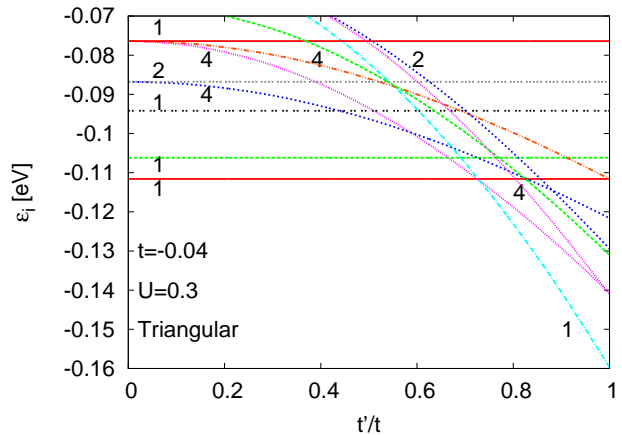}}}}
\caption{\label{fig:level}Low-lying levels for an isolated, triangular cluster
with $N_c=8$ as a function of $t'/t$ ($t'_{\rm cluster}/t_{\rm cluster}=1.57t'/t$).
The numbers show the level degeneracies. The Hubbard model parameters are $t=-.04$
eV and $U=0.3$ eV.
}
\end{figure}

To discuss the crossing of states in more detail, we show in Fig.  \ref{fig:level} some
low-lying states for different degree of geometrical frustration. For $t'/t<0.7$ the occupancies of
the $(\pi,0)$, $(0,\pi)$ and $(\pm \pi/2,\pm \pi/2)$ orbitals are the same and equal to unity.
The system uses these six orbitals to form a state where the Coulomb repulsion is strongly
reduced. As $t'$ is increased, the $(\pi/2,-\pi/2)$ and $(-\pi/2,\pi/2)$ orbitals are
lowered and the other orbitals in the space discussed above are raised. The occupancy of the orbitals,
however, is such that the energy is unchanged. At $t'/t>0.7$ this changes. The lowest
state is now one where occupancy of $(\pi/2,-\pi/2)$ and $(-\pi/2,\pi/2)$ is substantially
larger than unity ($\sim 1.5$) and the occupancy of the orbitals which move upwards is reduced.
The result is a more negative hopping energy. However, the system now has smaller possibilities
to correlate the electrons, and the Coulomb repulsion is larger. For $t'/t>0.7$ this is favorable.
For instance, for $t'/t=0.8$, the Coulomb energy is 0.093 (0.075) eV and the hopping energy
is -0.216 (-0.186) eV for the lowest state (the state that was lowest for smaller $t'$).

The lowest state is always non-degenerate, except around $t'/t=0.7$, where levels cross.
We will see below how the switch from a non-degenerate to a degenerate lowest cluster state,
coupled to a metallic bath, can lead to a switch from a non-Kondo to a bath-cluster Kondo
singlet formation. This is crucial for the pseudogap destruction and may play a role
for its disappearance as the geometrical frustration is increased in the organics.
However, this should be taken with caution since for $t'=t$ the  (non-degenerate) 
ground-state in the cluster is again well separated from higher states, 
as can be observed in Fig. \ref{fig:level}.  

\begin{table}
\caption{\label{table:6}Dependence of pairing correlations with $t'/t$ from ED on N$_c=8$ isolated clusters
with $t_{\rm cluster}=-0.0324$ ($t=-0.04$) eV and $U=0.3$ eV in the triangular lattice model for organics. }
\begin{tabular}{llll}
 t'/t & P$_s$ & P$_{d_{x^2-y^2}}$  &  P$_{d_{xy}}$  \\
\hline
0  &    0.19  & 0.39  & -0.5  \\
0.5  &   0.19  &  0.39    &  -0.5  \\
0.7  &    0.19  & 0.39  & -0.5  \\
0.8 &   0.014  &  -0.012    &  0.71  \\
1. &    0.015 & -0.0092  & 0.65  \\
\hline
\end{tabular}
\end{table}

\subsection{Double occupancy}\label{sec:3d}

\begin{figure}
\epsfig{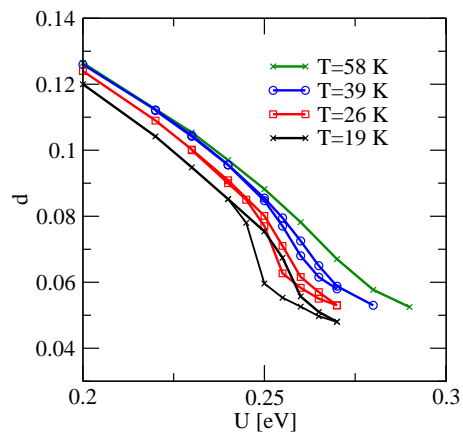}
\caption{(Color online) Dependence of double occupancy $d$ on $U$ 
in the half-filled Hubbard model on a square lattice, The parameters are $t=-0.04$ eV, 
$t'=0$ and the calculations are self-consistent.
}\label{fig:double}
\end{figure}
The double occupancy $d$ can be used as a measure of a metal-insulator 
transition.\cite{Kotliar} The transition shows up as a rapid drop in $d$ 
as a function of $U$ at the transition. Fig.~\ref{fig:double} shows the 
double occupancy for a nonfrustrated lattice. For large $T$ there is just 
a smooth drop with $U$. For smaller $T$, however, the drop is much more 
rapid, signaling a metal-insulator transition. The curve $d(U)$ shows 
hysteresis. As $U$ is increased, there is a strong drop at some $U_{c2}(T)$, 
where a metallic state cannot any longer be stabilized. As $U$ is then 
decreased, there is a rapid increase at some $U_{c1}(T)<U_{c2}(T)$, where 
the insulating state cannot be stabilized. This kind of hysteresis is
well-known from DMFT and DCA calculations.

\begin{figure}
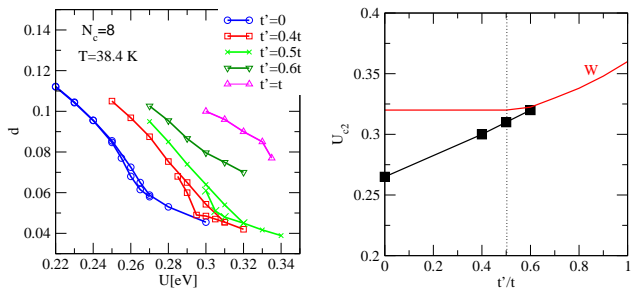

\epsfig{file=fig9a.eps,width=4cm,clip=}
\hskip0.2cm
\epsfig{file=fig9b.eps,width=4cm,clip=}
\caption{(Color online) 
Dependence of double occupancy $d$ on $U$ and geometrical frustration for a triangular lattice from self-consistent DCA.
An increase of the critical value $U_{c2}$ is found with $t'/t$ which is plotted in the right panel. 
This is compared with the dependence of the bare bandwidth $W$ on $t'/t$. The bandwidth increases from $W=0.32$eV for $t' \leq 0.5t$ to $W=0.36$ eV
for $t'=t$ with $|t|=0.04$ eV and $T=38.4$ K. 
\label{fig:doublefrust}}
\end{figure}

Geometrical frustration can be varied by changing $t'/t$. In the left part of Fig. \ref{fig:doublefrust} 
we show the behavior of the double occupancy with $U$  for different $t'/t$ and a 
triangular lattice. The critical values of $U$, $U_{c1}$ and $U_{c2}$, increase with 
$t'/t$ but the general behavior is similar up to $t'/t=0.5$. At $t'\sim 0.6 t$
there is a more rapid change in the absolute value of the double occupancy and its dependence
on $U$.  Such behavior is roughly correlated with a crossing in the cluster energy level structure 
discussed above (see Fig. \ref{fig:level}). As we will discuss below this change 
from non-degenerate to nearly degenerate cluster energy levels can lead 
to a qualitative change in the spectral properties of the cluster when coupled to the bath. 
As shown in the right part of the figure, the dependence of $U_{c2}$ with $t'/t$ 
follows the dependence of the bandwidth $W$ on $t'/t$ only for $t'/t > 0.5$. Below 
$t'/t<0.5$,  $U_{c2}$ increases with $t'/t$ despite the fact that the bandwidth 
remains constant at the value for the square lattice: $W=8|t|=0.32$ eV.

\subsection{Phase diagram}\label{sec:3p}

\begin{figure}
\epsfig{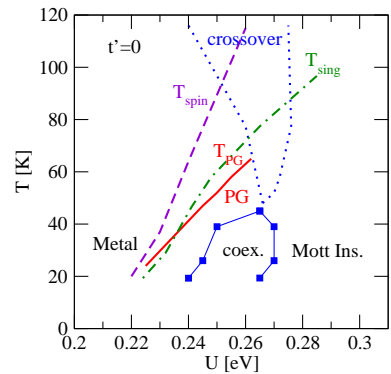} 
\caption{(Color online) Temperature vs. $U$ phase diagram for the half-filled Hubbard model from self-consistent DCA for 
the unfrustrated system ($t'=0$).
Metallic, Mott insulating and crossover regions are shown. The temperature scales $T_{\text{spin}}$ and $T_{\text{sing}}$
denote temperatures below which spin and $d$-wave pairing correlations, respectively,  become substantial.  
$T_{PG}$ indicates the temperature scale below which a pseudogap opens up in the $(\pi,0)$ spectral density. 
The parameters are $t=-0.04$ eV and $N_c=8$.
}\label{fig:phasedHubbard}
\end{figure}

\begin{figure}
\epsfig{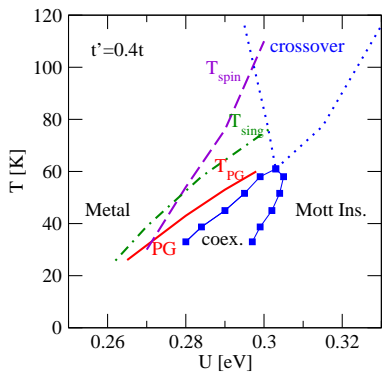} 
\caption{(Color online) The same as in Fig.~\ref{fig:phasedHubbard}, but for the frustrated 
anisotropic triangular lattice with $t'=0.4t$.
}\label{fig:phasedHubbardfrust}
\end{figure}

In Fig.~\ref{fig:phasedHubbard} we present a phase diagram of the 
Hubbard model for $t'=0$ and $t'=0.4t$ obtained from DCA on $N_c=8$ 
clusters. Based on the behavior of the double occupancy, discussed 
above, we have determined the critical values of $U$, $U_{c1}(T)$ 
and $U_{c2}(T)$. Between these two values there is a coexistence 
region, where both a metallic and an insulating phase can exist. 
The phase diagram also shows the pseudogap phase which occurs  
below the temperature,  $T_{PG}$. This is determined by 
from the spectral function $A({\bf K},\omega)$ at ${\bf K}=(\pi,0)$. 
A suppression of the weight at $\omega=0$ marks the onset of the 
pseudogap phase. A crossover region above the coexistence region is determined
from the spectra. In this region there is no proper quasiparticle at   
the chemical potential but also no proper Mott gap.                    
The quantity $T_{\rm spin}$ is obtained for a fixed $U$ as the $T$ 
where $\langle S_i^zS_j^z\rangle=-0.15$ for $i$ and $j$ nearest neighbor.
In Fig.~\ref{fig:phasedHubbardfrust}, showing results for a frustrated
model, the criteria was set to -0.20. In a similar way we have determined 
$T_{\rm sing}$ as the $T$ where $P_{x^2-y^2}=0.19$, slightly more than half 
the maximum value in Fig.~\ref{fig:pairing} of the pairing correlation.

The phase diagram displays metallic, Mott insulating,  'coexistence'  and 
pseudogap phases. The crossover region which separates the Mott insulating 
and metallic phases is shown.  The phase diagram with $t'=0.4t$ should be 
compared to the phase diagram\cite{Limelette} of the organic salt $\kappa$-(BEDT-TTF)$_2$Cu[N(CN)]$_2$Cl.

The $U_c(T)$ boundaries shown in Fig.~\ref{fig:phasedHubbard} display a 
positive slope in contrast to single site DMFT. This difference is attributed 
to short range non-local correlations which occur at sufficiently low temperatures 
and are neglected in DMFT.\cite{Liebsch09,Park08} Such behavior was also found 
previously on $N_c=4$ clusters.\cite{Liebsch09,Park08}  Magnetic correlations 
lead to a suppression of entropy at low $T$, $S \rightarrow 0$ as
$T \rightarrow 0$.   With increasing $T$, the system gains entropy by 
transforming into a metallic state since $S \propto T$ in the metal.\cite{Georges}
At higher temperatures the system can gain further entropy by transforming 
back into the paramagnetic insulating state with entropy $S \sim \ln(2)$. 
Hence, a reentrant behavior of the phase diagram occurs. The phase diagram 
resembles the experimental phase diagram with small degree of geometrical
frustration of the organic salts, such as $\kappa$-(BEDT-TTF)$_2$Cu[N(CN)]$_2$Cl.\cite{Kagawa} In particular, 
the positive slope of the $U_{c1}(T)$ and $U_{c2}(T)$ boundaries and the 
reentrant behavior agree with experiment.

\subsection{Effective mass enhancement and non-Fermi liquid behavior }

\begin{figure}
\epsfig{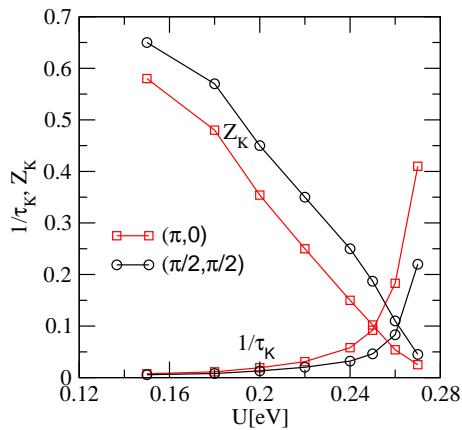}
\caption{(Color online) Quasiparticle effective mass and scattering rate 
dependence with $U$ from self-consistent DCA. The parameters are $t=-0.04$ eV, $t'=0$ and $T=38.4$ K.
}\label{fig:Zk}
\end{figure}

We introduce the quasiparticle weight $Z_{\bf k}=1/(1-d\Sigma({\bf k},\omega)/d\omega)$
and approximate $d\Sigma({\bf k},\omega)/d\omega\approx \Sigma({\bf k},i\omega_0)/(i\omega_0)$,
where $\omega_0=\pi/\beta$ with $\beta=300$ eV$^{-1}$ ($T=38.4$ K).
We also introduce the scattering rates, $1/\tau_{\bf k}$= -Im $\Sigma({\bf k},i\omega_0)$.
In Fig. \ref{fig:Zk} we show the dependence on $U$  of $Z_{\bf K}$ and $1/\tau_{\bf K}$ at
the patch momenta ${\bf K}$. $Z_{\bf K}$ is gradually suppressed with $U$.
At about $U \approx 0.22-0.24$ eV and $t'=0$  the  ${\bf K}=(\pi,0)$
and $(\pi/2,\pi/2)$ sectors display an enhancement of $1/\tau_{\bf K}$ suggesting
a non vanishing scattering rate at the Fermi surface signaling non-Fermi
liquid behavior.\cite{Kyung2,Civelli,Ferrero09,Millis,Liebscha} At this point 
$Z_{\bf k}$ looses the meaning of a quasiparticle weight.  
A momentum differentiation is clearly seen in $1/\tau_{\bf K}$ between
${\bf K}=(\pi/2,\pi/2)$ and ${\bf K}=(\pi,0)$. This differentiation grows
as the Mott metal-insulator transition is approached.

\begin{figure}
\epsfig{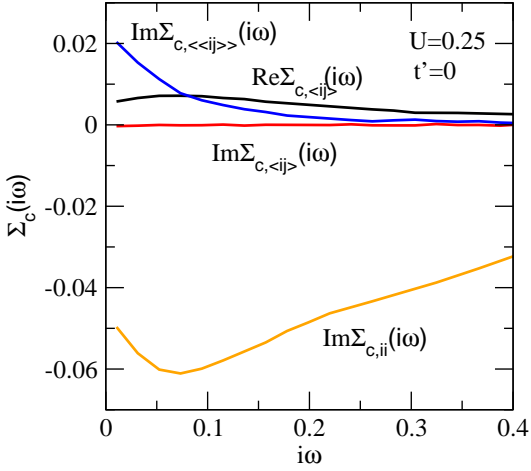}
\caption{(Color online) Cluster self-energy components in real space obtained from Fourier transforming
the self-consistent DCA self-energy in a $N_c=8$ cluster for $t=0.04$ eV and at $T=38.4$ K. The on-site 
($\Sigma_{ii}$), nearest neighbor ($\Sigma_{\langle ij \rangle}$) and next-nearest neighbor 
($\Sigma_{\langle\langle ij \rangle \rangle}$) self-energies are shown. The imaginary part of the 
next-nearest neighbor self-energy is enhanced as the Mott transition is approached
leading to differences in the $(\pi/2,\pi/2)$ and $(\pi,0)$ sectors at the Fermi surface.
\label{fig:selfener}
}
\end{figure}

It is useful to explore further the non-Fermi liquid behavior found at different
${\bf K}$-points on the Fermi surface
by analyzing the self-energy in real space. In Fig. \ref{fig:selfener} we show
the real space self-energy, $\Sigma_{c,ij}(i\omega_n)$. The imaginary part of the
on-site self-energy of the cluster, Im $\Sigma_{c,ii}(i\omega_n)$, shows
non-Fermi liquid behavior since it goes to a finite value as $\omega_n \rightarrow 0$.
In addition, the imaginary part of the off-diagonal self-energy between next-nearest
neighbor sites, Im $\Sigma_{c,\langle\langle ij \rangle\rangle}(i\omega_n)$,  shows a
significant enhancement in the pseudogap phase close to the Mott transition.

Expressing $\Sigma_c({\bf K},i\omega_n)$ in terms of its real space components
$\Sigma_{c,ij}(i\omega_0)$ through the Fourier transform\cite{Civelli}
\begin{equation}
\
\Sigma_c({\bf K},i\omega)={1 \over N_c} \sum_{i,j} e^{i{\bf K}\cdot({\bf R_i}-{\bf R_j})}\Sigma_{c,ij}(i \omega),
\end{equation}
we find that the self-energy in momentum space can be approximated by:
\begin{eqnarray}
&& \Sigma_c((\pi,0),i\omega)  \approx    \Sigma_{c,ii} (i\omega)-2\Sigma_{c,\langle\langle ij\rangle\rangle}(i\omega)
\nonumber \\ 
&& \Sigma_c((\pi/2,\pi/2),i\omega)  \approx  \Sigma_{c,ii} (i\omega),
\end{eqnarray}
since the real space contributions from the more distant neighbors is negligible.
Therefore, $\Sigma_c((\pi/2,\pi/2),i\omega)$ basically coincides with the on-site
self-energy. The $(\pi,0)$ self-energy is enhanced with respect to $\Sigma_c((\pi/2,\pi/2),i\omega)$
because of the additional off-diagonal contribution $\Sigma_{c,\langle\langle ij \rangle\rangle}(i\omega_n)$
not present at $(\pi/2,\pi/2)$ due to the phase factors.  The momentum differentiation
can then be related to the enhancement of the off-diagonal next-nearest neighbors
self-energy, $\Sigma_{c,\langle\langle ij \rangle\rangle}(i\omega_n)$.\cite{Civelli}
Such enhancement is also encountered in smaller $N_c=4$ clusters, but is missed by
single-site DMFT calculations which neglect the spatial dependence of the self-energy.
Earlier  CDMFT calculations on $N_c=4$ clusters find similar enhancement,\cite{Civelli}
although in such small clusters interpolation schemes are need to compute the self-energy
at $(\pi/2,\pi/2)$.  Also in the $N_c=3$ clusters a similar behavior of the off-diagonal
self-energy with $U$ arises close to the Mott transition. Even a cluster with $N_c=2$
gives momentum differentiation if the patches are chosen appropriately.\cite{Ferrero09}
In conclusion, non-Fermi
liquid behavior is found in the on-site self-energies of the cluster embedded in the
metallic host. The next-nearest neighbors self-energy are directly responsible for
the momentum differentiation.

\subsection{Importance of internal structure of cluster: first iteration DCA.}\label{sec:3f}

\begin{figure}
{\rotatebox{-90}{\resizebox{4.9cm}{!}{\includegraphics {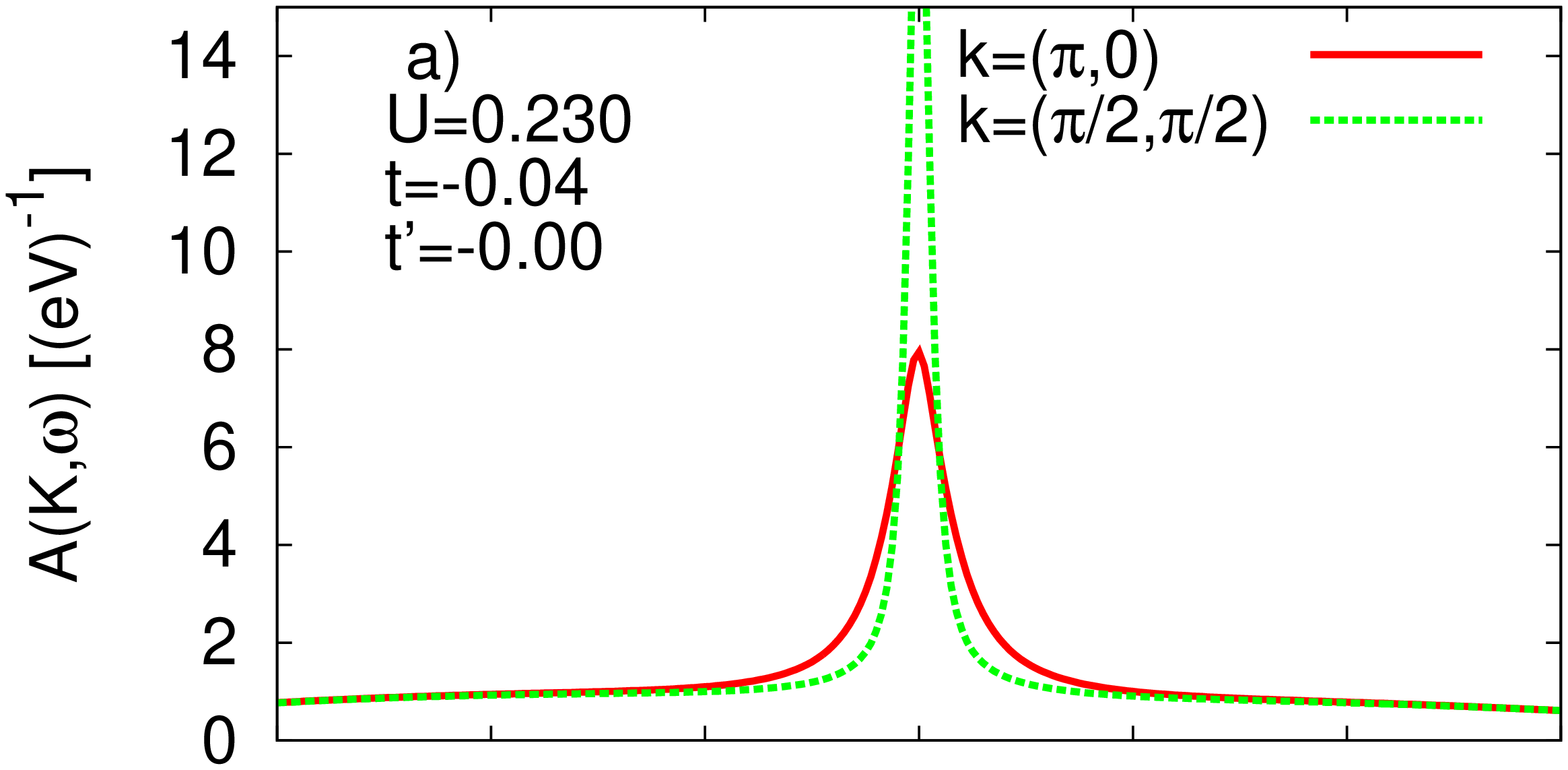}}}}
\vskip-1.87cm
{\rotatebox{-90}{\resizebox{4.9cm}{!}{\includegraphics {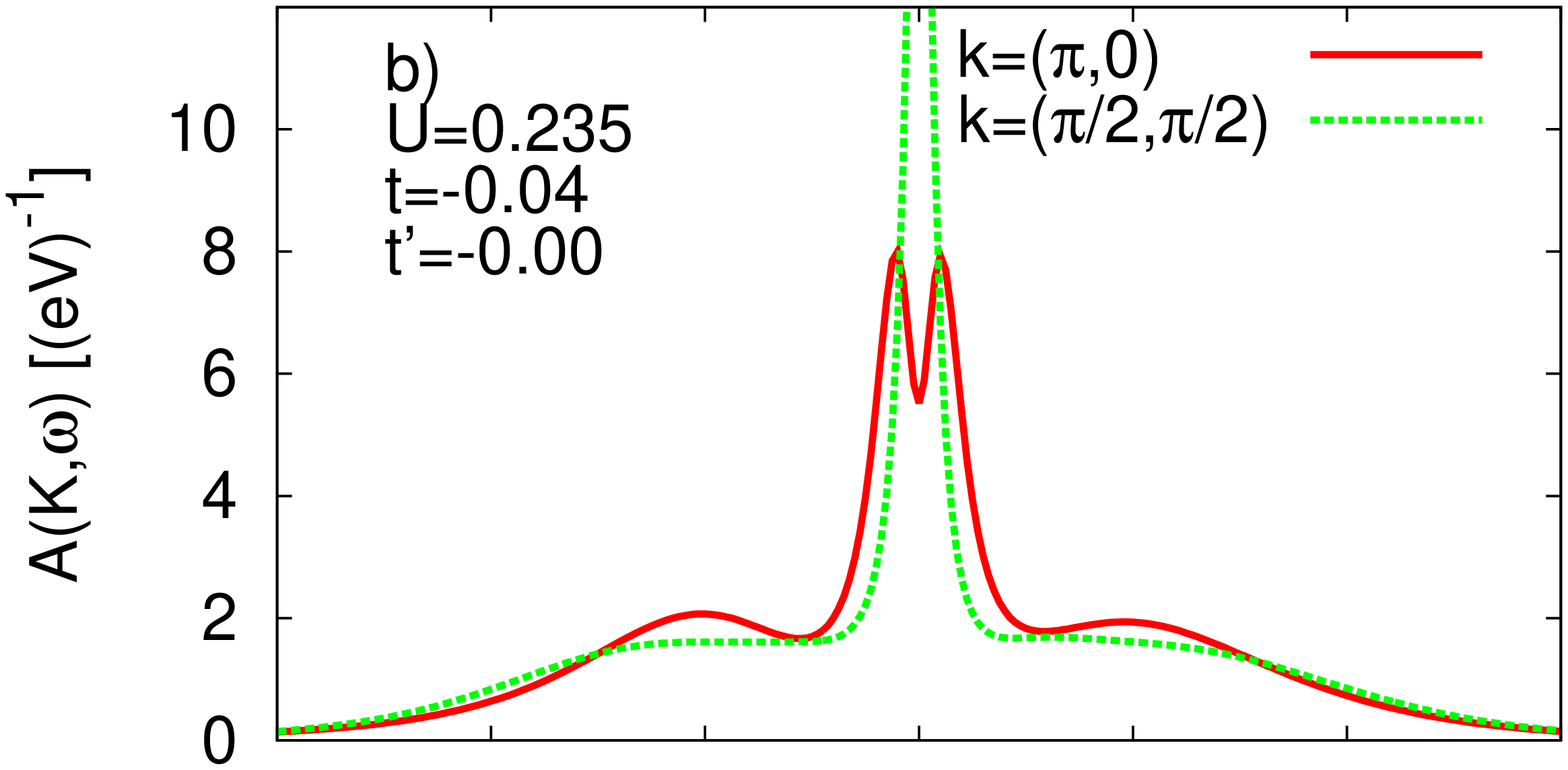}}}}
\vskip-1.88cm
{\rotatebox{-90}{\resizebox{4.9cm}{!}{\includegraphics {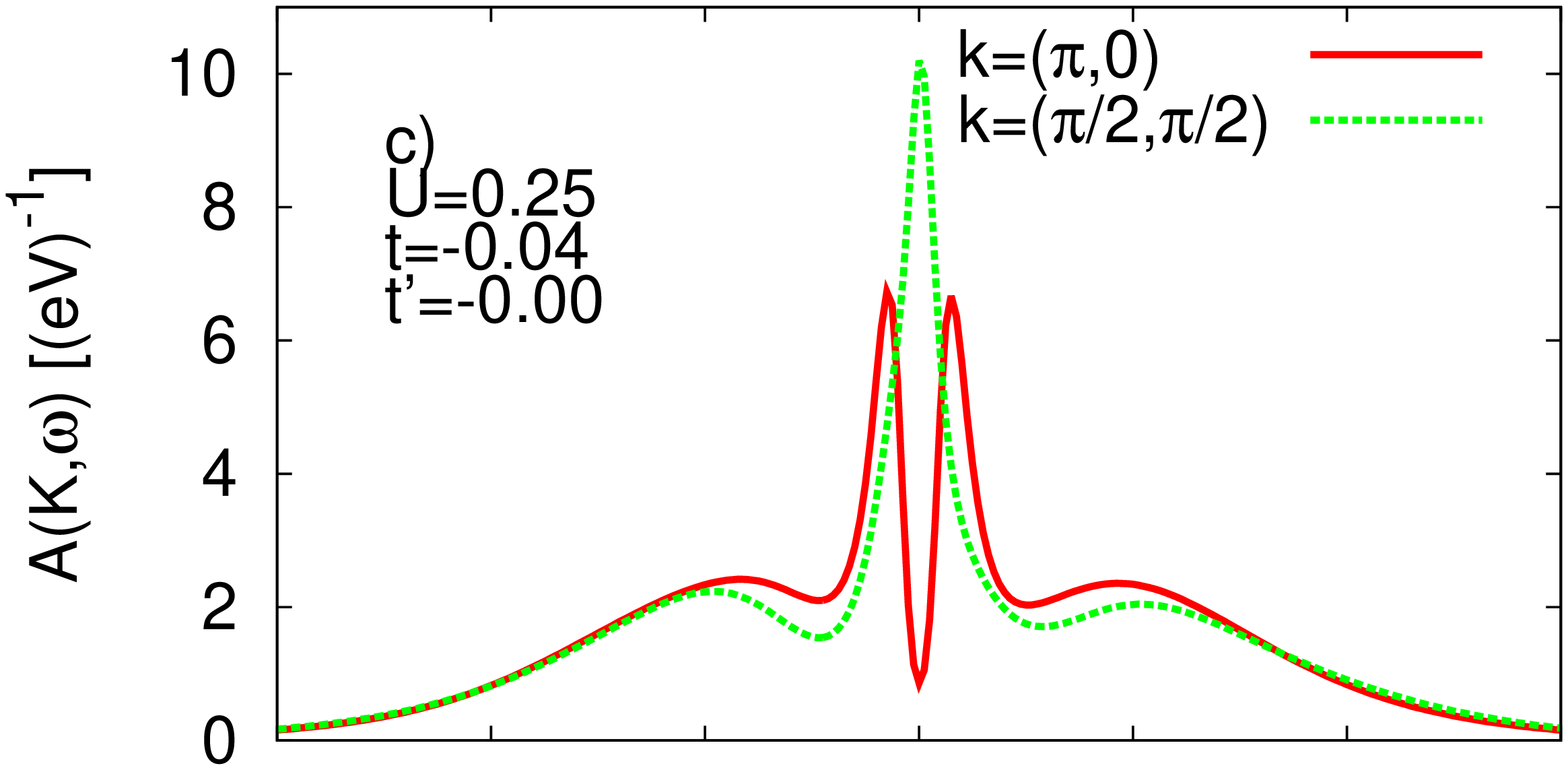}}}}
\vskip-1.89cm
{\rotatebox{-90}{\resizebox{4.9cm}{!}{\includegraphics {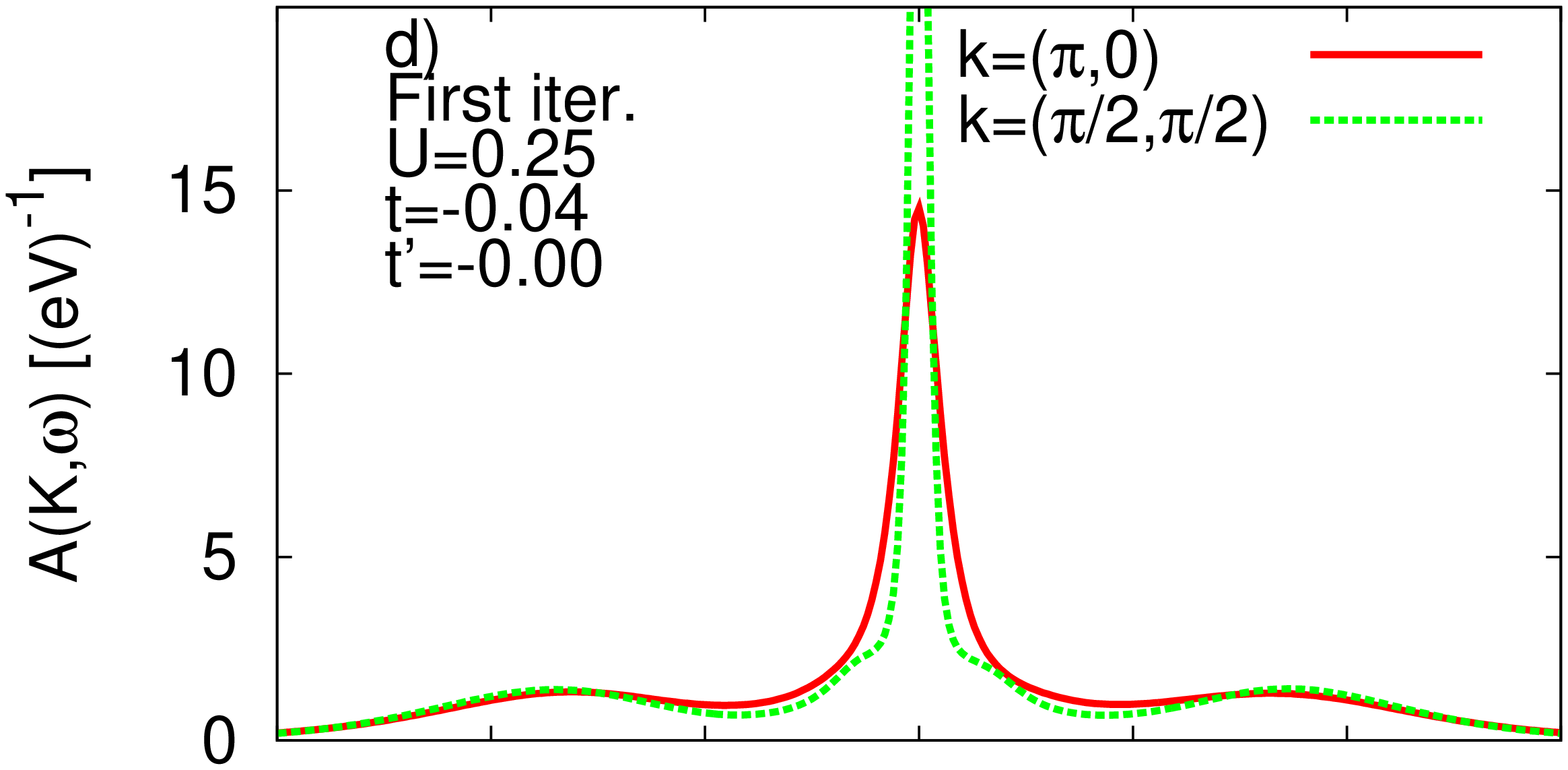}}}}
\vskip-1.88cm
{\rotatebox{-90}{\resizebox{4.9cm}{!}{\includegraphics {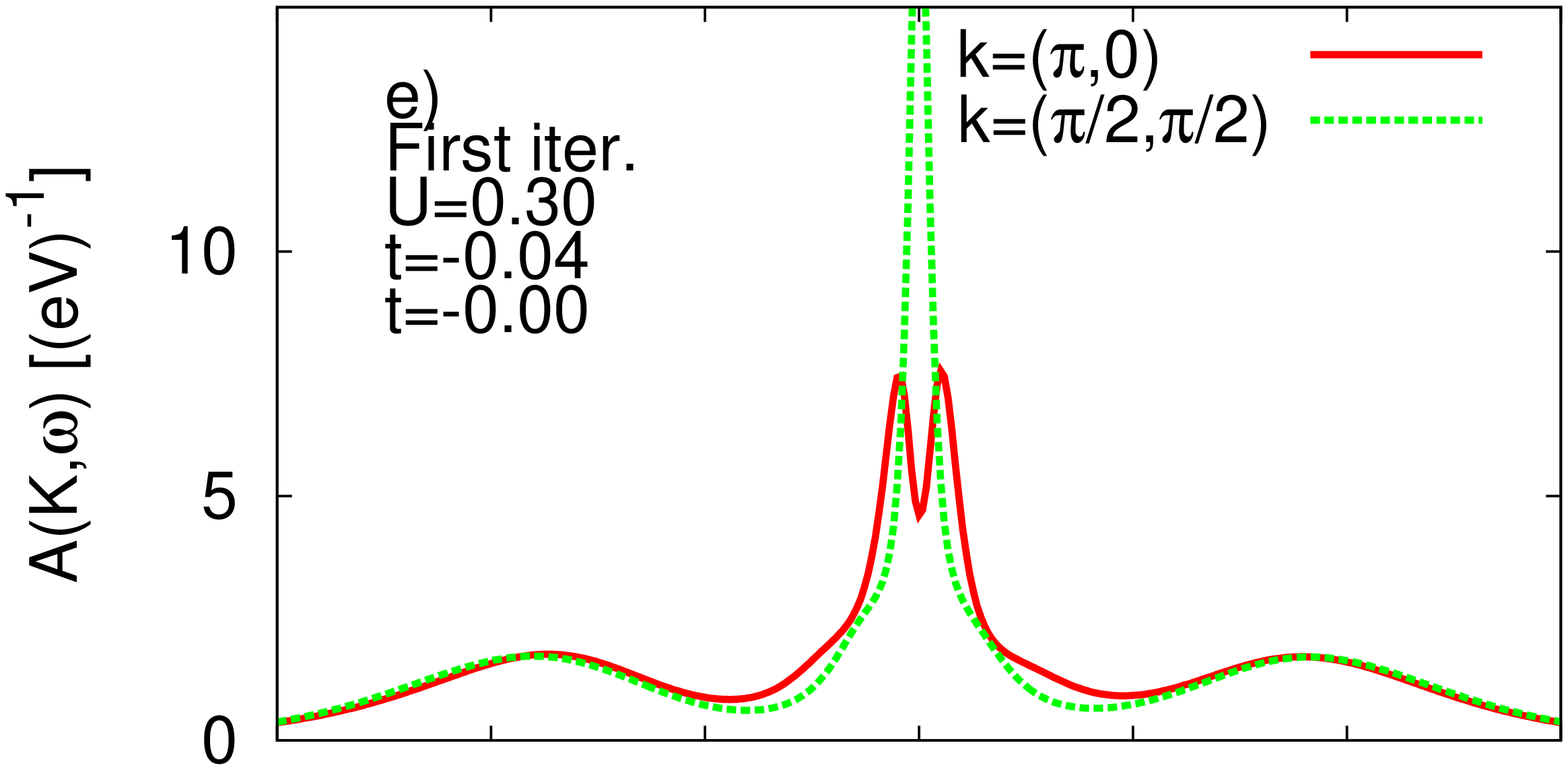}}}}
\vskip-1.37cm
\hskip0.18cm
{\rotatebox{-90}{\resizebox{4.75cm}{!}{\includegraphics {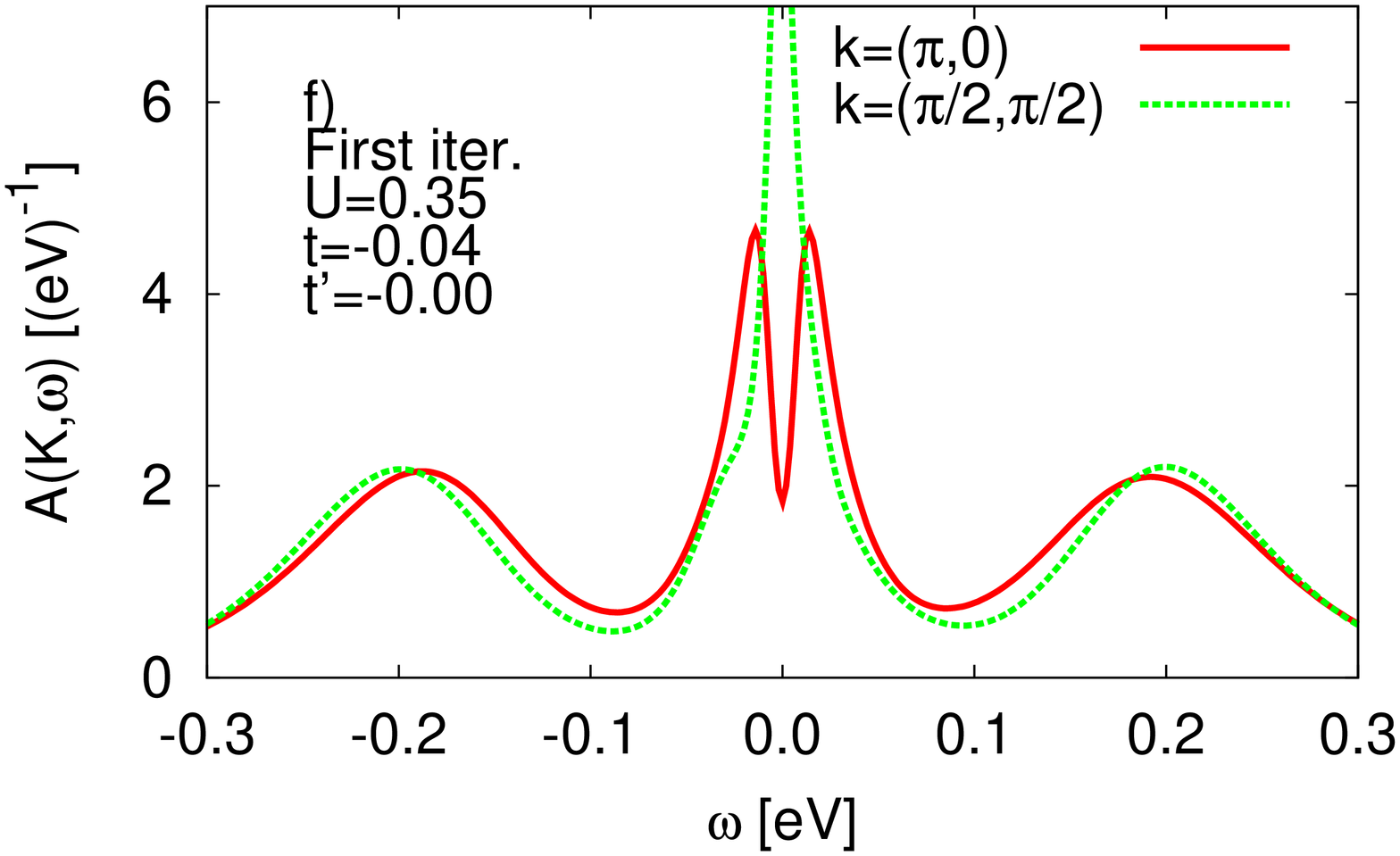}}}}
\caption{\label{fig:3.1}Spectral functions $A({\bf k}, \omega)$ for ${\bf k}=(\pi,0)$ and 
$(\pi/2,\pi/2)$. The parameters are $t=-0.04$ eV, $t'=0$, $N_c=8$ and $T=38$ K. 
Figs. a)-c) show self-consistent results, while d)-f) show results after one iteration.
The figures illustrate how in both cases a pseudogap opens for ${\bf k}=(\pi,0)$,
although for smaller values of $U$ for the self-consistent case.
}
\end{figure}
Fig.~\ref{fig:3.1} compares DCA calculations where the bath is determined self-consistently 
with  DCA calculations where the bath is determined from the noninteracting Hamiltonian,
i.e., using Eqs.~(\ref{eq:2.3}, \ref{eq:2.4}) with $\Sigma_c\equiv 0$. We refer to this as
the first iteration. The figure illustrates that a pseudogap is obtained also in the
first iteration, although for a larger $U$ than in the self-consistent calculation.
In the self-consistent case the bath can form a pseudogap itself and it is then 
difficult to separate how much the pseudogap in the spectral function is a result of 
these modifications of the bath and how much is due to other factors. In the first iteration 
the bath is fully metallic and the pseudogap formation is entirely due to the 
internal structure of the cluster and to different couplings to the bath for different 
${\bf k}$-vectors which also exist in the first iteration. As the results are iterated 
any tendency to pseudogap formation is enhanced by similar modifications in the bath.
This approach makes it easier to identify the driving force for pseudogap formation.

It is interesting to compare with dynamical mean-field theory (DMFT), i.e., $N_c=1$.
The self-consistent calculation describes the formation of a Mott insulator for appropriate
parameters, while the first iteration always gives a Kondo peak (although possibly with
very little weight) for a metallic bath. The occurrence of a pseudogap already in the
first iteration in DCA shows that the internal structure of the cluster is important 
for the formation of the pseudogap.

\begin{figure}
{\rotatebox{-90}{\resizebox{4.8cm}{!}{\includegraphics {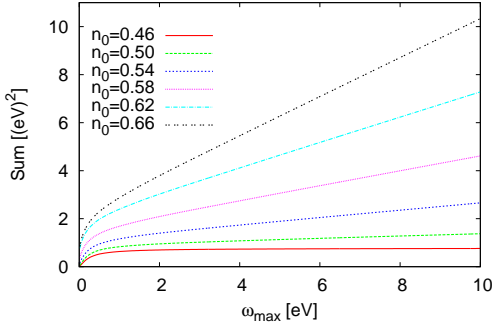}}}}                                        
\caption{\label{fig:3.2}
Sum over the absolute value of the self-energy squared. The parameters are
$t=-0.04$ eV, $t'=0.012$ eV, $N_c=8$, $U=0.3$ eV and $T=58$ K. The 
filling is $n=0.92$. The figure illustrates that the optimum $n_0=n/2=0.46$.
}
\end{figure}

In the Coulomb interaction a parameter $n_0$ is introduced [Eq.~(\ref{eq:2.1a})]. This has 
the effect of transferring weight form the self-energy to the one-particle part of the 
Hamiltonian. This plays a particular role when the calculation is stopped after the first 
iteration and the self-energy is neglected in generating the bath. As a measure of 
the importance of the self-energy we use 
\begin{equation}\label{eq:3.7a}
\sum_{i=1}^{N_c}\sum_{0<\omega_n<\omega_{\rm max}}|\Sigma_c({\bf K}_i,\omega_n)|^2.
\end{equation}
This quantity is shown in Fig.~\ref{fig:3.2}. As discussed below Eq.~(\ref{eq:2.1a}),
the choice $n_0=n/2$ leads to $\Sigma_c({\bf K}_i,\omega_n)\to 0$ for $\omega_n\to \infty$.
This is illustrated by the flatness of the curve for $n_0=0.46=(n/2)$ for large
$\omega_{\rm max}$. But this choice also reduces $\Sigma_c({\bf K}_i,\omega_n)$
substantially for small values of $\omega_n$. Therefore, $n_0=n/2$ is the 
optimum choice of $n_0$ in the first iteration calculation, where $\Sigma_c({\bf K}_i,\omega_n)$
is neglected when calculating the bath.

\subsection{Importance of coupling to bath}\label{sec:3h}

\begin{figure}
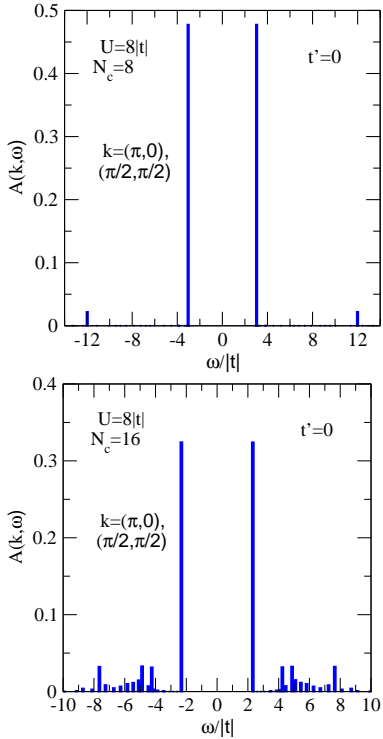

\epsfig{file=fig16a.eps,width=5cm,clip=} 
\epsfig{file=fig16b.eps,width=5cm,clip=} 
\caption{\label{fig:4.1}
Spectral function for isolated $N_c=8$ and $N_c=16$ clusters. 
The figure illustrates that the ${\bf k}=(\pi,0)$ and ${\bf k}=(\pi/2,\pi/2)$
spectra are identical. The parameters are $U=8|t_{\rm cluster}|$ and $t'=0$.}
\end{figure}

One might expect to see signs of a pseudogap already for an isolated cluster,
calculating the spectrum using exact diagonalization (ED). A finite cluster 
only has discrete peaks and we expect a gap. However, one might expect that 
either the gap would be smaller for ${\bf k}=(\pi/2,\pi/2)$ or that weight 
of the lowest binding energy peak would be larger for ${\bf k}=(\pi/2,\pi/2)$ 
than ${\bf k}=(\pi,0)$.  However, Fig.~\ref{fig:4.1} shows that the two spectra are
identical for $N_c=8$ (cluster 8A) and $N_c=16$ (cluster 16B), contradicting
this expectation. Actually, it has been shown\cite{Dagotto} that there is an
additional symmetry for these two clusters, which make the two spectra identical.

\begin{figure}
{\rotatebox{-90}{\resizebox{6.0cm}{!}{\includegraphics {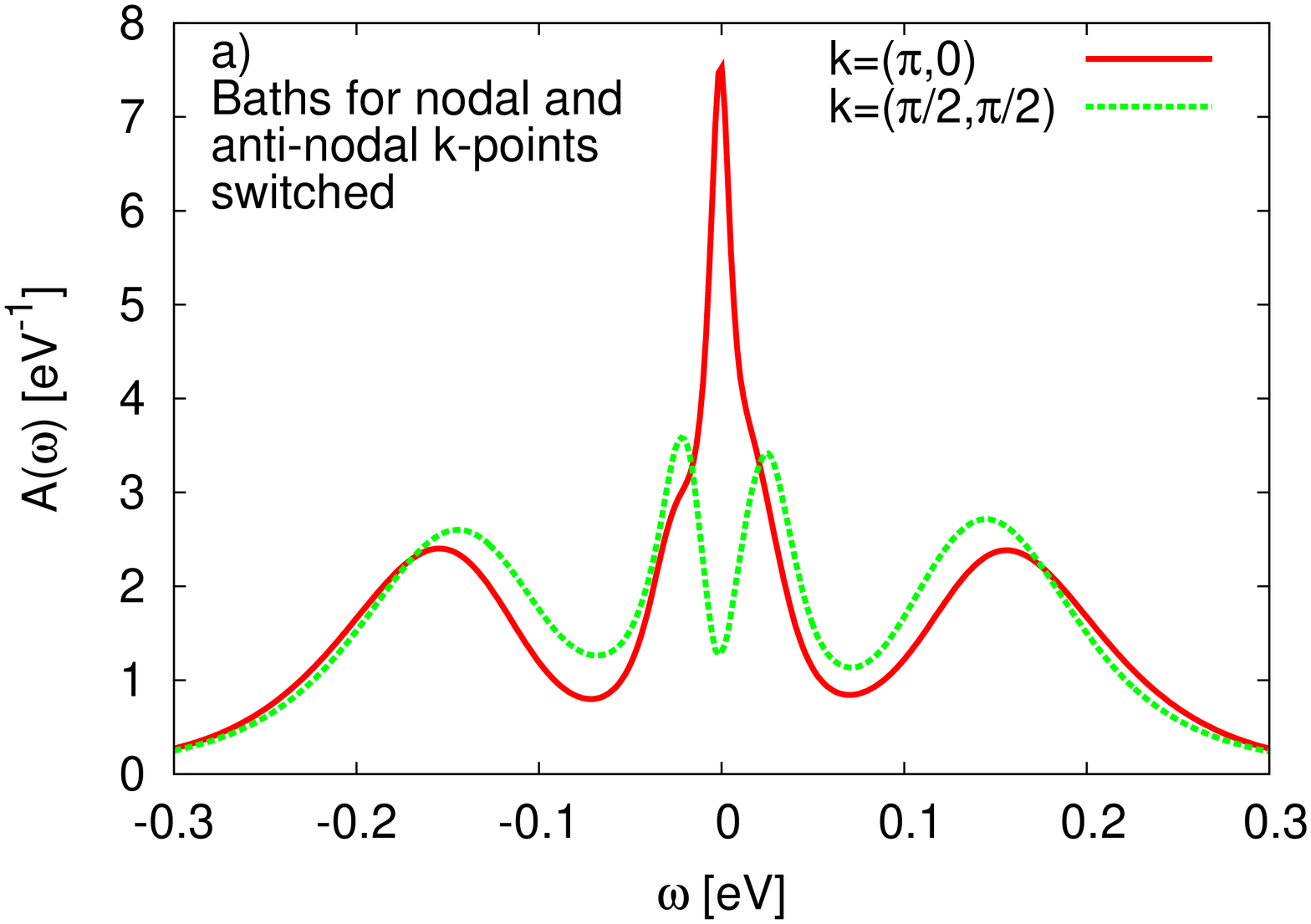}}}}

\hskip-0.3cm
{\rotatebox{-90}{\resizebox{6.2cm}{!}{\includegraphics {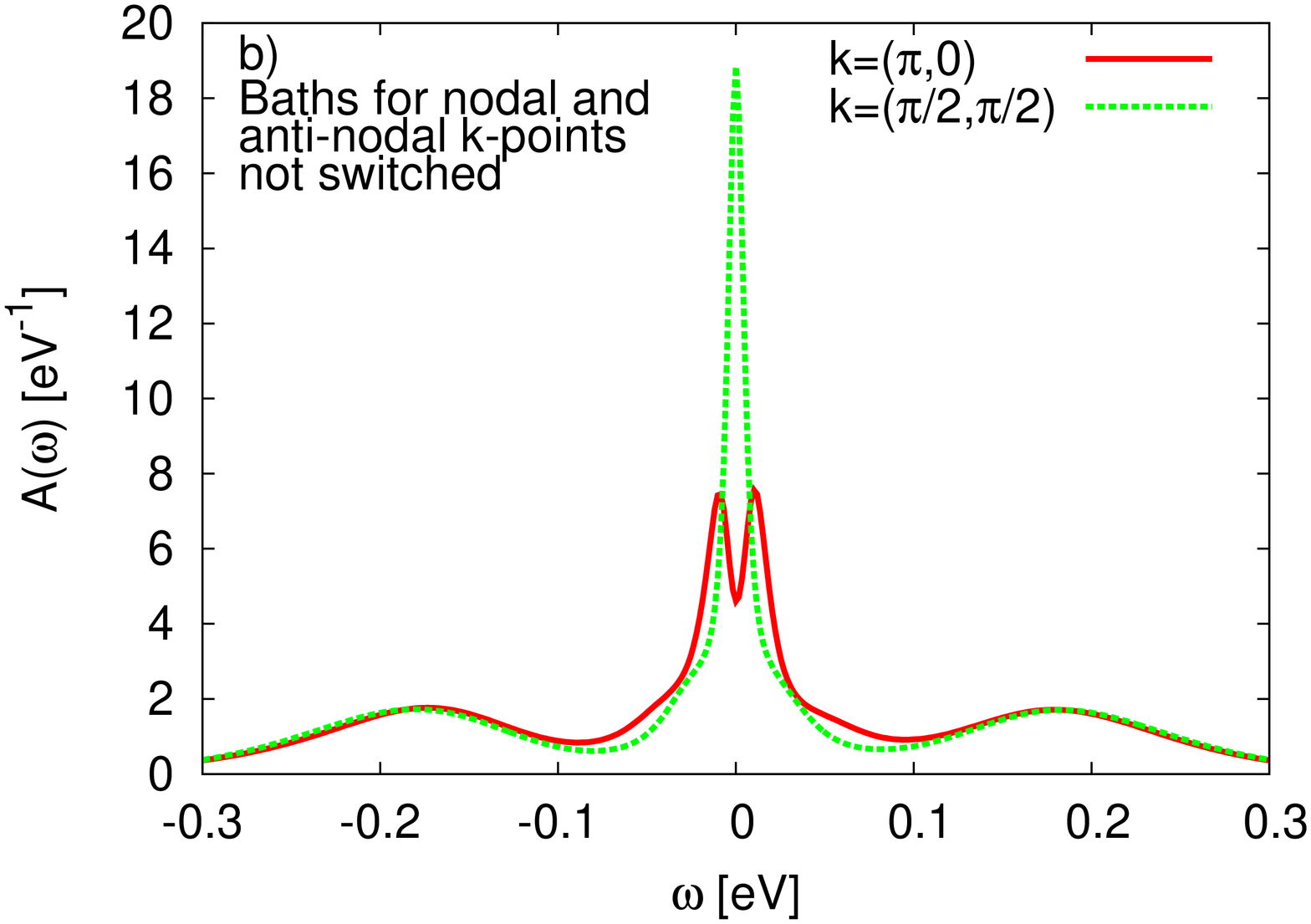}}}}
\caption{\label{fig:4.2}
Spectral function for $N_c=8$. In a) the baths
for the nodal and anti-nodal points have been switched, while in
b) this is not the case. The figure illustrates that switching the baths (a)
leads to a pseudogap for ${\bf k}=(\pi/2,\pi/2)$ instead of ${\bf k}=(\pi,0)$
as in b). The parameters are U=0.3 eV, t=-0.04 eV, $t'=0$ and $T=38.4$ K.
The results were obtained in the first iteration.}
\end{figure}

The pseudogap is, however, clearly observed in DCA calculations with $N_c=8$. The 
pseudogap must then also be related to the coupling to the bath. To test this 
we have performed DCA calculations where the couplings to the bath for the 
nodal and anti-nodal points have been switched. The results are shown in 
Fig.~\ref{fig:4.2}. Indeed, the pseudogap then appears at ${\bf k}=(\pi/2,\pi/2)$.
As discussed in Sec. \ref{sec:2} the coupling to the bath is stronger for 
${\bf k}=(\pi/2,\pi/2)$. Switching the coupling then leads to a weaker
coupling for this ${\bf k}$- point and it also leads to a pseudogap.

The numerical experiment in Sec.~\ref{sec:3f}, performing only one iteration, 
demonstrated the importance of the internal structure of the cluster. The additional 
experiment in Fig.~\ref{fig:4.2} demonstrates that the coupling to the bath
also plays a crucial role. 

For sufficiently large, $U>>t$, the excitation spectra of the cluster shows that the charge gap:
$\Delta_g=E_0(N+1)+E_0(N-1)-2E_0(N)$ differs from the gap that is extracted from photoemission.
For instance, for $U=30t$ and $N=8$ we find: $\Delta_g=24.5932|t|$, whereas the gap extracted
from the lowest energy peaks in $A({\bf k},\omega)$ is $26.588|t|$. The difference between the two gaps
comes from the fact that the lowest state with one hole is the Nagaoka state with a total spin $S=(N-1)/2$ for $U>>t$
which is $1.002t$ below the $S=1/2$ state.
Since the total spin should be conserved in the photoemission process, the $S=(N-1)/2$ state
has zero matrix element:
\begin{equation}
\langle \Psi_0(N-1)|c_{k\sigma}|\Psi_0(N) \rangle=0,
\end{equation}
and is not observed in photoemission while the $S=1/2$ state above it would have a non-zero
photoemission matrix element since it conserves the total spin of the system.

\subsection{Formulation in {\bf k}-space}\label{sec:3g}

The behavior of the spectral functions and pseudogap can be analyzed in a more transparent way 
using a momentum representation of the Hubbard model which we introduce here. 

The Coulomb part of the Hamiltonian is
\begin{equation}\label{eq:3g.1}
H_U={1\over 2}\sum_{{\bf k_1},{\bf k_2},{\bf k_3},{\bf k_4},\sigma,\sigma'}F_{{\bf k_1}{\bf k_2}{\bf k_3}{\bf k_4}}
c_{{\bf k_1}\sigma}^{\dagger}c_{{\bf k_2}\sigma'}^{\dagger}c_{{\bf k_4}\sigma'}^{\phantom \dagger}
c_{{\bf k_3}\sigma}^{\phantom \dagger}.
\end{equation}
Since the Coulomb interaction is on-site, the Coulomb integral $F$ takes the form
\begin{eqnarray}\label{eq:3g.2}
&&F_{{\bf k_1}{\bf k_2}{\bf k_3}{\bf k_4}}={U\over N^2}\sum_{l=1}^Ne^{-i({\bf k_1}+{\bf k_2}-{\bf k_3}-{\bf k_4})\cdot{\bf R}_l}\nonumber \\
&&= {U\over N}\delta_{{\bf k_1}+{\bf k_2}-{\bf k_3}-{\bf k_4}}.
\end{eqnarray}
We can then write the Hamiltonian as
\begin{eqnarray}\label{eq:3g.3}
&&H_U={U\over N}\sum_{\bf k}n_{{\bf k}\uparrow}n_{{\bf k}\downarrow}+
{ 1\over 2}{U\over N}\sum_{{\bf k}\ne{\bf k}'\sigma\sigma'}n_{{\bf k}\sigma}n_{ {\bf k}'\sigma^{}} \nonumber \\
&&+{U\over N}\sum^{'}_{{\bf k},{\bf k}',{\bf q \ne 0},\sigma \sigma'}c^{\dagger}_{{\bf k}\sigma}c^{\dagger}_{{\bf k}'\sigma'}
c_{{\bf k}'-{\bf q}\sigma'}^{\phantom \dagger}c_{{\bf k}+{\bf q} \sigma}^{\phantom \dagger} \nonumber
\end{eqnarray}
The total Hamiltonian is then
\begin{equation}\label{eq:3g.4}
H=H_U+\sum_{{\bf k}\sigma}\varepsilon_{\bf k}n_{{\bf k}\sigma},
\end{equation}
where $\varepsilon_{\bf k}$ depends on the model used.            

It is interesting to study the lowest states for for an isolated cluster with $N_c=4$. 
We label the ${\bf k}$-vectors
as ${\bf k}_1=(0,0)$, ${\bf k}_2=(\pi,0)$, ${\bf k}_3=(0,\pi)$ and ${\bf k}_4=(\pi,\pi)$.
Then the dominating configurations in the three lowest states are
\begin{eqnarray}\label{eq:4.5}
&&|1\rangle=a_1(|21,21\rangle-|31,31\rangle) \nonumber  \\                     
&&|2\rangle=a_2(|21,31\rangle-|31,21\rangle)   \\                     
&&|3\rangle=a_3(|21,21\rangle+|31,31\rangle).\nonumber                       
\end{eqnarray}
Here the first two numbers in each ket gives the two spin up electrons and the 
following two numbers the spin down electrons.
 For $|U/t|=5$ the coefficients are $a_1=0.64$, $a_2=0.61$ and $a_3=0.47$.
The terms in Eq.~(\ref{eq:4.5}) then represent 81, 74 and 45 per cent of the total weight. 

For $N_c=8$ we also include the four ${\bf k}$-vectors ${\bf k}_5=(\pi/2,\pi/2)$, 
${\bf k}_6=(-\pi/2,\pi/2)$, ${\bf k}_7=(\pi/2,-\pi/2)$ and ${\bf k}_8=(-\pi/2,-\pi/2)$.
The dominating terms in the lowest state of the isolated cluster is then
\begin{eqnarray}\label{eq:4.6}
&&|1\rangle=a(-|1368,1368\rangle+|1378,1378\rangle+|1356,1356\rangle \nonumber \\
&&-|1357,1357\rangle-|1278,1278\rangle+|1268,1268\rangle \\
&&+|1257,1257\rangle-|1256,1256\rangle), \nonumber
\end{eqnarray}
where $a=0.24$ for $U/t=5$. This corresponds to a total weight of about
0.44, meaning that there are many other terms which are not very much smaller.
However, we notice the tendency for both $N_c=4$ (in particular) and 
$N_c=8$ to have configurations where the same ${\bf k}$-vectors are 
occupied both for spin up and down. This plays an important role for the 
following discussion.

\section{Two-level model}\label{sec:4}

\begin{figure}
{\rotatebox{0}{\resizebox{4.0cm}{!}{\includegraphics {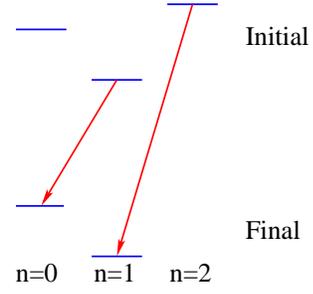}}}}
\caption{\label{fig:4.1a}Schematic level diagram for the model in Eq.~(\ref{eq:4.1}) 
in the absence of bath-cluster hopping. The number of electrons on the $c$ site is given by $n$.
The arrows show which configurations are connected in the photoemission process.  
An electron is removed from site $c$, and the initial configurations with $n=1$ and 2 are connected 
to final configurations with $n=0$ or 1, respectively.
}
\end{figure}

We first consider a very simple example, namely a two-level model. This model 
illustrates some important interference effects and the tendency to transfer 
spectral weight to the Fermi energy.\cite{GS,Delley} We introduce the model
\begin{eqnarray}\label{eq:4.1}
H_0&&=\varepsilon_b\sum_\sigma n_{b\sigma}+\varepsilon_c\sum_\sigma n_{c\sigma}
 \nonumber \\
&&+\sum_{ \sigma}V(c_{b\sigma}^{\dagger}c_{c\sigma}^{\phantom \dagger}
+c_{c\sigma}^{\dagger}c_{b\sigma})^{\phantom \dagger} +Un_{c\uparrow}n_{c\downarrow}.
\end{eqnarray}
Here $c^{\dagger}_{b\sigma}$ creates an electron on site $b$ with the spin $\sigma$,
and $n_{c\sigma}=c^{\dagger}_{c\sigma}c_{c\sigma}$. The electron can hop between 
sites $b$ and $c$ with a hopping integral $V(<0)$. On site $c$ there is a on-site 
Hubbard Coulomb integral $U$. In the language of a DCA calculation, site $c$ corresponds 
to a cluster site in a cluster with one atom and $b$ corresponds to a bath with just one 
level.The system has two electrons in the initial state.  We form three configurations, 
\begin{eqnarray}\label{eq:4.1a}
&&|0\rangle=c^{\dagger}_{{\rm b}\uparrow}c^{\dagger}_{b\downarrow}|{\rm vac}\rangle \nonumber \\
&&|1\rangle=(1/\sqrt{2})(c^{\dagger}_{{\rm b}\uparrow}c^{\dagger}_{c\downarrow}
+c^{\dagger}_{{\rm c}\uparrow}c^{\dagger}_{b\downarrow})|{\rm vac}\rangle \\
&&|2\rangle=c^{\dagger}_{{\rm c}\uparrow}c^{\dagger}_{c\downarrow}|{\rm vac}\rangle \nonumber
\end{eqnarray}
with zero, one or two electrons, respectively, on the $c$ site. Here $|{\rm vac}\rangle$
is a state with no electrons. The ground-state can then be written as
\begin{equation}\label{eq:4.2}
|\Phi\rangle=a_0|0\rangle+a_1|1\rangle+a_2|2\rangle,
\end{equation}
where $a_i$ are coefficients determined by the parameters. We now consider a photoemission 
process, where a spin down electron is removed from site $c$ of the ground-state, leading to final states 
with just one electron. This is illustrated in Fig.~\ref{fig:4.1a}. This process connects 
initial configurations with one or two electrons on the $c$ site to final configurations with zero or 
one electron, respectively, on the $c$ site.  The corresponding final states are                                              
\begin{equation}\label{eq:4.3}
|+\rangle=b_0|\tilde 0\rangle+b_1|\tilde 1\rangle, \hskip0.30cm |-\rangle=b_1|\tilde 0\rangle-b_0|\tilde 1\rangle,
\end{equation}
where $|\tilde 0 \rangle$ and $|\tilde 1\rangle$ have an electron on the $b$ or $c$ site,
respectively. For $V<0$ all coefficients are positive. The spectral weights are
\begin{equation}\label{eq:4.4}
|\langle +|c_{c\downarrow}|\Phi\rangle |^2=|{b_0a_1\over \sqrt{2}}+b_1a_2|^2, \hskip0.3cm
|\langle -|c_{c\downarrow}|\Phi\rangle |^2=|{b_1a_1\over \sqrt{2}}-b_0a_2|^2 
\end{equation}
Here $|\langle +|c_{c\downarrow}|\Phi\rangle |^2$ is the weight of the leading peak with
low binding energy and $|\langle -|c_{c\downarrow}
|\Phi\rangle |^2$ the weight of a satellite with large binding energy. The corresponding 
structures in an Anderson impurity model are the Kondo peak and the Hubbard side band, 
respectively. For $U\gg|t|$ and $\varepsilon_c \approx -U/2$, 
$a_1$ and $b_1$ are large and all other coefficients are small. 
We might then expect most of the weight in the Hubbard side band. However, there is constructive 
interference for the low binding energy peak and destructive interference for the high energy peak. 
The result is a transfer of weight towards the Fermi energy.\cite{GS,Delley} This just reflects that the 
lowest initial and final states are both bonding. This effect is important. 
To illustrate this let us consider the case when there is now hopping in the final state.
Then $b_1=1$ and $b_0=0$, and the peaks in the spectrum gives direct information
about the weights $a_1^2$ and $a_2^2$ of the initial configurations $|1\rangle$ and $|2\rangle$,
respectively. For instance, we use $\varepsilon_b=0$, $\varepsilon_c=-2$, $V=-1$ and $U=4$.
With $b_1=1$ and $b_0=0$, we would then find the weights of the low and high energy peaks to be 
0.14 and 0.36, respectively. This simply reflects that the parameters are such that the
$n=1$ configuration has most of the weight (0.72) in the initial state. This couples 
to the final configuration $n=0$, which corresponds to the excited state.
 Properly taking into account interference, on the other hand, 
reverses the weights and leads to the weights 0.33 and 0.17 for the low and high binding 
energy peaks, respectively. These effects are crucial for building up spectral weight
in the neighborhood of the Fermi energy. However, the model considered in Eq.~(\ref{eq:4.1})
cannot explain a pseudogap. To do this we need to generalize to a four-level model, where 
the cluster (here site $c$) has an internal structure. This is done in the next section.

\section{Four-level model}\label{sec:5}

\begin{figure}
{\rotatebox{-90}{\resizebox{6.0cm}{!}{\includegraphics {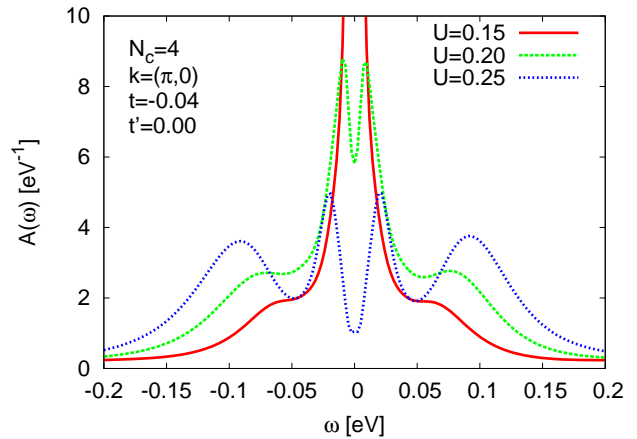}}}}
\caption{\label{fig:5.0}Spectral function for a DCA calculation with a four site cluster as a 
function of $U$. The parameters are $t=-0.04$ eV, $t'=0$ and $T=29$ K. The results were 
obtained after the first iteration.
}
\end{figure}

A DCA calculation for $N_c=4$ shows a pseudogap, even in the first iteration,
as is shown in Fig.~\ref{fig:5.0}. We can therefore use such a cluster to study the origin
of the pseudogap. Actually, we can make even more simplifications. The four site cluster
has only four wave vectors ${\bf K}$. Two of these,
${\bf K}=(0,0)$ and ${\bf K}=(\pi,\pi)$, have one-particle energies which are fairly far
from the Fermi energy at half-filling. We then expect the main physics to be
determined by ${\bf K}_1=(\pi,0)$ and ${\bf K}_2=(0,\pi)$. For simplicity we then study
a model with a cluster with just these two levels.

An alternative two-level cluster model was studied by Ferrero {\it et al.}.\cite{Ferrero09}
They introduced a two patch model in momentum space, which simulates the $(\pi,0)$ and
$(\pi/2,\pi/2)$ points on the Fermi surface. This model allows a study of the important
${\bf k}$-dependence of the pseudogap, and it was interpreted in terms of an orbital
selective Mott transition.\cite{Liebsch03,Koga04} However, such a model oversimplifies
the real situation in the interesting intermediate $U$-regime where the pseudogap opens
up in the $(\pi,0)$, $(0,\pi)$ but not in the $(\pm \pi/2,\pm \pi/2)$ sector. We find below that for such values
of $U$ there is a strong correlation between the $(\pi,0)$ and $(0,\pi)$ levels,
which is important for the opening of the pseudogap. This effect is not available
in Ref.~\onlinecite{Ferrero09} when there is a pseudogap for $(\pi,0)$ but a peak
for $(\pi/2,\pi/2)$, since the $(\pi/2,\pi/2)$ level then primarily couples to its bath and
not to $(\pi,0)$.  This effect is included in the four-level model below at the price of
not being able to describe the important ${\bf k}$-dependence. For this purpose we
later study a $N_c=8$ model, which has the proper ${\bf k}$-dependence, and show that
the essential parts of the physics in the four-level model survive in the $N_c=8$
calculation.

\begin{figure}
{\rotatebox{0}{\resizebox{!}{3cm}{\includegraphics {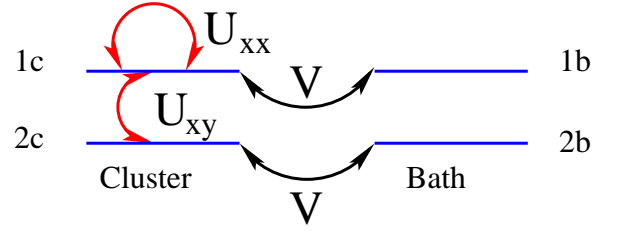}}}}
\caption{\label{fig:5.0a}Schematic picture of the four-level model.
}
\end{figure}

The cluster ${\bf k}$-states are each connected to their own baths. For simplicity,
we assume that each bath has just one level. This leads to a four-level model with
two two-fold degenerate orbitals. This model shows some important effects related to
the pseudogap and is relatively easy to analyze. The one-particle part is given by
\begin{eqnarray}\label{eq:5.1}
H_0&&=\varepsilon_c\sum_{i=1}^2\sum_\sigma n_{ic\sigma}+\varepsilon_b\sum_{i=1}^2\sum_\sigma n_{ib\sigma}
 \nonumber \\
&&+V\sum_{i \sigma}  (\psi_{ic\sigma}^{\dagger}\psi_{ib\sigma}^{\phantom \dagger}+\psi_{ib\sigma}^{\dagger}\psi_{ic\sigma}^{\phantom \dagger}).
\end{eqnarray}
where $c$ refers to cluster levels and $b$ to bath levels.
We add a Coulomb interaction with multiplet effects on the $c$ site
\begin{eqnarray}\label{eq:00}
&&H_U=U_{xx}\sum_{i=1}^2n_{ic \uparrow}n_{ic \downarrow}+U_{xy}\sum_{\sigma\sigma'}n_{1c\sigma}n_{2c\sigma'} \nonumber \\
&&+K(\psi_{1c\uparrow}^{\dagger}\psi_{1c\downarrow}^{\dagger}\psi_{2c\downarrow}^{\phantom \dagger}\psi_{2c\uparrow}^{\phantom \dagger}+h.c.)   \\
&&-K\sum_{\sigma}\psi^{\dagger}_{1c\sigma}\psi^{\dagger}_{2c\sigma}\psi_{2c\sigma}^{\phantom \dagger}\psi_{1c\sigma}^{\phantom \dagger} 
-K\sum_{\sigma}\psi^{\dagger}_{1c-\sigma}\psi^{\dagger}_{2c\sigma}\psi_{2c-\sigma}^{\phantom \dagger}\psi_{1c\sigma}^{\phantom \dagger}. \nonumber
\end{eqnarray}
The fourth term is a diagonal term favoring parallel spins, i.e., a Hund's rule coupling. 
The interaction term, $H_U$, can be expressed in the standard Kanamori form used in the context of multiorbital
band models by taking: $U_{xx}=U$ and $U_{xy}=U-2K$. Here we take a different approach.
As discussed in Appendix \ref{sec:A.1}, the level structure of the four-site cluster can be simulated by using
$U_{xx}<U_{xy}$. The lowest state of the isolated site $c$ is then a singlet.
Alternatively, we can use $U_{xx}>U_{xy}$, which leads to a triplet ground-state
for site $c$. As we will see, the physics is quite different in the two cases.
We write $U_{xx}=U-\Delta U$ and $U_{xy}=U+\Delta U$, using $\Delta U=0.03U$,
$K=0.1U$ and $V=-0.02$ eV.  We consider the symmetric case where 
$\varepsilon_b=0$ and $\varepsilon_c=-3U/2+(K-\Delta U)/2$. This model is shown schematically in Fig.~\ref{fig:5.0a}. 

The interaction part $H_U$ of our four-level model differs from 
the two-orbital Anderson model introduced\cite{Leo04} in the context of A$_3$C$_{60}$ since
pair-hopping terms are not present in that model 
so that spin rotational symmetry is broken.  
We note that the pair-hopping term is important for inducing a ground state of the type in Eq. (\ref{eq:4.5}) 
and (\ref{eq:4.6}) containing doubly-occupied sectors. 
Previous studies in the context of multiband models for transition metal oxides and fullerides
focus on the effect of ferromagnetic Hunds coupling-type exchange between electrons on different orbitals
with the same \cite{Leo04,deMedici,Werner} or different bandwidths\cite{Liebsch03,Silke} 
on the Mott transition. We note that the Hubbard model leads to particular
Coulomb interaction terms which in general differ from multiband models.

We first study the isolated $c$ site, which has the $S_z=0$ two-electron configurations
\begin{eqnarray}\label{eq:5.2}
|a\rangle=c^{\dagger}_{1c\uparrow}c^{\dagger}_{1c\downarrow}|\rm{vac}\rangle \nonumber \\
|b\rangle=c^{\dagger}_{2c\uparrow}c^{\dagger}_{2c\downarrow}|\rm{vac}\rangle \nonumber \\
|c\rangle=c^{\dagger}_{1c\uparrow}c^{\dagger}_{2c\downarrow}|\rm{vac}\rangle  \\
|d\rangle=c^{\dagger}_{2c\uparrow}c^{\dagger}_{1c\downarrow}|\rm{vac}\rangle  \nonumber
\end{eqnarray}
and the eigenenergies and states
\begin{eqnarray}\label{eq:5.3}
&&E_{1-}=U_{xx}-K;\hskip0.5cm |1-\rangle ={1\over \sqrt{2}}(|a\rangle-|b\rangle) \nonumber \\
&&E_{2-}=U_{xy}-K;\hskip0.5cm |2-\rangle ={1\over \sqrt{2}}(|c\rangle-|d\rangle)    \\         
&&E_{1+}=U_{xx}+K;\hskip0.5cm |1+\rangle ={1\over \sqrt{2}}(|a\rangle+|b\rangle) \nonumber \\
&&E_{2+}=U_{xy}+K;\hskip0.5cm |2+\rangle ={1\over \sqrt{2}}(|c\rangle+|d\rangle) \nonumber   
\end{eqnarray}
$|2-\rangle$ is a triplet, while the other states are singlets. We can also form triplet states
with $S_z=\pm 1$. 

For the sake of comparison, we note that our highest non-degenerate singlet state, $|2+\rangle$,
is equivalent to the lowest eigenstate of the two-orbital model \cite{Leo04} when the
exchange coupling, $K<0$, and the degenerate triplet $|2-\rangle$ to the lowest eigenstate 
when $K>0$. 

\begin{figure}
{\rotatebox{-90}{\resizebox{6.0cm}{!}{\includegraphics {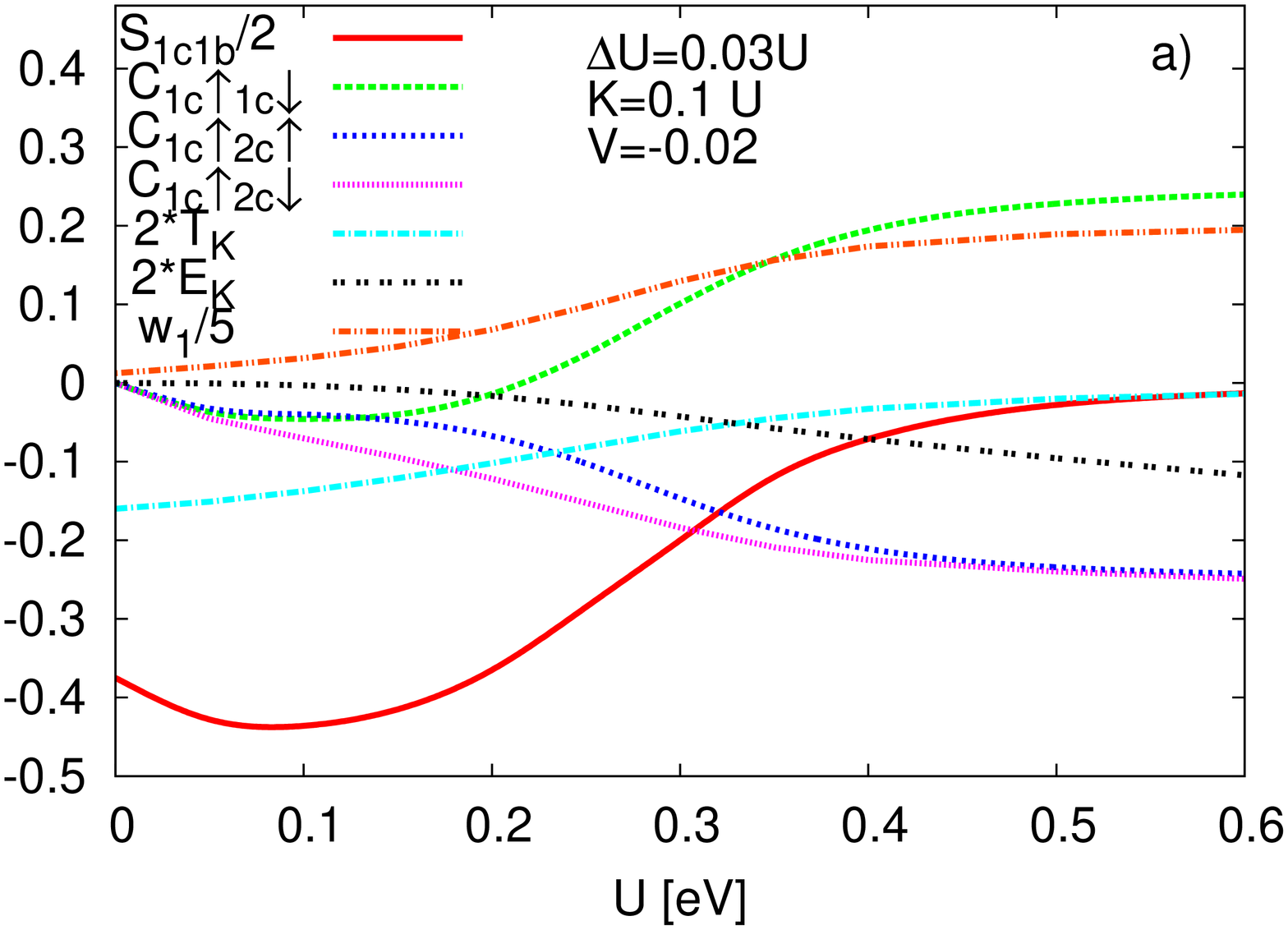}}}}                                        
{\rotatebox{-90}{\resizebox{6.0cm}{!}{\includegraphics {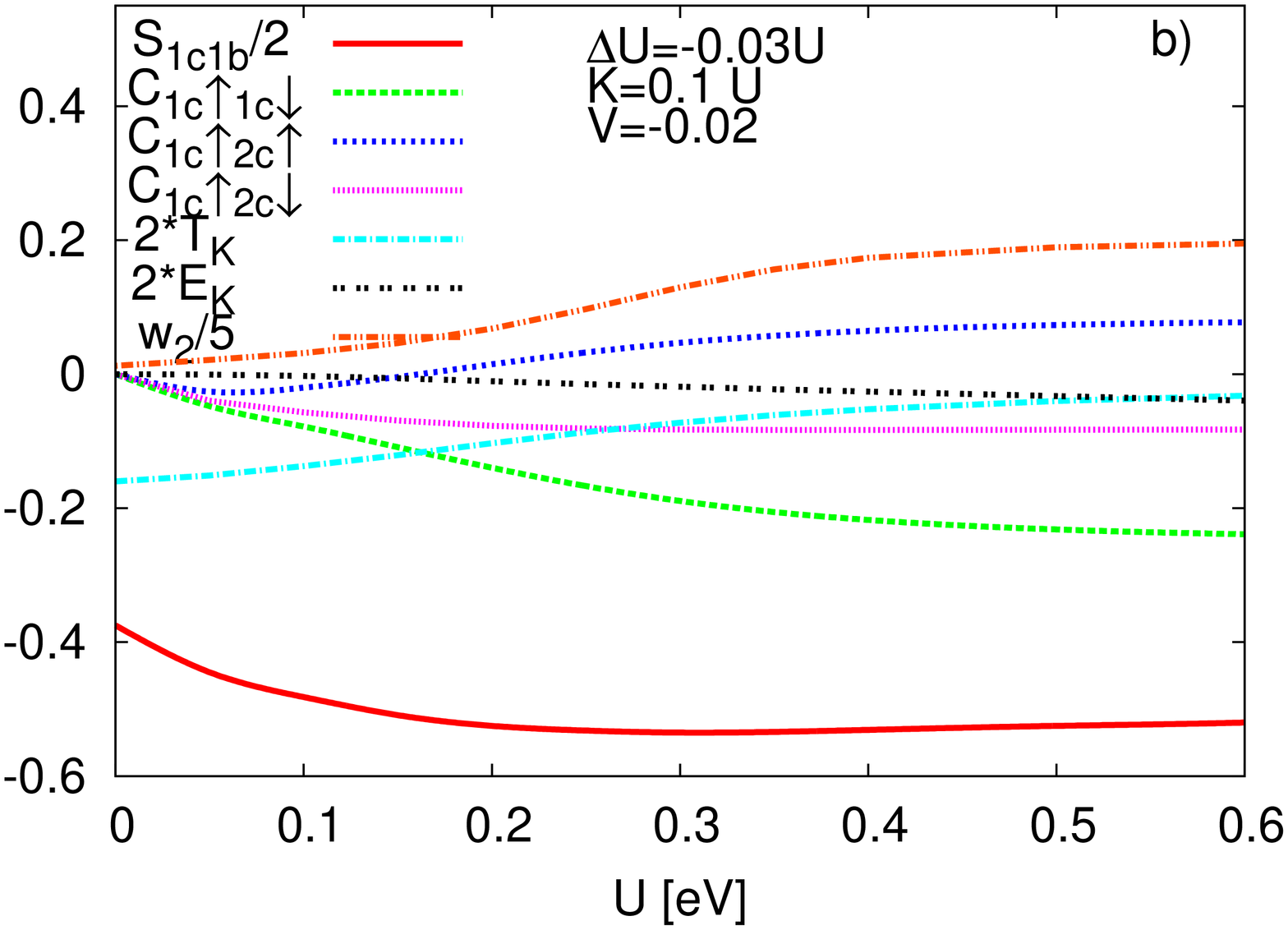}}}}                                        
\caption{\label{fig:5.1}Correlation functions $C_{1 \uparrow 1\downarrow}$, 
$C_{1\uparrow 2\uparrow}, C_{1 \uparrow2\downarrow}$, $S_{1c1b  }$ [Eq.~(\ref{eq:5.4})], hopping energy
$T_K$, ''off-diagonal'' Coulomb energy $E_K$ and weight ($w_1$ or $w_2$) of the lowest $c$ state.
In a) $\Delta U=0.03U$ and in b) $\Delta U=-0.03U$. In a) the lowest $c$ state is a singlet and in b) a triplet.
The parameters are  $K=0.1U$ and 
$V=-0.02$ eV.
}
\end{figure}

To study the nature of the ground-state, we calculate the correlation
functions $C_{i\sigma j\sigma'}$ between occupation numbers  defined in
Eq.~(\ref{eq:3.1a}) and the spin correlation function
\begin{equation}\label{eq:5.4}
S_{1c1b}=\langle \Phi| {\bf S}_{1c}\cdot {\bf S}_{1b}|\Phi \rangle,                               
\end{equation}
where $|\Phi\rangle$ is the ground-state and ${\bf S}_{1x}$ is the spin operator of 
level 1 on the site $x$.
$S_{1c1b}$ measures the spin correlation between level $1c$ and $1b$.                 
This term is expected to be important if the system is in a Kondo like state. We also
calculate the kinetic (hopping) energy $T_K$ and the ''off-diagonal'' Coulomb energy $E_K$,
due to terms connecting configurations with different occupation numbers on the site $c$. 
Finally, we calculate the weights $w_1$ and $w_2$ in the ground-state of the $c$ states 
$|1-\rangle$ and $|2-\rangle$, respectively. For $\Delta U>0(<0)$ we expect $w_1$ ($w_2$) 
to approach unity for $U\to \infty$, since hopping between the $b$ and $c$ sites is then 
completely suppressed. In Fig.~\ref{fig:5.1}a ($\Delta U>0$) we show $w_1$ and in 
Fig.~\ref{fig:5.1}b ($\Delta U<0$) $w_2$. Observe that both quantities have been divided 
by a factor of five.

We first consider $\Delta U>0$ in Fig.~\ref{fig:5.1}a. For $U$ not too large, $|T_K|$ and 
$|S_{1c1b}|$ are large,  implying that spin correlation and coupling 
between the sites are important. The correlation function 
$C_{1c\uparrow 1c\downarrow}$ is somewhat negative for small $U$ and becomes more negative 
as $U$ increases up to $U\sim 0.075$. This implies that the system starts to form a $s=1/2$ 
spin in each level $i$ on site $c$. This is consistent with a ($s=1/2$) Kondo effect between 
the level $i$ ($i=1, 2$) on the $c$ site and the same level in the bath ($b$). As $U$ is 
increased further, $C_{1c\uparrow 1c\downarrow}$ turns positive and $C_{1c\uparrow 2c\uparrow }$ 
and $C_{1c\uparrow 2c\downarrow}$ become more negative. This means that the site $c$ starts 
approaching the ground-state of the isolated $c$ site, where these correlation functions 
approach the values 0.25, At the same time the spin-flip term and the kinetic energy are 
reduced and the ''off-diagonal'' Coulomb energy $E_K$ then becomes more important.
Thus there is a competition between two different effects.
For small values of $U$ hopping is important and it is favorable to let each
level of $c$ form a Kondo like state with its bath level. The two $c$ levels
are rather weakly correlated and $|E_K|$ is small.  As $U$ is increased,
hopping is reduced and the gain from the Kondo like effect becomes smaller.
Then it becomes more favorable to let the system go into the lowest state of the
site $c$, where $|E_K|$ is rather large. That the system approaches 
the ground-state of site $c$ is illustrated by the weight $w_1$ approaching the 
value unity and the correlation functions approaching the values appropriate 
for the state $|1-\rangle$ in Eq.~(\ref{eq:5.3}).

We next consider the case when $\Delta U<0$ in Fig.~\ref{fig:5.1}b. In this
case the lowest state on site $c$ is a triplet. This triplet can couple 
to the bath in an $S=1$ Kondo effect. Thus the competition between hopping 
and $c$ site correlation is much less severe than for $\Delta U>0$, for which
the lowest state is a singlet which does not allow Kondo like effects.
The spin correlation term in Eq.~(\ref{eq:5.4}) is constructed to show the Kondo 
effect of the type found for $\Delta U>0$. However it nevertheless 
illustrates that important spin correlations are taking place for $\Delta U<0$ and all
values of $U$. In this case there is therefore no cross over from a Kondo
like system to a non Kondo system, but rather from two $s=1/2$ Kondo
systems to a $S=1$ Kondo system. While $C_{1c \uparrow 2c \uparrow}$ becomes very 
negative for large $U$ in Fig.~\ref{fig:5.1}a, it stays positive in Fig.~\ref{fig:5.1}b for large $U$,
illustrating the triplet formation on the $c$ site. Table~\ref{table:5.1} shows the weights
of the states in Eq.~(\ref{eq:5.3}) on the $c$ site in the ground-state. The state 
$|2-\rangle$ is a $S_z=0$ triplet state. We therefore show the total weight of 
the triplet states (``Triplet''), including the $S_z=\pm 1$. The table also shows 
the overlap between the ground-states for the actual value of $U$ and for $U=0$.      

\begin{widetext}

\begin{table}
\caption{\label{table:5.1}Weights of the different states of the $c$ cluster (Eq. \ref{eq:5.3}) in the ground state 
of the four-level model, overlap (``over'') to the noninteracting state, 
results for the different terms entering Eq.~(\ref{eq:6.7}) and weights of the low-binding energy peaks 
in photoemission. The estimated peak weights are obtained through the approximate expression Eq.~(\ref{eq:6.7}) which
are compared with the exact calculated weights. A near cancellation between Terms 1 and 2 leads to the vanishingly small
calculated photoemission weight of Peak 1 associated with the pseudogap.  
The parameters are $V=-0.02$ eV, $\Delta U=\pm 0.03U$, $K=0.1U$, $U_{xx}=U-\Delta U$ and $U_{xy}=U+\Delta U$.}
\begin{tabular}{cccccccccccccccc}
\hline
\hline
$U$ & $\Delta U$ &  \multicolumn{4}{c}{States}& Over.   & Term 1 & Term 2 & Term 3 & \multicolumn{2}{c}{Peak 1} & Peak 2 & \multicolumn{2}{c}{Peak 3} &Peak 4 \\
         &          & $|1-\rangle$ &  Triplet      & $|1+\rangle$  & $|2+\rangle$ &  & & & & Est & Calc &  Calc & Est & Calc & Calc \\
\hline
2.0 & .06  & 1.00   & 0.000           & 0.000          & 0.000        &.42 & -.00592 & .00597    & .00023       & 2.6$\times$10$^{-7 }$   & 1.3$\times$10$^{-7 }$& .00049        &  .00014   & .00023    & .00028     \\
1.0 & .03  & 1.00   & 0.002           & 0.000          & 0.000        &.44 & -.0117  & .0121   & .00195          &  1.9$\times$10$^{-5 }$  & 0.9$\times$10$^{-5 }$& .0213         &  .00056   & .00102    & .00116     \\
0.5 & .015 & 0.95   & 0.035           & 0.012          & 0.001        &.57 & -.0211  & .0265     & .0193         &  .0019           & .0008         & .012          &  .0023    & .0057     & .0055      \\
0.25& .0075& 0.49   & 0.348           & 0.024          & 0.016        &.89 & -.0161  & .0786    & .141           &  .12             & .052          & .044          &  .0090    & .029      & .019      \\
0.10& .003 & 0.16   & 0.36            & 0.072          & 0.060        &.99 & -.0032  & .234    & .422           & 1.16             & .31           & .005          &  .056     & .017      & .007       \\
0.00& .000 & 0.06   & 0.19            & 0.063          & 0.063        & 1  &         &                &            &                  & .500          & .000          &           & .000      & .000       \\
2.0 & -.06  & .000    & 1.00            & .000           &.000          &.71 & -.00206& .0105     & .0087        & .00066           & .00065         & .00052        &     & .00076    & .00095     \\
1.0 & -.03  & .002    & 0.99            & .000           &.000          &.73 & -.00397& .0211     & .0193         & .0031            & .0029          & .0020         &   & .0031     & .0037      \\
0.5 & -.015 & .018    & 0.94            & .001           &.003          &.80 & -.00682& .0432     & .0516         & .0195            & .0151          & .0072         &    & .013      & .013       \\
0.25& -.0075& .084    & 0.75            & .015           &.021          &.91 & -.00765& .0925     & .152          & .151             & .086           & .014          &     & .075      & .022       \\
0.1& -.008 & .108    & 0.42            & .055           &.066          &.99 & -.00280 & .248     & .441         & 1.27             & .30            & .005          &     & .023      & .007       \\
\hline
\end{tabular}
\end{table}

\end{widetext}

In the limit of very large $U$ and for $\Delta U>0$, the ground-state takes the form
\begin{equation}\label{eq:5.5}
|\Phi\rangle={1\over 2}(c^{\dagger}_{1c\uparrow}c^{\dagger}_{1c\downarrow}-
c^{\dagger}_{2c\uparrow}c^{\dagger}_{2c\downarrow})
(c^{\dagger}_{1b\uparrow}c^{\dagger}_{1b\downarrow}-
c^{\dagger}_{2b\uparrow}c^{\dagger}_{2b\downarrow})|{\rm vac}\rangle.
\end{equation}
The $c$ part of the wave function then has singlet character.
If instead $\Delta U<0$ and $U$ is very large, we obtain
\begin{eqnarray}\label{eq:5.6}
&&|\Phi\rangle={1\over \sqrt{3}}[c^{\dagger}_{1c\uparrow}c^{\dagger}_{2c\uparrow}
c^{\dagger}_{1b\downarrow}c^{\dagger}_{2b\downarrow} 
-{1\over 2}(c^{\dagger}_{1c\uparrow}c^{\dagger}_{2c\downarrow}+c^{\dagger}_{1c\downarrow}c^{\dagger}_{2c\uparrow})\nonumber \\
&&\times (c^{\dagger}_{1b\uparrow}c^{\dagger}_{2b\downarrow}+c^{\dagger}_{1b\downarrow}c^{\dagger}_{2b\uparrow})
+c^{\dagger}_{1c\downarrow}c^{\dagger}_{2c\downarrow}
c^{\dagger}_{1b\uparrow}c^{\dagger}_{2b\uparrow}]|{\rm vac}\rangle.
\end{eqnarray}
Here the $c$ part has triplet character. These results confirm the discussion above.

\begin{figure}
{\rotatebox{-90}{\resizebox{6.0cm}{!}{\includegraphics {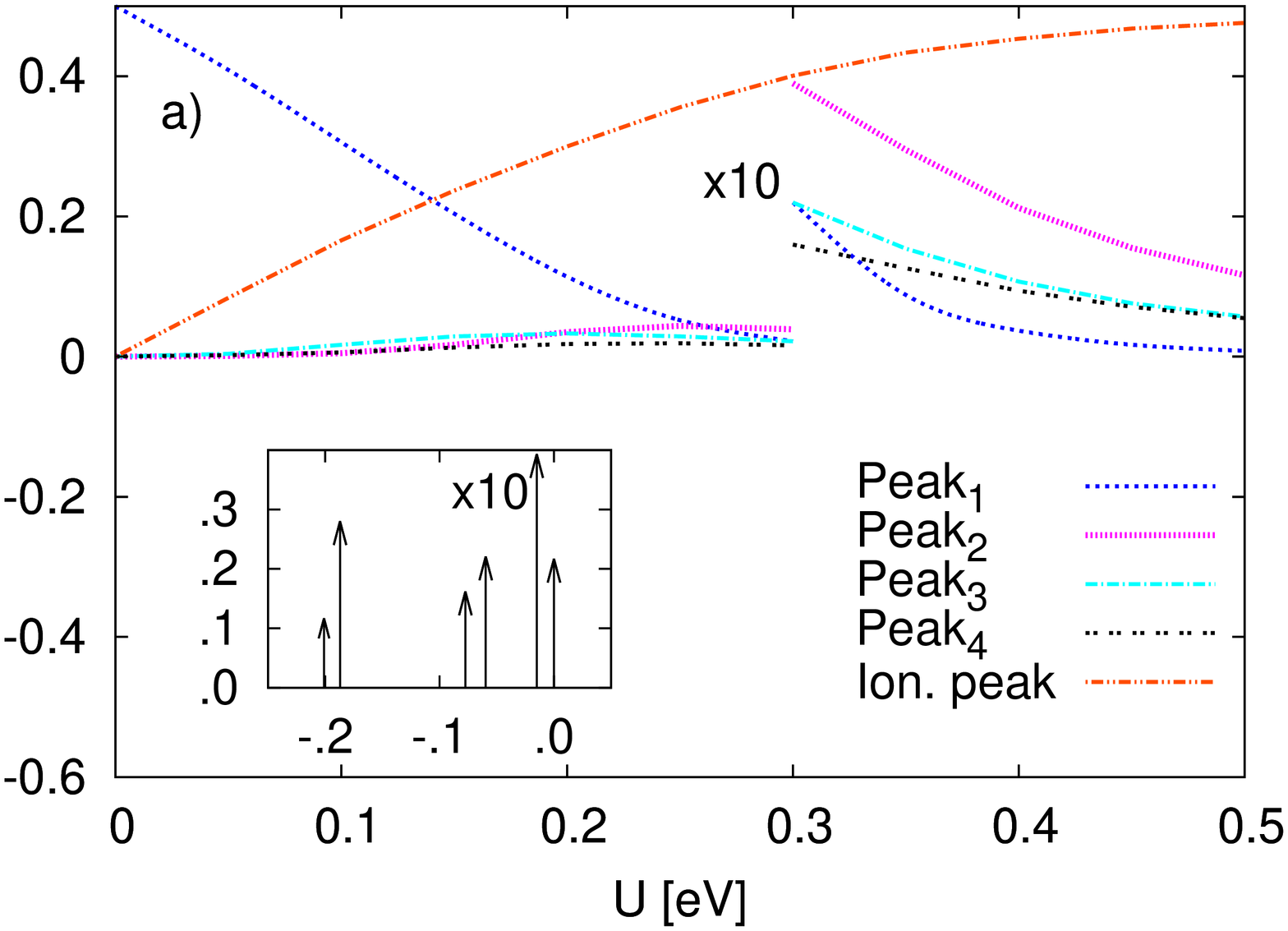}}}}                                
{\rotatebox{-90}{\resizebox{6.0cm}{!}{\includegraphics {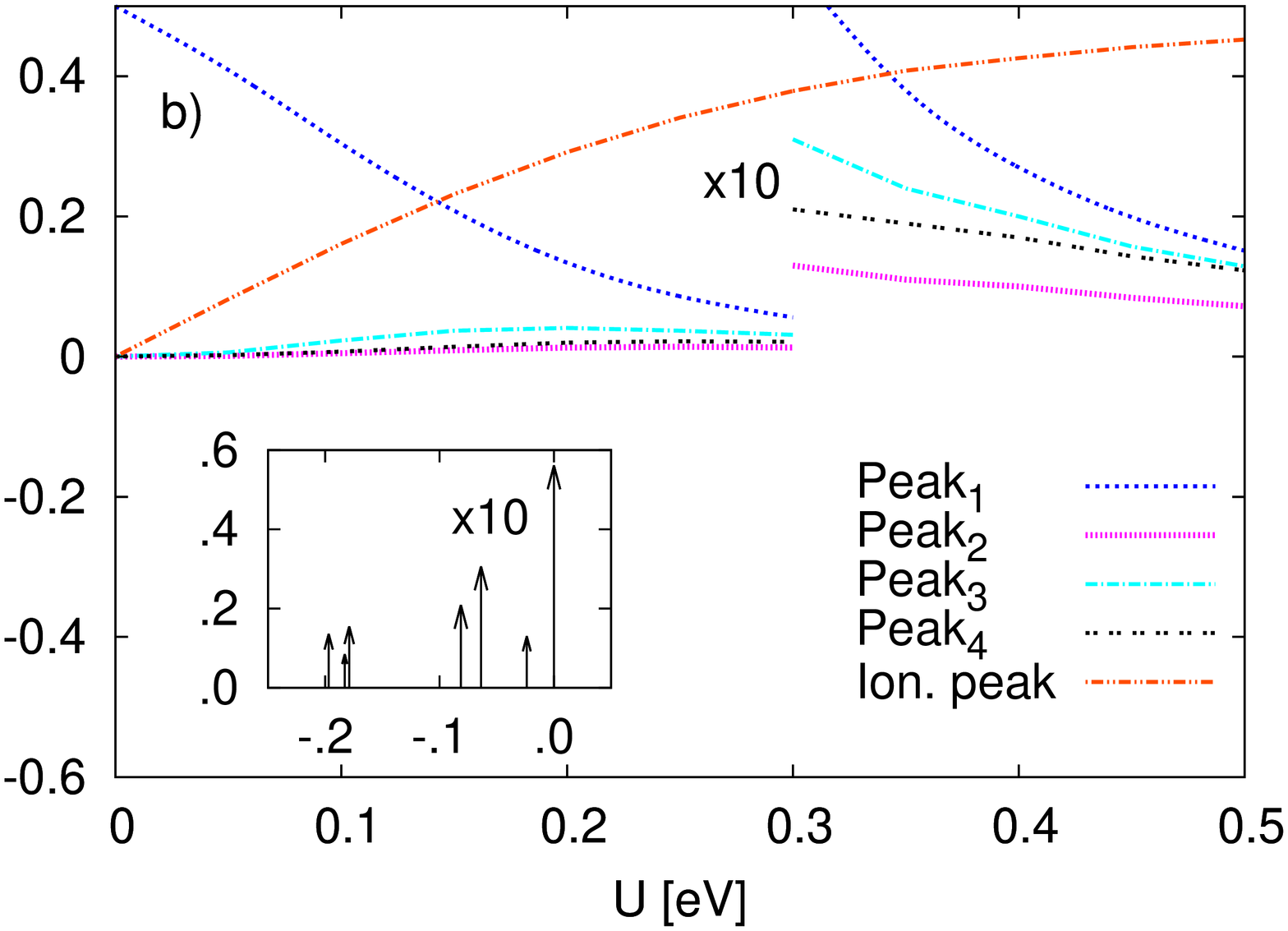}}}}                                
\caption{\label{fig:5.1a}
Spectral weights for the four-level model. The curves show the weights of different peaks as 
a function of $U$. The inset shows the photoemission spectrum for $U=0.3$ eV.
The four low-binding energy peaks have been multiplied by a factor of 10 in the insets and 
main figures.
In a) $\Delta U=0.03U$ and in b) $\Delta U=-0.03U$. In a) the lowest $c$ state is a singlet and in b) a triplet.
The parameters are  $K=0.1U$ and 
$V=-0.02$ eV.  }
\end{figure}
The photoemission part of the spectral function has four peaks close to the Fermi
energy and in addition peaks approximately at $U/2$ binding energy (for the symmetric 
case), as shown by the insets of Fig.~\ref{fig:5.1a}. The four peaks close to $\omega=0$
correspond to final states with a large weight of the neutral states on the $c$ site.                                 
For $\Delta U>0$, the peak with the second lowest binding energy corresponds to a final
state which has mainly triplet character on the cluster, while for the other three
low-lying peaks the character is mainly singlet. The $x^2-y^2$ pairing correlation
$P_{x^2-y^2}$ in Eq.~(\ref{eq:7.4}) is large for the lowest neutral cluster state and
zero for the others. The peaks corresponding to the exact neutral states then
results from a process where a singlet $x^2-y^2$ has been broken up. For $\Delta U<0$,
on the other hand, the lowest neutral state is a triplet.

The peaks at roughly $-U/2$ corresponding to a Hubbard band. The total
weight of the photoemission spectrum is 0.5. The calculated weights of these peaks are
shown in the main part of Fig.~\ref{fig:5.1a}. For small values of $U$ most of the weight is
in the leading peak. As $U$ is increased, the Hubbard band peak grows and the leading peak
is reduced. This is along the lines discussed in Sec.~\ref{sec:4} for the two-level model.
For $\Delta U>0$, however, as $U$ is further increased the leading peak also looses
weight relative to the other peaks with small binding energy. The difference can be orders
of magnitude. This corresponds to a pseudogap. For $\Delta U<0$ and $U$ not too large, on the other hand,
the lowest binding energy peak has a larger weight than other low binding energy peaks,
and there is no pseudogap. The difference between the two cases is crucial for understanding
the pseudogap, and it is analyzed in the next section.

\section{Interference}\label{sec:6}

In this section we analyze why there is a pseudogap for $\Delta U>0$ but 
not $\Delta U<0$ in the four-level model. The derivation below applies to 
the case when there is an infinite bath and we make no particular assumptions 
about the cluster. We consider a case where $U>>|V|$, where $V$ is the hopping
to the bath. The ground-state wave function can then conveniently 
be described in terms of configurations with an integer number of electrons on 
the cluster. We consider an integer average filling $n_0$. We treat photoemission processes 
where an electron with the quantum number $\nu$ is removed from the cluster. 
Here $\nu$ is a combined index containing a spin index and other indices, 
e.g., a ${\bf K}$ index. The part of the wave function corresponding to 
$n_0$ electrons can be split in two pieces
\begin{equation}\label{eq:6.1}
|\Phi_{n_0}\rangle=|0\rangle+|1\rangle,
\end{equation}
where
\begin{equation}\label{eq:6.2}
c_{\nu}|0\rangle=0 \hskip0.3cm {\rm and} \hskip0.3cm c^{\dagger}_{\nu}c^{\phantom \dagger}_{\nu}|1\rangle=|1\rangle.
\end{equation}
We assume there is hopping between the bath and the cluster determined by
\begin{equation}\label{eq:6.3}
H_0=\sum_{\nu}\sum_{\varepsilon}(V_{\nu\varepsilon}^{\phantom \dagger}c^{\dagger}_{\nu}c^{\phantom \dagger}_{\nu \varepsilon}+{\rm h.c.})
\end{equation}
In the limit of very weak hopping, we approximate the part of the
ground-state corresponding to $n_0+1$ electrons as
\begin{eqnarray}\label{eq:6.4}
|\Phi_{n_0+1}\rangle=-\sum_{\nu'}\sum_{\varepsilon}{V_{\nu' \varepsilon}\over \Delta E_+-\varepsilon}
c^{\dagger}_{\nu'}c^{\phantom \dagger}_{\nu'\varepsilon}|\Phi_{n_0}\rangle,
\end{eqnarray}
where we have approximated the energy difference between cluster states with one extra
electron (hole) and the lowest neutral state as $\Delta E_+$ ($\Delta E_-$).
We write
\begin{equation}\label{eq:6.4a}
|\Phi\rangle \approx |\Phi_{n_0-1}\rangle+|\Phi_{n_0}\rangle+|\Phi_{n_0+1}\rangle.
\end{equation}
This approach looks like nondegenerate perturbation theory, and for the case of a nondegenerate
ground-state of the isolated cluster this is also the case. However, in Kondo problem treated below,
the states $|0\rangle$ and $|1\rangle$, which are assumed to be known, in general contain more information than can be obtained
from perturbation theory, and therefore the treatment goes beyond perturbation theory. For the 
Kondo problem this is, nevertheless, a crude approximation.
We now apply the operator $c_{\nu}$
\begin{eqnarray}\label{eq:6.5}
&&c_{\nu}(|\Phi\rangle)=c_{\nu}|1\rangle-\sum_{\varepsilon}{V_{\nu \varepsilon}\over \Delta E_+-\varepsilon}
c^{\phantom \dagger}_{\nu\varepsilon}|0\rangle \nonumber \\
&&-\sum_{\nu'(\ne \nu)}\sum_{\varepsilon}{V_{\nu' \varepsilon}\over \Delta E_+-\varepsilon}
c_{\nu }c^{\dagger}_{\nu'}c^{\phantom \dagger}_{\nu'\varepsilon}|1\rangle.
\end{eqnarray}
The term $c_{\nu}|\Phi_{n_0-1}\rangle$ can be neglected for $U\gg |V|$.
We form a final state
\begin{eqnarray}\label{eq:6.6}
&&|\tilde{0}\rangle=c_{\nu\varepsilon_F}(|0\rangle+|1\rangle)\\
&&-\sum_{\varepsilon}{V_{\nu \varepsilon}\over \Delta E_-+\varepsilon}
c^{\dagger}_{\nu\varepsilon}c^{\phantom \dagger}_{\nu}c_{\nu\varepsilon_F}|1\rangle, \nonumber \\
&&-\sum_{\nu'(\ne\nu)}\sum_{\varepsilon}{V_{\nu' \varepsilon}\over \Delta E_-+\varepsilon}
c^{\dagger}_{\nu'\varepsilon}c^{\phantom \dagger}_{\nu'}c_{\nu\varepsilon_F  }^{\phantom \dagger}|\Phi_{n_0}\rangle \Big) \nonumber. 
\end{eqnarray}
where $\varepsilon_F=0$ is the Fermi energy. This state is closely related to the initial ground-state
except for a hole at the Fermi energy. We use this as an approximation to the lowest final state.
We could also have added a term corresponding to $n_0+1$ electrons on the cluster, but this would not have contributed to the 
matrix elements below.
Then the amplitude for the photoemission process is
\begin{eqnarray}\label{eq:6.7}
&&\langle \tilde{0} |c_{\nu}|\Phi\rangle=
\sum_{\varepsilon}V_{\nu\varepsilon}({\langle 1| c_{\nu\varepsilon_F}^{\dagger}c_{\nu\varepsilon}^{\phantom \dagger}|1\rangle \over \Delta E_-+\varepsilon}
-{\langle 0| c_{\nu\varepsilon_F}^{\dagger}c_{\nu\varepsilon}^{\phantom \dagger}|0\rangle \over \Delta E_+-\varepsilon})
\\
&&-\sum_{\nu'(\ne\nu)\varepsilon} V_{\nu' \varepsilon}({\langle 0|c^{\dagger}_{\nu\varepsilon_F }c^{\dagger}_{\nu'}c^{\phantom \dagger}_{\nu'\varepsilon }c^{\phantom \dagger}_{\nu}|1\rangle\over \Delta E_-+\varepsilon} 
+{\langle 0|c^{\dagger}_{\nu\varepsilon_F }c^{\phantom \dagger}_{\nu} c^{\dagger}_{\nu'}c^{\phantom \dagger}_{\nu'\varepsilon }|1\rangle\over \Delta E_+-\varepsilon}).  \nonumber   
\end{eqnarray}
If $(|0\rangle-|1\rangle)$ is an excited state of the cluster, we can form a final state analogous to 
Eq.~(\ref{eq:6.6}) and obtain an expression for the weight similar to Eq.~(\ref{eq:6.7}), except for 
some signs. In cases where there is a strong negative interference for the coupling to the lowest final 
state there might be a strong positive interference for the coupling to this satellite.

\begin{figure}
{\rotatebox{-0}{\resizebox{7.0cm}{!}{\includegraphics {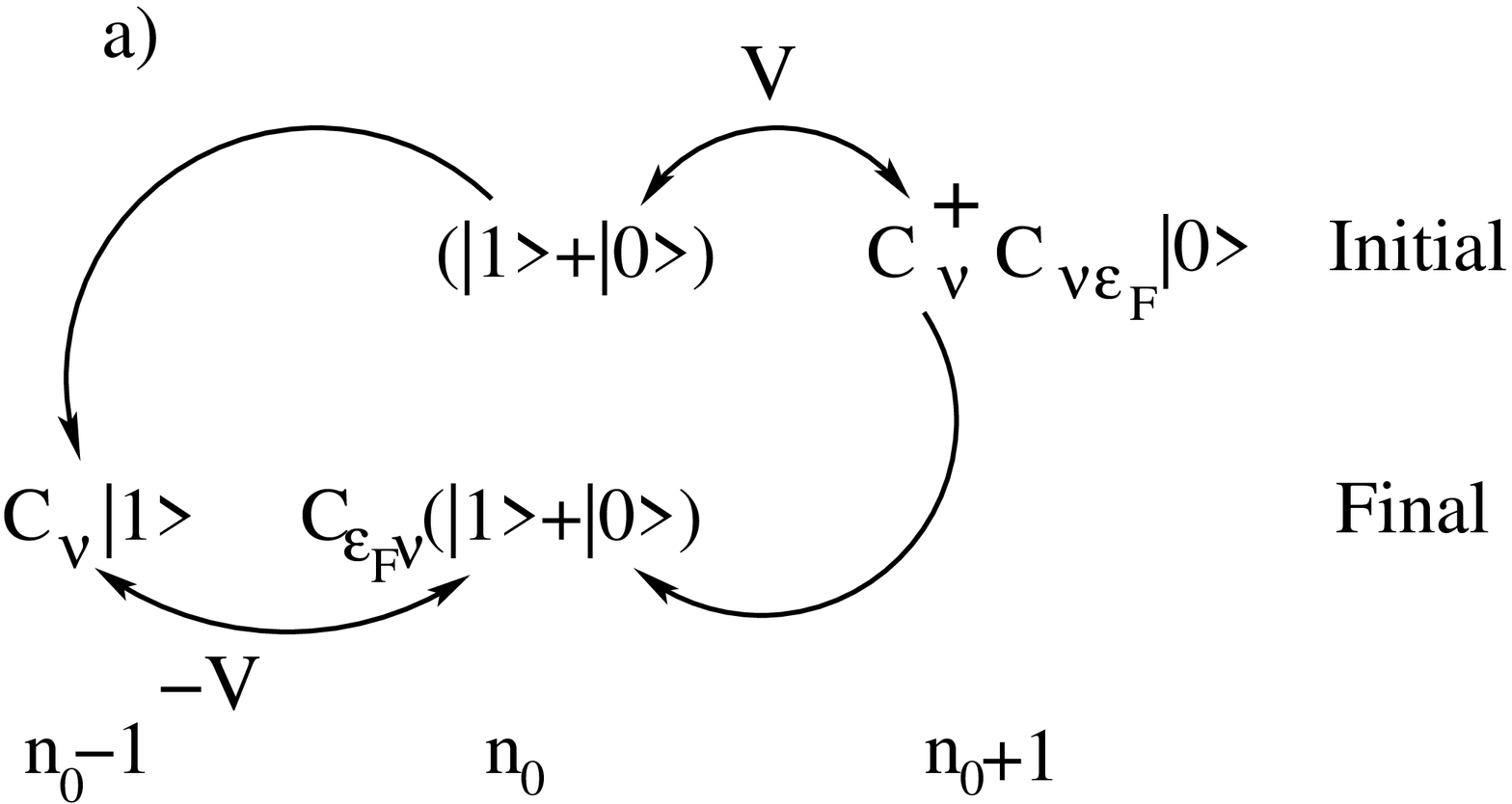}}}}                                        
{\rotatebox{-0}{\resizebox{7.0cm}{!}{\includegraphics {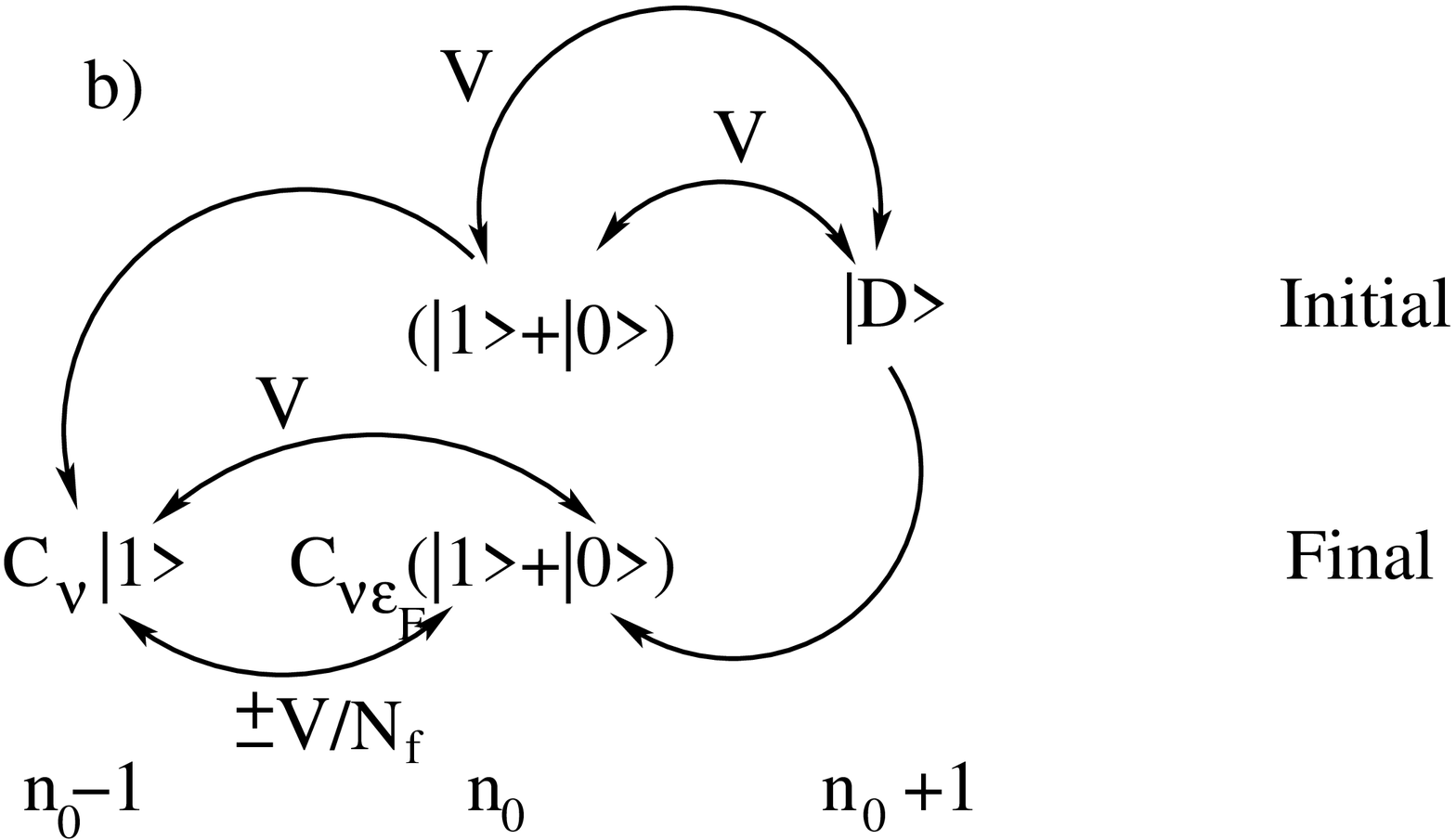}}}}                                        
\caption{\label{fig:6.1}Schematic illustration of initial and final states 
for the case when the lowest state $|0\rangle+|1\rangle$ of the isolated cluster
 is a) nondegenerate or b) degenerate in the $U>>|V|$ limit. An electron with the quantum number $\nu$ is removed in a 
photoemission process. $|0\rangle$ contains no electron $\nu$ while $|1\rangle$ does.
Figure a) (nondegenerate case) illustrates the negative interference between two channels for reaching 
the final state, corresponding to the two first terms in Eq.~(\ref{eq:6.7}).
Figure b) (degenerate case) illustrate how spin-flip terms (third and fourth 
terms in Eq.~(\ref{eq:6.7})) interfere positively with path going through double occupancy. D
stands for doubly occupied states. The negative amplitude 
between $n_0-1$ and $n_0$ sectors of the final state is suppressed by $1/N_f$ 
with respect to positive spin-flip contributions leading to overall positive interference 
between different paths.  
}
\end{figure}

We can now discuss the problem of a cluster coupling to a large bath in the limit of very weak coupling. 
We first consider the case when a good zeroth order approximation is given by a direct product
of the lowest cluster state and the bath filled up to the Fermi energy (no holes present), having the case in mind when
the lowest cluster state is nondegenerate. The Ansatz in 
Eq.~(\ref{eq:6.6}) for the lowest final state should then also be appropriate for a very large bath.  
To leading (zeroth) order in $V/U$ the factor $\langle 1| c_{\varepsilon_F\nu}^{\dagger}c_{\varepsilon\nu}^{\phantom \dagger}|1\rangle 
=\delta_{{\varepsilon_F},\varepsilon}\langle 1|1\rangle$ in the first term in Eq.~(\ref{eq:6.7}) and similar results
for the second term. Similarly the term 
$c^{\dagger}_{\varepsilon_F \nu}c^{\phantom \dagger}_{\varepsilon \nu'}$ ($\nu\ne \nu'$)
acting on the bath in the third and fourth terms in Eq.~(\ref{eq:6.7}) give zero contribution. 
To leading order in $V/U$ the third  and fourth terms are then zero and for the symmetric case the
first and second term cancel. To leading order the lowest binding energy peak then has zero weight.
This is schematically illustrated in Fig.~\ref{fig:6.1}a. The decisive point is that the matrix elements
\begin{eqnarray}\label{eq:6.7a}
&&\langle 0|Hc^{\dagger}_{\nu}c_{\nu \varepsilon_F}|0\rangle=V_{\nu \varepsilon_F} \\
&&\langle 1| c^{\dagger}_{\nu} H c_{\nu \varepsilon_F}|1\rangle=-V_{\nu \varepsilon_F} \nonumber
\end{eqnarray}
have different signs. Then the two paths of reaching the final state have destructive interference.

If $(|0\rangle-|1\rangle)$ is an excited state of the isolated cluster, there is instead constructive 
interference for the final state $c_{\varepsilon_F\nu}(|0\rangle-|1\rangle)$, since the first and second 
terms in Eq.~(\ref{eq:6.7}) now have the same sign.

This kind of scenario may be expected when the lowest cluster state is nondegenerate and the bath 
is infinite. It is interesting to compare with the case $\Delta U>0$ for the four-level 
model, since the isolated cluster ground-state is then nondegenerate. For this finite system the 
ground-state of the cluster plus bath [Eq.~(\ref{eq:5.5})] is more complicated than suggested above, 
even for $U\to \infty$. The model nevertheless illustrates the ideas above.  Table~\ref{table:5.1} 
shows the calculated weight of the lowest binding energy peak (``Peak 1'') and the estimate according 
to Eq.~(\ref{eq:6.7}). Indeed both weights are extremely small for a large value of $U$. 
In table \ref{table:5.1} the terms occurring in Eq. (\ref{eq:6.7}) are explicitly 
evaluated showing the cancellation between the first and second terms explaining the vanishing 
weight of ``Peak 1''. The estimate Eq.~(\ref{eq:6.7}) is only correct to leading order in $V/U$, and to this order the weight 
is zero. The weight is then determined by higher order terms, which are not correct in Eq.~(\ref{eq:6.7})
and, in addition, Eq.~(\ref{eq:6.7}) refers to an infinite system. Nevertheless, 
the weight is of the right order of magnitude. 
Table~\ref{table:5.1} also shows the weight the peak corresponding to the final state $(|0\rangle-|1\rangle)$
(``Peak 3'') together with an estimate analogous to Eq.~(\ref{eq:6.7}). For very large values of $U$ this 
estimate is fairly accurate, given that it was done assuming an infinite bath. 
The destructive interference at the lowest binding energy 
can also be understood in terms of phase shift arguments.\cite{Ferrero07} However, 
the formalism used here contains additional information about the spectral weights
of higher energy excitations determining the pseudogap.

The case $\Delta U<0$ in the four-level model is quite different.
In this case the ground-state is Kondo-like. To obtain a better understanding of this case,
we study a simple model of Ce compounds,\cite{GS} based on the Anderson impurity model.
\begin{eqnarray}\label{eq:6.8}
&&H=\sum_{\nu=1}^{N_f}\int_{-B}^B\varepsilon c^{\dagger}_{\nu\varepsilon }c^{\phantom \dagger}_{\nu \varepsilon} d\varepsilon
+\varepsilon_f\sum_{\nu=1}^{N_{f}}c^{\dagger}_{\nu}c^{\phantom \dagger}_{\nu} \nonumber\\
&&+V\sum_{\nu=1}^{N_f}\int d\varepsilon(c^{\dagger}_{\nu\varepsilon } c^{\phantom \dagger}_{\nu}+{\rm h.c.})+U\sum_{\nu <\nu'}n_{\nu}n_{\nu'} \nonumber 
\end{eqnarray}
Here the impurity has a $N_f$-fold  degeneracy, including orbital and spin degeneracies.
For the ground-state we  have the basis states
\begin{eqnarray}\label{eq:6.9}
&&|\phi_0\rangle=\Pi_{\varepsilon}^{\rm occ}\Pi_{\nu=1}^{N_f}c^{\dagger}_{\nu\varepsilon }|{\rm vaccum}\rangle \nonumber \\
&&|\varepsilon \rangle={1\over \sqrt{N_{f}}}\sum_{\nu=1}^{N_{f}}c^{\dagger}_{\nu}c^{\phantom \dagger}_{\nu\varepsilon }|\phi_0\rangle \\
&&| {\varepsilon\varepsilon' }\rangle={1\over \sqrt{N_{f}(N_{f}-1)}}
\sum_{\nu,\nu'}
c^{\dagger}_{\nu}c_{\nu\varepsilon }^{\phantom \dagger}c^{\dagger}_{\nu'}c_{\nu'\varepsilon}^{\phantom \dagger}|\phi_0\rangle  \nonumber
\end{eqnarray}
For the states $| {\varepsilon\varepsilon' }\rangle$ we require that $\varepsilon>\varepsilon'$ to avoid overcompleteness.
In the limit of a very large $N_f$, the ground-state is written as\cite{GS}
\begin{equation}\label{eq:6.10}
|\Phi\rangle=A\Big(|\phi_0\rangle+\int_{-B}^0 d\varepsilon a(\varepsilon)|\varepsilon \rangle+\int_{-B}^0 d\varepsilon \int_{-B}^\varepsilon d\varepsilon'  b(\varepsilon,\varepsilon' )|\varepsilon \varepsilon'\rangle\Big).
\end{equation}
We remove an electron with quantum number $\nu$. The relevant final state basis states are
\begin{eqnarray}\label{eq:6.11}
&&|\tilde{\varepsilon}\rangle=c_{\nu \varepsilon}|\phi_0\rangle \nonumber \\
&&|\tilde{\varepsilon\varepsilon'1 }\rangle={1\over \sqrt{N_f-1}}\sum_{\nu'\ne \nu} c^{\dagger}_{\nu'}c^{\phantom \dagger}_{\nu'\varepsilon'}c^{\phantom \dagger}_{\nu \varepsilon }
|\phi_0\rangle \\
&&|\tilde{\varepsilon\varepsilon'2 }\rangle= c^{\dagger}_{\nu}c^{\phantom \dagger}_{\nu\varepsilon'}c^{\phantom \dagger}_{\nu \varepsilon}
|\phi_0\rangle,
\end{eqnarray}
where $\varepsilon>\varepsilon'$ for the state $|\tilde{\varepsilon\varepsilon'2 }\rangle$. To leading order in $(1/N_f)$ $|\tilde{\varepsilon\varepsilon'2 }\rangle$ can be neglected,
and to this order we can form a final state 
\begin{equation}\label{eq:6.13}
|\varepsilon_{F} {\rm final}\rangle=c_{\nu \varepsilon_{F}} |\Phi\rangle.
\end{equation}
The overlap to the ground-state is then
\begin{eqnarray}\label{eq:6.14}
&&\langle \varepsilon_{F} {\rm final}|c_{\nu}|\Phi\rangle \\
&&={A^2\over \sqrt{N_f}}\Big(a(\varepsilon_F)
+\sqrt{N_f-1 \over N_f}\int_{-B}^0 d\varepsilon a(\varepsilon)b(\varepsilon_F,\varepsilon)\Big),\nonumber 
\end{eqnarray}
where $\epsilon_F=0$ is assumed. Importantly, this expression shows positive interference, in contrast to what we found for
the $\Delta U>0$ four-level model. For instance, if $V<0$ all coefficients $A$, $a$ and $b$
are positive. If the states $|\tilde{\varepsilon\varepsilon'2 }\rangle$ are considered, it is not possible to write 
a final state in the simple form (\ref{eq:6.13}), and there are then also contributions to $|\langle \varepsilon_{F} {\rm final}|c_{\nu}|\Phi\rangle|^2$
with negative interference. These contributions are, however, one order higher in $(1/N_f)$.

We analyze the result as above, and form the states $|0\rangle$ and $|1\rangle$ as defined above[Eq.~(\ref{eq:6.1})].
\begin{eqnarray}\label{eq:6.15}
&&|0\rangle={A\over \sqrt{N_f}}\int_{-B}^0 d\varepsilon a(\varepsilon)\sum_{\nu'(\ne \nu)}c^{\dagger}_{\nu'}c^{\phantom \dagger}_{\nu'\varepsilon } |\phi_0 \rangle \nonumber \\
&&|1\rangle={A\over \sqrt{N_f}}\int_{-B}^0 d\varepsilon a(\varepsilon)c^{\dagger}_{\nu}c^{\phantom \dagger}_{\nu \varepsilon } |\phi_0 \rangle.
\end{eqnarray}
It turns out that the second to fourth terms in Eq.~(\ref{eq:6.7}) give the leading contributions 
in $(1/N_f)$, and therefore we first consider the second term in the limit when $-\varepsilon_f\gg B$.
Since $|0\rangle$ has no holes with quantum number $\nu$ below $\varepsilon_F$,
this term only contributes for $\varepsilon=\varepsilon_F$. Then we have
\begin{eqnarray}\label{eq:6.15a}
&&-\sum_{\varepsilon}V {\langle 0| c_{\nu\varepsilon_F}^{\dagger}c_{\nu\varepsilon}^{\phantom \dagger}|0\rangle \over U+\varepsilon_f-\varepsilon_F} \\
&&=-{A^2(N_f-1)\over N_f}{V\over U+\varepsilon_f-\varepsilon_F}\int_{-B}^0 d\varepsilon a(\varepsilon)^2. \nonumber
\end{eqnarray}
This corresponds to the process $|0\rangle \to |D\rangle \to c_{\nu\varepsilon_F}|0\rangle$ in Fig.~\ref{fig:6.1}.
The matrix element of the Hamiltonian is proportional to $V$ and the removal of the electron $\nu$ does not involve a minus sign. 
Next we consider the third term in Eq.~(\ref{eq:6.7}).
\begin{eqnarray}\label{eq:6.17}
&&-\sum_{\nu'(\ne \nu)}\int_{-B}^0d\varepsilon V{\langle 0| c^{\dagger}_{\nu\varepsilon_F }
c^{\dagger}_{\nu'}c^{\phantom \dagger}_{\nu'\varepsilon }c^{\phantom \dagger}_{\nu}
|1\rangle \over -\varepsilon_f+\varepsilon} \nonumber \\ 
&&=-{A^2(N_f-1)V \over N_f}\int_{-B}^0d\varepsilon {1 \over -\varepsilon_f+\varepsilon }a(\varepsilon) a(\varepsilon_F)  
\end{eqnarray}
This term corresponds to the transition $|1\rangle \to c_{\nu}|1\rangle \to  c_{\nu\varepsilon_F}|0\rangle$.
It describes spin flips and it has no correspondence for the nondegenerate case. It shows how the initial cluster neutral state
with an electron $\nu$ is connected to a final cluster neutral state with an electron $\nu'\ne \nu$ via an intermediate state with no electron on the cluster.
In the same way the fourth term describes a spin-flip term connecting initial and final neutral states via a doubly occupied state. 
As in Eq.~(\ref{eq:6.15a}), the matrix element 
of the Hamiltonian is in both cases proportional to $V$ and the removal of the electron $\nu$ does not involve a minus sign. These two terms interfere 
positively with each other and with the second term.

Finally, we consider the first term in Eq.~(\ref{eq:6.7}).
\begin{eqnarray}\label{eq:6.16}
&&\int_{-B}^0d\varepsilon{V\over -\varepsilon_f+\varepsilon}\langle 1|c^{\dagger}_{\nu \varepsilon_F}c_{\nu \varepsilon } |1\rangle  \nonumber \\
&&={A^2 V\over N_f}\Big({1 \over -\varepsilon_f+\varepsilon_F}\int_{-B}^{0^-} d\varepsilon a^2(\varepsilon)- \\
&&\int_{-B}^{0^-} d\varepsilon{1 \over -\varepsilon_f+\varepsilon}
a(\varepsilon)a(\varepsilon_F)\Big) \nonumber
\end{eqnarray}
The state $c_{\nu \varepsilon_F}|1\rangle$ corresponds to the final basis state $|\tilde{\varepsilon\varepsilon'2 }\rangle$.
Coupling to the $n_0-1$ state, the cluster electron can now fill one out of two holes for $\varepsilon\le\varepsilon_F$. This differs
from the states we considered in the nondegenerate state, where there was only one hole. If the hole at $\varepsilon_F$ is filled,
the sign is the same as in the nondegenerate case, and a negative interference with the following three terms in Eq.~(\ref{eq:6.7})
is obtained, due to the commutations of fermion operators. If instead the other hole is filled, a constructive interference is obtained.
Both terms in Eq.~(\ref{eq:6.16}) are one order smaller in $(1/N_f)$ 
than the following terms in Eq.~(\ref{eq:6.7}). Therefore there is in total a strong constructive interference.

This is illustrated in
Table~\ref{table:5.1} for the case $\Delta U<0$. For large values of $U$ the theory above describes
accurately the weight of the lowest binding energy peak. In this case $|0\rangle-|1\rangle$ is not a
proper excited state. Actually, it is not even orthogonal to $|0\rangle+|1\rangle$. Therefore 
we have no simple estimate of the weight of the third peak in this case.

\begin{figure}
{\rotatebox{-90}{\resizebox{6.0cm}{!}{\includegraphics {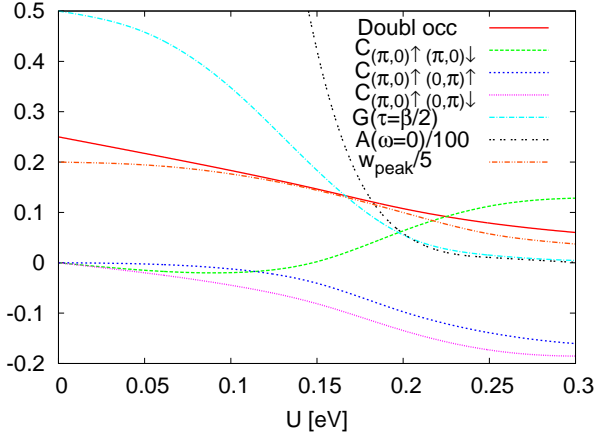}}}}
\caption{\label{fig:6.2}DCA results for a four site cluster.  
The double occupancy, $C_{(\pi,0) \uparrow (\pi,0)\downarrow}$, 
$C_{(\pi,0) \uparrow (0,\pi)\uparrow}$, $C_{(\pi,0) \uparrow (0,\pi)\downarrow}$ 
[defined as in Eq.~(\ref{eq:3.1a})], $G(\tau=\beta/2)$, $A(\omega=0)/100$
and the weight $w_{\rm peak}$ of the peak within 0.05 eV of the Fermi 
energy are shown. The parameters are $t=-0.04$ eV, $t'=0$ and $T=29$ K. 
The results were obtained after the first iteration.
}
\end{figure}

We can now analyze a $N_c=4$ DCA calculation in the light of the results above. Fig.~\ref{fig:6.2}
shows some results. We first consider the correlation functions $C_{(\pi,0) \uparrow (\pi,0)\downarrow}$, 
$C_{(\pi,0) \uparrow (0,\pi)\uparrow}$, $C_{(\pi,0) \uparrow (0,\pi)\downarrow}$, defined as in 
Eq.~(\ref{eq:3.1a}). The correlation function $C_{(\pi,0) \uparrow (\pi,0)\downarrow}$ shows 
how initially for small $U$ there is a negative correlation between spin up and spin down 
electrons in the $(\pi,0)$ orbital. This corresponds to a Kondo like coupling of each orbital
to its bath and a large amplitude $A(\omega=0)$ of the peak at the Fermi energy.
This is also illustrated by 
\begin{equation}\label{eq:6.19}
G(\tau=\beta/2)=\int_{-\infty}^{\infty} {e^{-\beta \omega/2} \over 1+e^{-\beta \omega}}
A(\omega)d\omega,
\end{equation}
which gives an average of $A(\omega)$ around $\omega=0$ over an energy range of the order 
of $\pi T$. As $U$ increases, $C_{(\pi,0) \uparrow (\pi,0)\downarrow}$ turns positive and
at the same time $C_{\rm (\pi,0) \uparrow (0,\pi)\uparrow}$ and 
$C_{\rm (\pi,0) \uparrow (0,\pi)\downarrow}$ become very negative. This is the correlation
 also observed for the four-level model in Fig.~\ref{fig:5.1} and it corresponds to the formation
of a cluster singlet where the main configuration is
\begin{equation}\label{eq:6.18}
{1\over \sqrt{2}}[c^{\dagger}_{(\pi,0)\uparrow}c^{\dagger}_{(\pi,0)\downarrow}
-c^{\dagger}_{(0.\pi)\uparrow}c^{\dagger}_{(0,\pi)\downarrow}]
c^{\dagger}_{(0,0)\uparrow}c^{\dagger}_{(0,0)\downarrow}|{\rm vaccum}\rangle,
\end{equation}
equivalent to the ground-state of the four-level model, except for the 
additional double occupation of  ${\bf k}=(0,0)$. At this point the peak weight 
drops dramatically, shown by both $A(\omega=0)$ and $G(\tau=\beta/2)$. 

The figure also shows the weight $w_{\rm peak}$ of the peak integrated over a
range $\pm 0.05$ eV around the Fermi energy. This corresponds to 
the drop described in Sec.~\ref{sec:4} for the two-level model. As $U$ is
increased the double occupancy is reduced. As described in Fig.~\ref{fig:4.1a}
the coupling to neutral final cluster states is then weaker and there
is less weight close to the Fermi energy. Correspondingly the interaction
in the final states between charged and neutral final configurations is
weaker. This reduces the weight $w_{\rm peak}$. This is, however, just
a part of the physics. More important are the interference effects discussed 
in this section. When the dominating part of the ground-state starts to become
a nondegenerate state on the cluster, interference effects move weight from
the Fermi energy to side bands relatively close to the Fermi energy (much
closer than the Hubbard side bands) but away from the Fermi energy.
This is illustrated in Fig.~\ref{fig:6.2}, by the much faster drop
by $G(\tau=\beta/2)$ and, in particular, $A(\omega=0)$.

\section{Eight site cluster}\label{sec:7}
\subsection{Unfrustrated case}

\begin{figure}
{\rotatebox{-90}{\resizebox{6.0cm}{!}{\includegraphics {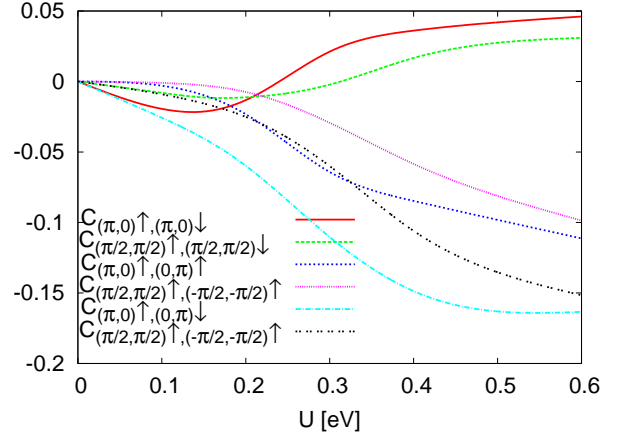}}}}
\caption{\label{fig:7.2} Correlation functions $C_{ij}$ defined in analogy to 
Eq.~(\ref{eq:3.1a}) as a function of $U$ for an eight site cluster. The parameters 
are $t=-0.04$ eV, $t'=0$ and $T=29$ K. The results were obtained in the first iteration.
}
\end{figure}
As seen in, eg., Fig.~\ref{fig:3.1}, the embedded eight-site cluster gives
a pseudogap for ${\bf K}=(\pi,0)$ but no pseudogap for ${\bf K}=(\pi/2,\pi/2)$
for intermediate values of $U/W$. This is also the case for the first iteration.
We now discuss the physics of this in relation to the results of the previous sections.
Fig.~\ref{fig:7.2} shows correlation functions defined as in Eq.~(\ref{eq:3.1a}).
We first consider $C_{(\pi,0)\uparrow (\pi,0)\downarrow}$ (full red curve).
In analogy with $C_{(\pi,0)\uparrow,(\pi,0)\downarrow}$ in Fig.~\ref{fig:6.2} for the four-site cluster,
the curve first turns negative, indicating that a Kondo state is formed. For $U$ 
of the order of 0.24 eV it turns positive, indicating the formation of a localized state in 
the cluster, again in analogy with the four-site cluster. The correlation of 
${\bf K}=(\pi,0)$ $ \uparrow$  to ${\bf K}=(0,\pi)$ $\uparrow$  or $\downarrow$ is strongly
negative, also as for the four-site cluster. We next focus on the correlation between 
${\bf K}=(\pi/2,\pi/2)$ $\uparrow$ and ${\bf K}=(\pi/2,\pi/2)$ $\downarrow$ (dotted green curve).
This curve behaves in a similar way as the corresponding curve for ${\bf K}=(\pi,0)$ 
but it is displaced towards higher values of $U$. This is easy to understand in terms of the 
stronger coupling to the bath for ${\bf K}=(\pi/2,\pi/2)$ than for ${\bf K}=(\pi,0)$,
illustrated in, e.g., Fig.~\ref{fig:2.2}. This stronger coupling makes it favorable 
to keep a Kondo like state for ${\bf K}=(\pi/2,\pi/2)$ up to a larger value of $U$.
For still large values of $U$, however, it is more favorable to form a localized cluster state 
also for this ${\bf K}$ value. 

\begin{figure}
{\rotatebox{-90}{\resizebox{6.0cm}{!}{\includegraphics {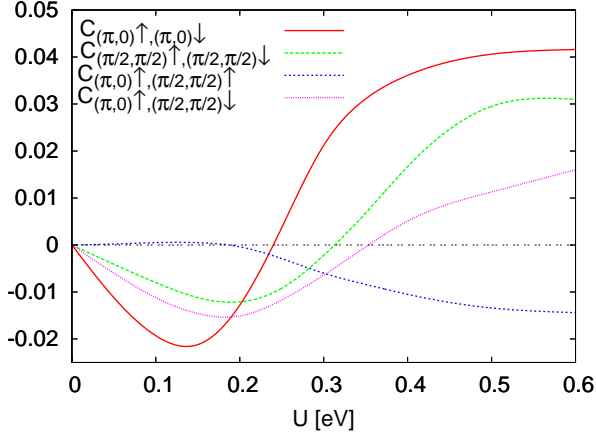}}}}
\caption{\label{fig:7.2a} Correlation functions $C_{ij}$ between the $(\pi,0)$ and 
$(\pi/2,\pi/2)$ states [Eq.~(\ref{eq:3.1a})] as a function of $U$ for an eight site cluster. 
The parameters are $t=-0.04$ eV, $t'=0$ and $T=29$ K. The results were obtained in the first iteration.
}
\end{figure}

For the four-site cluster, only the $(\pi,0)$ and $(0,\pi)$ levels are available
at the Fermi energy. It is then not surprising that these form a localized state for
larger values of $U$, as in the four--level model. For the eight-site cluster, on 
the other hand, also the $(\pm \pi/2,\pm \pi/2)$ levels are available.
One may then ask to what extent the $(\pi,0)$ and $(0,\pi)$ levels can form a 
localized state without involving the $(\pm \pi/2,\pm \pi/2)$ levels. This is illustrated 
in Fig.~\ref{fig:7.2a}. For $U\le 0.24$ eV, where $C_{(\pi,0)\uparrow (\pi,0)\downarrow}$
is negative, there is essentially no correlation between $(\pi,0)$ and
$(\pm \pi/2,\pm \pi/2)$ for equal spin, while there is a negative coupling for 
different spins. The Kondo-like states in the $(\pi,0)$ and $(\pm \pi/2,\pm \pi/2)$ 
channels are therefore not completely independent as was also observed for the
Kondo-like states in the $(\pi,0)$ and $(0,\pi)$ channels. For $0.24 <U<0.31$ eV , where
the localized state in the $(\pi,0)-(0,\pi)$ space starts to form, the unequal spin 
correlation function $C_{(\pi,0)\uparrow (\pi/2,\pi/2) \downarrow}$ becomes small.
Here the correlation between $(\pi,0)$ and $(\pm \pi/2,\pm \pi/2)$ is indeed rather
weak, and a similar localized state forms as for the four-site cluster. For $U>0.31$ eV
a localized state starts to form also in the $(\pm \pi/2,\pm \pi/2)$ space.
Now the correlation between the two spaces is substantial for both equal
and unequal spins. This is discussed below. 

Given the large difference in the coupling to the bath,
it may seem surprising that the localization of the $(\pi/2,\pi/2)$ electrons does
not happen for an even larger value of $U$. To understand this we discuss the
gain in correlation energy when electrons localize on the cluster.

\begin{figure}
{\rotatebox{-90}{\resizebox{6.0cm}{!}{\includegraphics {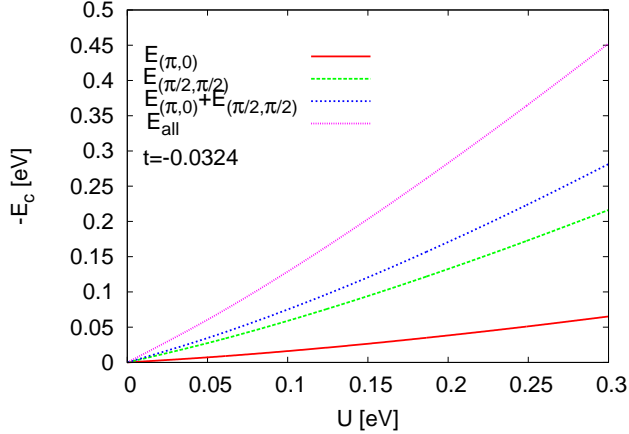}}}}
\caption{\label{fig:7.3}Correlation energy for an isolated eight site cluster.
Results are shown for the case when the $(\pm \pi/2,\pm \pi/2)$ states are 
not allowed to participate in the correlation [referred to as ``$(\pi,0)$''] or the 
$(\pi,0)$ and $(0,\pi)$ states are not allowed to correlate [referred to as
``$(\pi/2,\pi/2)$''] or when all states are involved in the correlation
(referred to as ``all''). The difference between E$_{all}$ and 
E$_{(\pi,0)}$+E$_{(\pi/2,\pi/2)}$ curves quantifies the energy gain 
from the correlation between $(\pi,0)$ and $(\pm \pi/2,\pm \pi/2)$ electrons.
The parameters are $t_{\rm cluster}=-0.0324$ eV, $t'=0$ and $T=0$.
}
\end{figure} 
Fig.~\ref{fig:7.3} shows correlation energies for an isolated eight site cluster
as a function of $U$. The curve $E_{(\pi,0)}$ was obtained by including just four 
electrons and putting the $(\pm \pi/2,\pm\pi/2)$ levels at very high energies. 
This corresponds to nominally occupy the ${\bf K}=(0,0)$ level doubly and 
then allowing the remaining two electrons to correlate in the $(\pi,0)$ and 
$(0,\pi)$ states. Then the $(\pm \pi/2,\pm\pi/2)$ levels do not participate in 
the correlation process. This may crudely represent the situation when the 
$(\pm \pi/2,\pm\pi/2)$ electrons form Kondo states with the bath but the 
$(\pi,0)$ and $(0,\pi)$ electrons have localized. The latter electrons can 
then correlate internally, while the $(\pm \pi/2,\pm\pi/2)$ electrons primarily 
optimize the interaction with the bath and correlate less efficiently with the  
$(\pi,0)$ and $(0,\pi)$ electrons, as discussed above. Although there are nominally two electrons
in the ${\bf K}=(0,0)$ level, these electrons are allowed to fully correlate with 
the other two electrons, and the ${\bf K}=(\pi,\pi)$ orbital also participates. 
For the curve $E_{(\pi/2,\pi/2)}$ we have instead put the $(\pi,0)$ and $(0,\pi)$
at very high energy and we now include six electrons, nominally two electrons 
in the ${\bf K}=(0,0)$ level and the remaining four electrons nominally in 
the $(\pm \pi/2,\pm\pi/2)$ orbitals. $E_{\rm all}$ is obtained by calculating 
the correlation energy without any constraints for eight electrons. This would 
correspond to all eight electrons being roughly localized in the cluster.

We also show $E_{(\pi,0)}+E_{(\pi/2,\pi/2)}$. This sum involves some 
double counting, since in one case the orbitals ${\bf K}=(0,0)$ and
${\bf K}=(\pi,\pi)$ can fully correlate with the $(\pi,0)$ and $(0,\pi)$
electrons and in the other with the $(\pm \pi/2,\pm\pi/2)$ electrons.
Nevertheless, this curve is much lower than $E_{\rm all}$. This shows 
that a lot of correlation energy is gained by allowing the $(\pi,0)$ and 
$(0,\pi)$ electrons to correlate with the  $(\pm \pi/2,\pm\pi/2)$ electrons,
which is not included in the sum $E_{(\pi,0)}+E_{(\pi/2,\pi/2)}$. 
This is also illustrated by the substantial correlation between $(\pi,0)$ and 
$(\pm \pi/2,\pm \pi/2)$ in Fig.~\ref{fig:7.2a} for large $U$. This gives a
 strong tendency for all electrons to localize simultaneously, since 
then particularly much energy can be gained. The  difference in the 
values of $U$ where the curves in Fig. \ref{fig:7.2a} turn upwards for the $(\pi,0)$ and 
$(\pm \pi/2,\pm\pi/2)$ channels is therefore substantially smaller than 
one would expect from the differences in the couplings (factor of 3-4).  
Nevertheless, the upturn does happen for a larger $U$ in the $(\pm \pi/2,\pm\pi/2)$ 
channel, and therefore there is a pseudogap in $(\pi,0)$ but not in 
$(\pm \pi/2,\pm\pi/2)$ for intermediate values of $U$.

\begin{figure}
{\rotatebox{-90}{\resizebox{4.4cm}{!}{\includegraphics {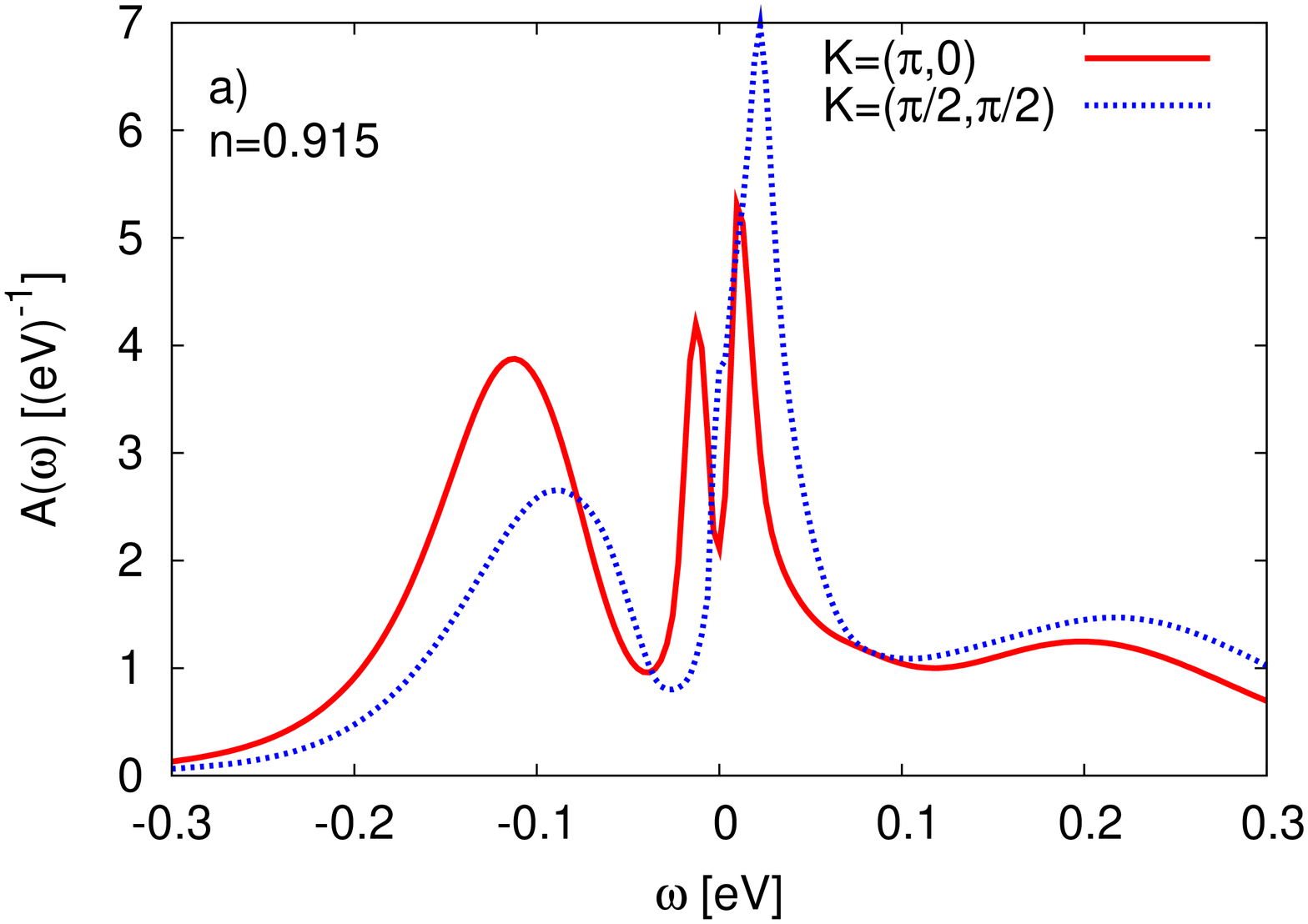}}}}
\hskip-0.6cm
{\rotatebox{-90}{\resizebox{4.4cm}{!}{\includegraphics {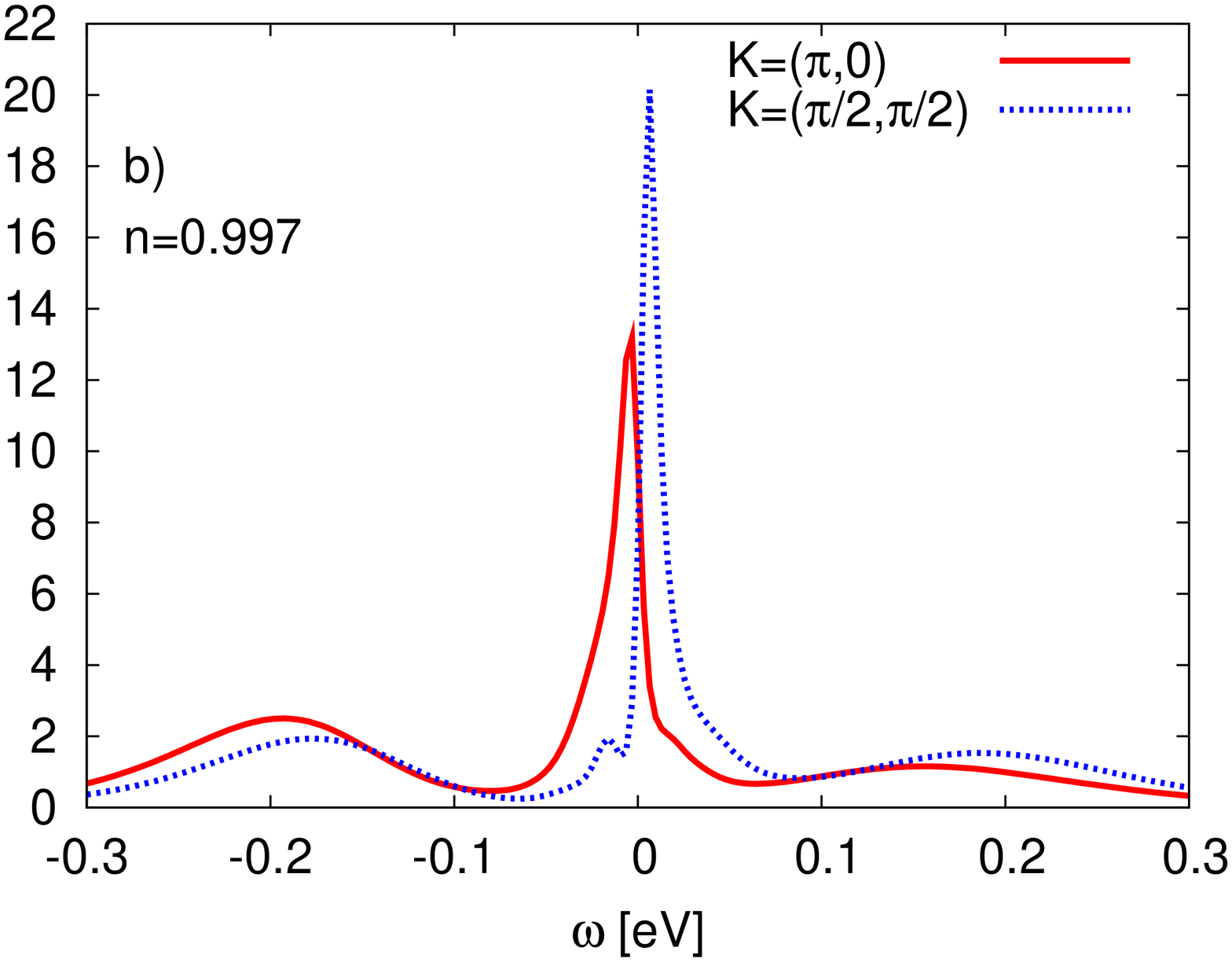}}}}
\hskip-0.6cm
{\rotatebox{-90}{\resizebox{4.4cm}{!}{\includegraphics {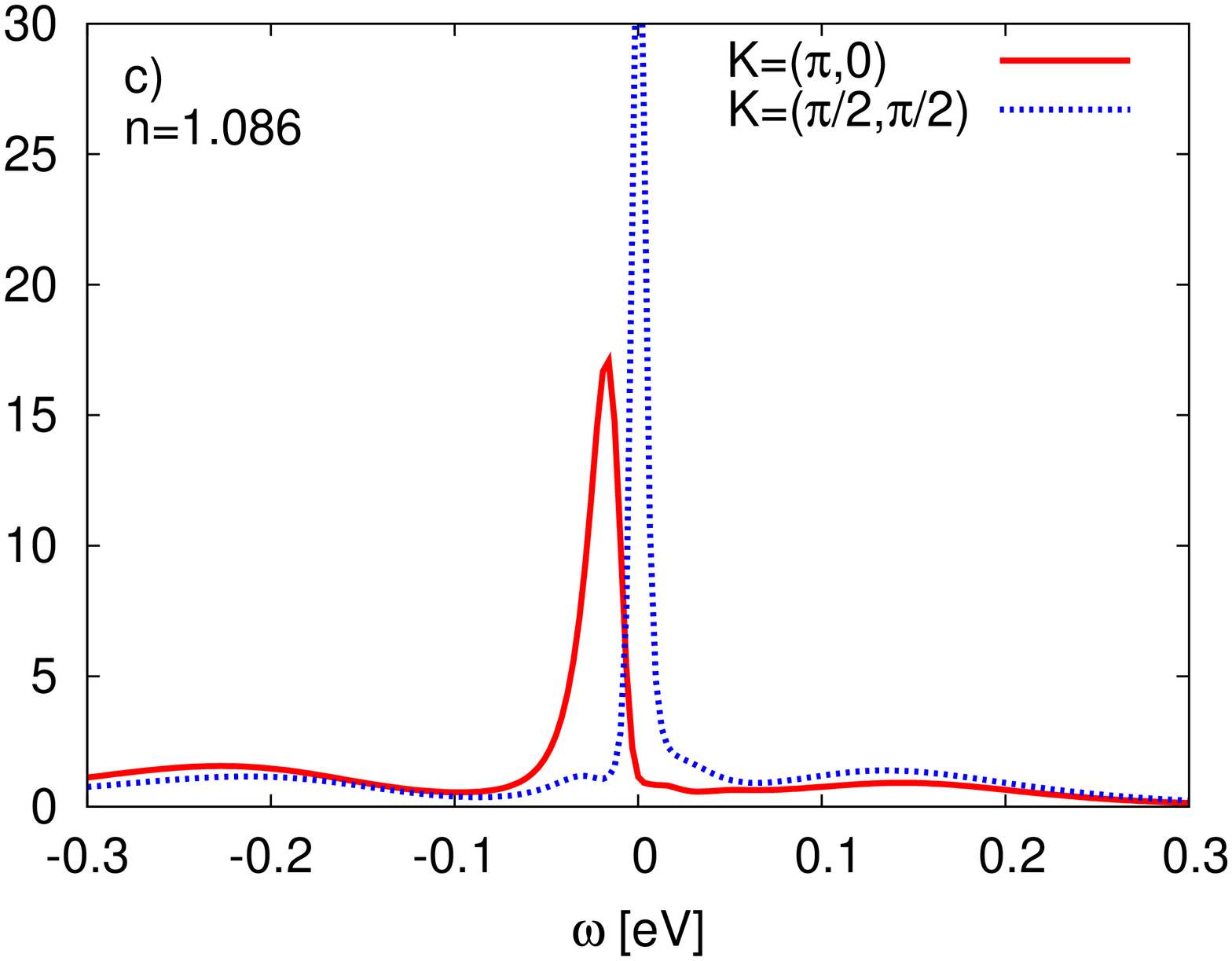}}}}
\caption{\label{fig:7.4}Spectral function for different fillings. The parameters are $t=-0.04$ eV, $t'=0.12$ eV,
$U=0.32$ eV and $T=38.4$ K. First iteration results are plotted.
}
\end{figure} 

\subsection{Frustrated case}\label{sec:7a}
 
So far we have considered the unfrustrated case when $t'=0$. We now put in frustration, $t'=-0.3t$,
which is more appropriate for cuprates. The results are shown in Fig.~\ref{fig:7.4} for
different fillings. Fig.~\ref{fig:7.4}a shows a hole-doped case (doping 8.5$\%$). This 
shows a clear pseudogap for ${\bf K}=(\pi,0)$. For the approximately half-filled case ($n=0.997$) 
(Fig.~\ref{fig:7.4}b) there is no pseudogap. Increasing the filling further to $n=1.086$ 
(Fig.~\ref{fig:7.4}c) the filling of the $K=(\pi,0)$ level is substantially increased, 
but otherwise the spectral function is not changed much and no pseudogap is formed. 
These results were obtained in the first iteration. Self-consistent calculations also
show a pseudogap for the hole-doped case but not for the electron-doped case, but 
there are differences in the doping dependence. For instance, in the self-consistent case the 
pseudogap becomes very pronounced for small hole-dopings.

This difference between electron- and hole-doping has been observed 
experimentally.\cite{Armitage} Experimentally, the electron-doped sample is antiferromagnetic
for the parameters considered here,
while the calculations show that the $(\pi,0)$ pseudogap also disappears for the paramagnetic 
calculation.\cite{Macridin} It is then particularly interesting to study the paramagnetic state,
to see why just a small change in the chemical potential removes the $(\pi,0)$ pseudogap.

\begin{figure}
{\rotatebox{-90}{\resizebox{6.0cm}{!}{\includegraphics {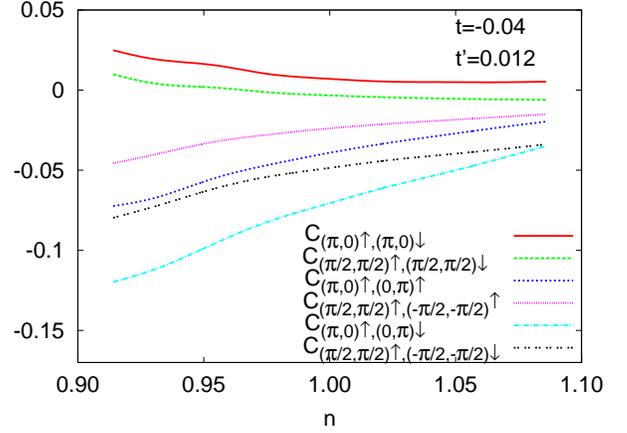}}}}
\caption{\label{fig:7.5} Correlation functions $C_{ij}$ defined in analogy to 
Eq.~(\ref{eq:3.1a}) as a function of the filling $n$ for a frustrated eight site cluster. The parameters 
are $U=0.32$ eV, $t=-0.04$ eV,    $t'=0.012$ eV and $T=38.4$ K. The 
parameter $n_0$ [Eq.~(\ref{eq:2.1a})]was chosen as $n/2$. The results were obtained after 
the first iteration.
}
\end{figure}

To further illustrate the difference between hole- and electron-doped systems, we show 
correlation functions in Fig.~\ref{fig:7.5}
also studied earlier for $t'=0$. For the hole-doped system, the $C_{(\pi,0) \uparrow (\pi,0) \downarrow}$
correlation function has a fairly large positive value, indicating that a localized state has formed in
the $(\pi,0)$-$(0,\pi)$ space. As the filling is increased, this correlation function is reduced,
indicating that the tendency to form a localized cluster state is weakened. The 
correlation function $C_{(\pi/2,\pi/2)\uparrow (\pi/2,\pi/2) \downarrow}$ 
is weakly positive for a substantial hole doping, but apparently not enough to cause a pseudogap.
This function is also reduced with increasing filling and becomes negative for electron-doping.
We now analyze the reasons for the difference between hole-doped and electron-doped 
systems.

\subsubsection{Exact diagonalization}\label{sec:7aa}
To obtain additional understanding we perform exact diagonalization (ED) calculations.
We analyze the doping dependence of the $N_c=8$ clusters coupled to a a bath described 
by $N_b=8$ bath states. This coupling was held fixed as the chemical potential is varied. 
We have evaluated spin correlations and hybridization energies between
cluster and bath electrons to quantify the strength of antiferromagnetic correlations
associated with the Kondo effect.  
We evaluate the hybridization energy from the expression:
\begin{equation}\label{eq:7.1}
\langle V({\bf K}) \rangle=\sum_{\tilde {\bf k} \sigma} V({\bf K})  (\langle c^\dagger_{{\bf K}, \sigma} 
c_{{\bf K}+\tilde {\bf k}, \sigma} \rangle +   \langle c^\dagger_{{\bf K}+ \tilde {\bf k},\sigma} c_{{\bf K},\sigma} \rangle),
\end{equation} 
where $c^{\dagger}_{{\bf K}\sigma}$ and $c^{\dagger}_{{\bf K}+\tilde {\bf k}\sigma}$ creates an electron on the cluster
or in the bath, respectively. Here the sum over $\tilde {\bf k}$ is reduced to one value. We also calculate the 
correlation between a spin $S_z({\bf K})$ in the cluster and the spin of its bath level $s_z({\bf k})$.

For the half-filled unfrustrated case, $t'=0$, the various spin correlations and hybridization energies 
are given in Table \ref{table:7.1}. 
It is found that the hybridization energy is smaller for $(\pi,0)$ than $(\pi/2,\pi/2)$
as it should, since the coupling is smaller for $(\pi,0)$. The spin correlation $\langle S_z({\bf K})s_z({\bf k})\rangle$
is negative, as for the four-level model in Fig.~\ref{fig:5.1}. This correlation is more negative for $(\pi/2, \pi/2)$
than for $(\pi,0)$, due to the stronger cluster-bath coupling.                                                      
Increasing $U$ leads to a suppression of the spin correlations as one 
would expect, since the Kondo couplings are suppressed and it becomes more favorable to correlate the electrons inside 
the cluster. This supports the conclusions above.
 
\begin{table}
\caption{Bath-cluster spin correlations and hybridization energies from ED calculations on $N_b=N_c=8$
clusters at half-filling ($n=1$). The $z$-component of the spin of a bath (cluster) electron in, for instance, 
the $(\pi,0)$ channel is denoted by  $s_z(\pi,0)$ [$S_z(\pi,0)$]. The cluster bath couplings are 
$V(\pi,0)=0.008$ eV  and $V(\pi/2,\pi/2)= 0.021$ eV.
We take $t=-0.04$ eV, $t'=0$ for $U=0.25$ eV and $U=0.32$ eV.  }
\label{table:7.1}
\begin{tabular}{llllll}
\hline
\hline
$U$ &$\mu$ & \multicolumn{2}{c} {$\langle S_z({\bf K})s_z({\bf k})\rangle$ }  & \multicolumn{2}{c} { $\langle V({\bf K})  \rangle $}   \\

 & & $(\pi,0)$ & $({\pi\over 2},{\pi\over 2})$  & $(\pi,0)$   &  $({\pi\over 2},{\pi\over 2})$ \\
\hline
0.25 & 0.125 &     -0.0020  & -0.0051 & -0.0021   &  -0.0097\\
0.32 &  0.160 &   -0.0012  &  -0.0030     &-0.0014   &    -0.0066        \\
\hline
\end{tabular}
\end{table}

The dependence of the spin correlations on doping is summarized in 
Table \ref{table:7.2} for the frustrated case ($t'=-0.3t$).
In the $(\pi,0)$ channel the cluster-bath antiferromagnetic spin correlations
and the hybridization energy increase with the filling, moderately on the hole-doped side and more strongly
on the electron-doped side. In the $(\pi/2, \pi/2)$ channel,
on the other hand, these quantities vary little with filling in the range considered,
with a tendency to a minimum at $n=1$.

\begin{table}
\caption{Bath-cluster spin correlations and hybridization energies from exact diagonalization calculations on $N_b=N_c=8$
frustrated clusters.  There is one bath level per cluster orbital
at the Fermi energy with the bath cluster couplings fixed at $V(\pi,0)=0.008$ eV  and $V(\pi/2,\pi/2)= 0.021$ eV.
We also show $J_K({\bf K})$ calculated according to Eq.~(\ref{eq:7.2}).
The parameters are $t=-0.04$ eV, $t'=-0.3t$ and $U=0.25$ eV.
}
\label{table:7.2}
\begin{tabular}{llllllll}
\hline
\hline
$\mu$ & $n$ & \multicolumn{2}{c} {$\langle S_z({\bf K})s_z({\bf k})\rangle$ }  & \multicolumn{2}{c} { $\langle V({\bf K})  \rangle $} &
\multicolumn{2}{c}{${J_K({\bf K}) /     |V({\bf K})|^2}$}  \\

 & & $(\pi,0)$ & $({\pi\over 2},{\pi\over 2})$  & $(\pi,0)$   &  $({\pi\over 2},{\pi\over 2})$ &$(\pi,0)$&$({\pi\over 2},{\pi\over 2})$  \\
                \hline
                0.085 &  0.96   &  -0.0042    &    -0.0068    &  -0.0036  &    -0.012  & 16.1 & 17.8  \\
                        0.105  &   0.98     &  -0.0038   &    -0.0063  & -0.0035   & -0.011 & 16.9 & 16.4 \\
                        0.125 &  1.00  &   -0.0048       &    -0.0065 & -0.0043   &  -0.011 & 18.8 & 16.0 \\
                        0.135  &  1.02    &  -0.0064     &      -0.0063   &  -0.0047  &  -0.012  & 20.4 & 16.1 \\
                        0.145  &   1.04     &  -0.0084   &    -0.0071  & -0.0055   & -0.013 & 22.7 & 16.4 \\
                        \hline
                        \end{tabular}
                        \end{table}

\begin{figure}
{\rotatebox{-90}{\resizebox{5.0cm}{!}{\includegraphics {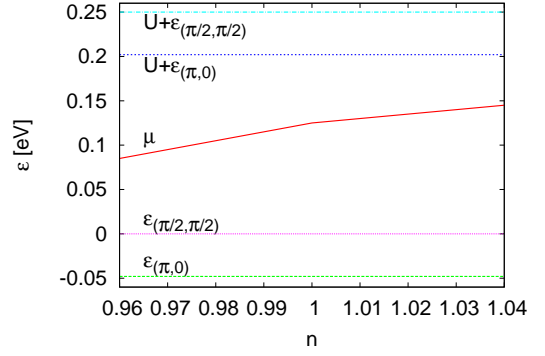}}}}
\caption{\label{fig:7.6}Energy levels and chemical potential relevant for the
calculation of the Schrieffer-Wolff exchange coupling, $J_K$.
}
\end{figure}

In order to rationalize the doping dependence of the antiferromagnetic Kondo correlations,  
we have estimated the Kondo couplings obtained from a Schrieffer-Wolff transformation of 
the Anderson Hamiltonian:
\begin{equation}\label{eq:7.2}
J_K({\bf K})=|V({\bf K})|^2 \left( {1 \over \epsilon({\bf K})+ U-\mu}+{1\over \mu - \epsilon({\bf K}) } \right). 
\end{equation}
This estimate should be relevant when the correlation between different ${\bf K}$ levels on the
cluster is weak and the coupling to the bath dominates.
The results are shown in Table~\ref{table:7.2}. $J_K(\pi,0)$ increases in going from the hole-doped
to the electron-doped side, as found for the spin correlations. $J_K(\pi/2,\pi/2)$ varies little 
with a weak minimum at $n=1$. This is also in line with the results for the spin correlations.
The hybridization energy varies in a similar way as the spin correlations. This is not
surprising, since the Schrieffer-Wolff exchange coupling, $J_K$, is derived from hopping between the cluster and 
the bath, as can also be seen from Eq.~(\ref{eq:7.2}). The variation of $J_K$ with filling
can be understood from the energies shown in Fig.~\ref{fig:7.6}. For lower range of fillings 
considered here, the chemical potential $\mu$ is almost in the middle between the levels 
$\varepsilon({\bf K})$ and $U+\varepsilon({\bf K})$ for ${\bf K}=(\pi,0)$. As the filling 
is increased $\mu$ moves upwards and gets much closer to $U+\varepsilon({\bf K})$. This 
favors a large $J_K$ [Eq.~(\ref{eq:7.2})]. These arguments show that there is a substantial difference 
for the frustrated lattice and ${\bf K}=(\pi,0)$ between the hole-doped and electron-doped
systems. It is also interesting that if the sign of $t'$ had been the opposite, as for the organics,
this effect would have favored a pseudogap for the electron-doped systems. For ${\bf K}=(\pi/2,\pi/2)$ 
$\mu$ is roughly in the middle for $n=1$ and is closer to one of the two levels for other fillings, 
giving a minimum in $J_K$ at $n=1$. In these ED calculations we held the coupling between the 
cluster and the bath fixed as $\mu$ was varied. This is an approximation, as can be seen from 
Eq.~(\ref{eq:2.5}).  Below we study $\mu$ dependence of the coupling. 

\subsubsection{$\mu$ dependence of the coupling strength}\label{sec:7ab}

\begin{figure}
{\rotatebox{-90}{\resizebox{6.0cm}{!}{\includegraphics {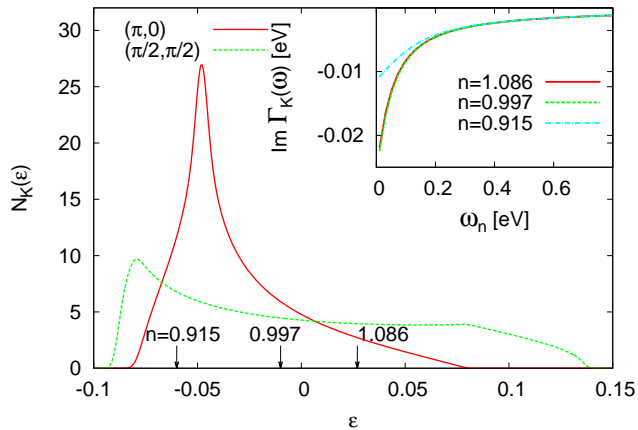}}}}
\caption{\label{fig:7.7}The ${\bf K}=(\pi,0)$ and $(\pi/2,\pi/2)$ sector density
of states $N_{\bf K}(\varepsilon)$. The arrows show the values of $-\Delta=-Un_0+\mu$ [Eq.~(\ref{eq:2.3a})] for
the different values of the filling $n$. The inset shows Im $\Gamma_{\bf K}(\omega_n)$  
for different values of $n$. The parameters are $t=-0.04$ eV, $t'=0.012$ eV, $U=0.32$ eV
and $T=38.4$ K. First iteration results were used.
}
\end{figure}

To discuss the dependence of the cluster-bath coupling $\Gamma_{\bf K}(\omega_n)$ 
on $\mu$ (or $n$) we use Eq.~(\ref{eq:2.4a}). As before we consider the first iteration 
when $\Sigma_c\equiv 0$.
Then                                 
\begin{equation}\label{eq:7.3ab}
 {\rm Im} \Gamma_{\bf K}(\omega_n)= { {\rm Im} G_0(\omega_n,{\bf K})
\over [{\rm Re} G_0(\omega_n,{\bf K})]^2+[{\rm Im} G_0(\omega_n,{\bf K})]^2}+\omega_n.
\end{equation}
From Eqs.~(\ref{eq:2.3}, \ref{eq:2.4}) it follows that Im $G_0(\omega_n\to 0,{\bf K})\sim N_{\bf K}(-\Delta)$,
where $N_{\bf K}(\varepsilon)$ is the sector density of states..
Fig.~\ref{fig:7.7} shows the  $N_{\bf K}(\varepsilon)$, and the arrows 
mark the positions of $-\Delta$ for different values of the filling  $n$. For $n=0.915$ it is clear 
that Im $G_0[\omega_n,(\pi,0)]$ is large. Since $-\Delta$ is very asymmetrically located with respect 
to the center of gravity of $N_{\bf K}(\varepsilon)$ also $|{\rm Re} G_0[\omega_n,(\pi,0)]|$ is large. Then it
follows from Eq.~(\ref{eq:7.3ab}) that Im $\Gamma_{(\pi,0)}(\omega_n\to 0)$ is small.
For $n=0.997$ and, in particular, $n=1.086$, Im $G_0[\omega_n,(\pi,0)]$ is  smaller. 
$|{\rm Re} G_0[\omega_n,(\pi,0)]|$ is also smaller. This leads to a larger Im $\Gamma_{(\pi,0)}(\omega_n\to 0)$,
where numbers happen to combine in such a way that Im $\Gamma_{(\pi,0)}(\omega_n\to 0)$
is very similar for $n=0.997$ and $n=1.086$.
This is shown by the inset in Fig.~\ref{fig:7.7}. The curves for $n=0.997$ and $n=1.086$ fall
almost on top of each other. The specific numbers for $G_0$ for $\omega_n=0.0105$ eV are 
$(-22.0,-31.8)$, $(14.8,-18.8)$ and $(14.1,-9.2)$ for $n=0.915$, 0.997, 1.086, respectively. 
For large $\omega_n$, $\Gamma_{\bf K}$ is determined by the second moment of the sector density 
of states [Eq.~(\ref{eq:2.6})]. Therefore Im $\Gamma_{\bf K}(\omega_n)$ is very similar 
for all fillings in this limit.

We are now in the position to understand the behavior of the ${\bf K}=(\pi,0)$ spectrum for
the frustrated system as a function of filling.  Starting out at small fillings ($n\sim 0.92$), both
$J/V^2$ and Im $\Gamma_{\bf K}(\omega_n \to 0)$ are small for ${\bf K}=(\pi,0)$, favoring a pseudogap.           
For filling one, $J/V^2$ increases somewhat and Im $\Gamma_{\bf K}(\omega_n \to 0)$ 
increases substantially. At the same time the pseudogap disappears. For still larger filling ($n=1.086$),
$J/V^2$ increases substantially and Im $\Gamma_{\bf K}(\omega_n \to 0)$ stays approximately constant for ${\bf K}=(\pi,0)$.
This further suppresses the pseudogap. This is supported by the results in Figs.~\ref{fig:7.4}, \ref{fig:7.5}. 

\section{Character of the pseudogap}\label{sec:8}

To study the character of the pseudogap we have studied the ground-state and excited states 
of a neutral isolated cluster. We have used the hopping integral $t_{\rm cluster}=-0.0324$ eV,
corresponding to $t=-0.04$ eV in DCA for $N_c=8$. In the pseudogap phase the electrons
in the $(\pm \pi/2,\pm \pi/2)$ levels are primarily correlated with the electrons
in their baths and rather uncorrelated with electrons in the $(\pi,0)$ and $(0,\pi)$ levels
(see Fig.~\ref{fig:7.2a}). Here we then assume that the two sets of electrons are completely 
uncorrelated and focus on the properties of the electrons in the $(\pi,0)$ and $(0,\pi)$ levels.
For this purpose we put the $(\pm \pi/2,\pm \pi/2)$ at some extremely high energy and then
treat an isolated molecule with just four electrons. This then correspond to the nominally 
two electrons in the $(0,0)$ level and two electrons in the $(\pi,0)$ and $(0,\pi)$ levels.
The four electrons nominally in the $(\pm \pi/2,\pm \pi/2)$ are not included in the calculation.
This then approximately describes the pseudogap situation where the $(\pm \pi/2,\pm \pi/2)$
electrons are delocalized and the $(\pi,0)$ and $(0,\pi)$ electrons are localized.
The results are shown in Table~\ref{table:7.3}. For the dominating configurations for the 
states shown in Table~\ref{table:7.3} the level $(0,0)$ is doubly occupied. The filling of
the $(\pi,0)$ and $(0,\pi)$ levels is similar as in Eq.~(\ref{eq:5.3}) for the four lowest states.  
We have also calculated the pairing correlation $P_d$ defined in Eq.~(\ref{eq:7.4}) for 
the $x^2-y^2$ symmetry. 

We now consider photoemission, where an electron is emitted from the cluster. The spectral 
weight close to the Fermi energy then corresponds to low-lying final states, which are primarily 
composed of neutral cluster states. We can think of this as an electron hopping in from the bath 
after the photoemission process to obtain a neutral cluster. Since the lowest state 
has $S=0$, removing an electron couples to $S=1/2$ states. An electron hopping in from the bath 
then couples to $S=0$ or $S=1$ states. The relevant final states should then primarily be composed 
of such states. The peak to peak size of pseudogap in Fig.~\ref{fig:3.1} is approximately 0.02 eV
(for $U=0.3$ eV).  
This then approximately corresponds to a transition to the second state in Table~\ref{table:7.3},
giving a gap of about 2(0.1744-0.1632)=0.022 eV. The lowest state has large $d_{x^2-y^2}$
pairing fluctuations, while the first excited state has a negative $P_d$. We can then interpret 
the pseudogap as resulting from the break up of a local $d_{x^2-y^2}$ singlet state. 
We note that these are very short-range correlations, which may not be observable in experiment.

\begin{table}
\caption{\label{table:7.3}Energy $E$, spin $S$  and superconductivity 
correlation $P_d$ for low-lying states with $S_z=0$ for an isolated cluster with $N_c=8$ and $T=0$. 
The $(\pm \pi/2,\pm \pi/2)$ levels are artificially put at very high energy and effectively
not part of the calculation. The number of electrons is four.
The parameters are $U=0.3$, $t_{\rm cluster}=-0.0324$ eV and $t'=0$.
}
\begin{tabular}{ccc}
\hline
\hline
$E$ & $S$ &  $P_d$  \\
\hline
-0.1744 & 0 &  0.69 \\
-0.1632 & 1 & -0.19 \\
-0.1160 & 0 & -0.19 \\
-0.0824 & 0 & -0.25 \\
\hline
\end{tabular}
\end{table}

Here we have considered ratios of $U/|t|\sim 10$, close to a Mott transition, 
where the splitting of the states in Table~\ref{table:7.3} depend rather weakly 
on $U/|t|$ for fixed $t$. These splittings are much larger than the  
values for the pseudogap in Ref. \onlinecite{Ferrero07}. However, the latter were 
obtained for much larger values of $U/t$. For such large values of $U/|t|$ the 
splittings in Table~\ref{table:7.3} would go as $1/U$. 

\begin{table}
\caption{\label{table:7.4}Energy $E$, spin $S$, degeneracy Deg. and superconductivity
correlation $P_d$ for low-lying states with $S_z=0$ for an isolated cluster with $N_c=8$ and $T=0$.
All levels are included and there are eight electrons.
Deg. refers to the degeneracy of $S_z=0$ states. $P_d$ for the state with $E=-0.0867$ is the
average over the six degenerate states. The parameters are $U=0.3$, $t_{\rm cluster}=-0.0324$ eV and $t'=0$.
}
\begin{tabular}{cccc}
\hline
\hline
$E$ & $S$ & Deg.  & $P_d$  \\
\hline
-0.1116 & 0 & 1 & 0.39 \\
-0.1062 & 1 & 1 & 0.30 \\
-0.0941 & 2 & 1 & 0.12 \\
-0.0867 & 1 & 6 & 0.17 \\
-0.0764 & 0 & 9 & 0.11 \\
\hline
\end{tabular}
\end{table}

As a comparison, Table~\ref{table:7.4} shows a calculation where the $(\pm \pi/2,\pm \pi/2)$
levels are not suppressed and the number of electrons is eight, as in a neutral cluster.
This is more relevant for the large $U$ situation, when all electrons on the cluster are localized.
Since there are more levels, the energy splitting of the states is smaller. The difference 
in $P_d$ is also smaller. At finite hopping to the bath and finite $T$, there is a finite 
occupation of many cluster states.  For small $T$ and large $U$,  however, the lowest cluster 
state dominates. The lowest cluster state has RVB character\cite{Tang} as discussed previously. This agrees with the results in 
Fig.~\ref{fig:spinRVB}, showing that the spin susceptibility $S(\pi,\pi)$ for large $U$ behaves 
as expected for an RVB state. The value, $P_d=0.39$ is slightly larger than the result found in 
Fig.~\ref{fig:pairing} for large $U$ for a cluster in a bath. This is reasonable, since there is 
also a slight mixture of excited states with smaller $P_d$ for the embedded cluster. 

\section{Implications of a pseudogap}\label{sec:9}

In the previous section we have seen that the pseudogap can be understood as 
corresponding to the break up of a (very short-range) $d_{x^2-y^2}$ singlet state.  We can then 
reverse the statement and ask if a pseudogap implies the presence of a state 
with large superconductivity fluctuations. From the arguments above we would expect 
that this is not the case. In Sec.~\ref{sec:6} we argued that if the lowest
cluster state is nondegenerate the system ought to show a pseudogap for sufficiently
large $U$ (in the first iteration). To test this we would then like to find
a system where the lowest cluster state is nondegenerate, but $P_{x^2-y^2}$ is small.
We therefore consider a system where an antiferromagnetic field is applied 
to the cluster sites, but not to the bath. Thus we add
\begin{equation}\label{eq:8.1}
H_{\rm AF}= \sum_i \gamma_i(n_{i\uparrow}-n_{i\downarrow}),
\end{equation}
to the Hamiltonian in Eq.~(\ref{eq:2.1}). Here $\gamma_i=\gamma$ if $i$ belongs
to one sublattice and $\gamma_i=-\gamma$ otherwise. We expect the field to be bad for 
$x^2-y^2$ pairing. We have chosen a small value of $\gamma=0.1t=0.004$ eV. 
For $U=0$ this has a very small effect on the solution for the $N_c=8$ cluster. 

\begin{figure}
{\rotatebox{-90}{\resizebox{6.0cm}{!}{\includegraphics {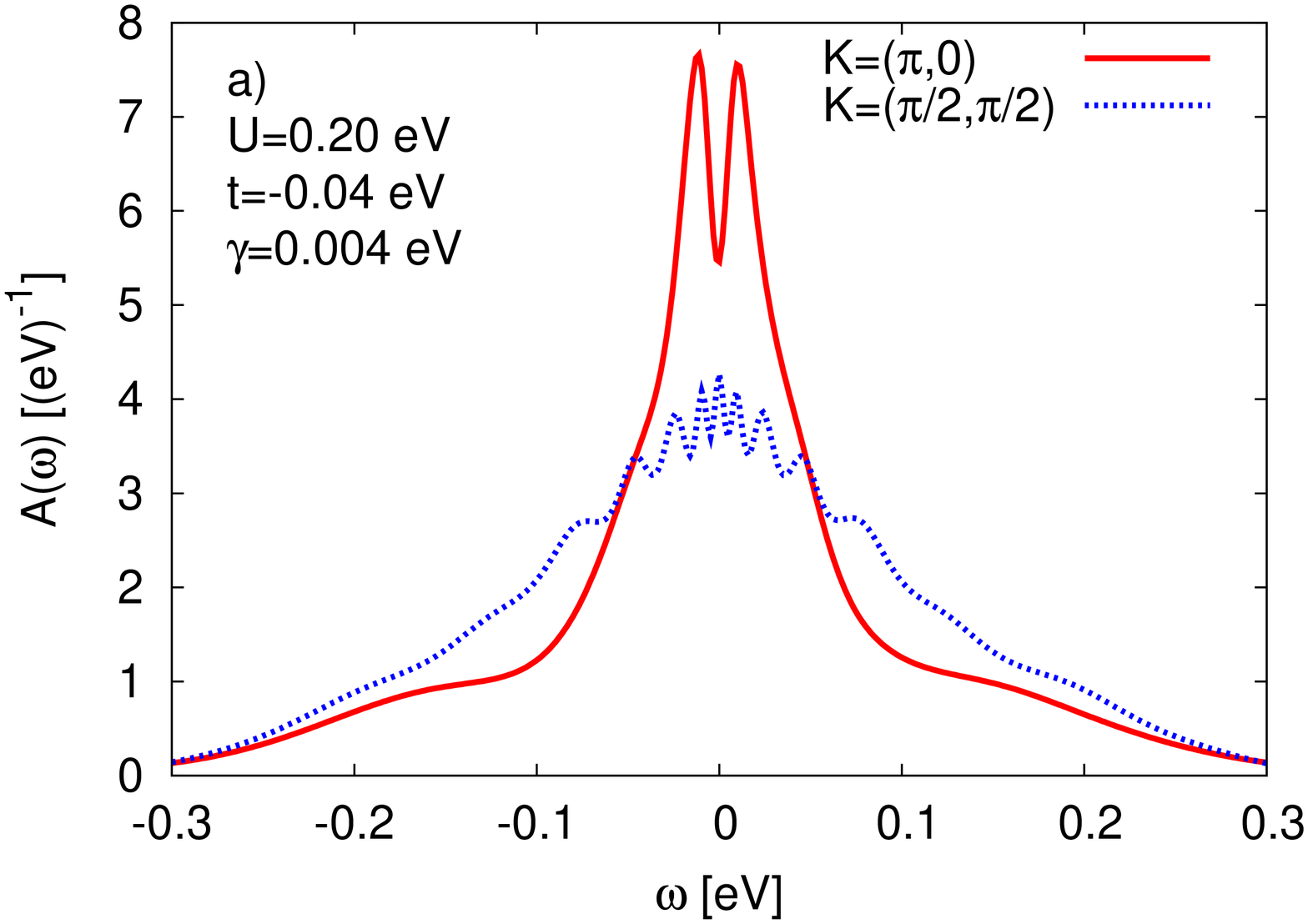}}}}
{\rotatebox{-90}{\resizebox{6.0cm}{!}{\includegraphics {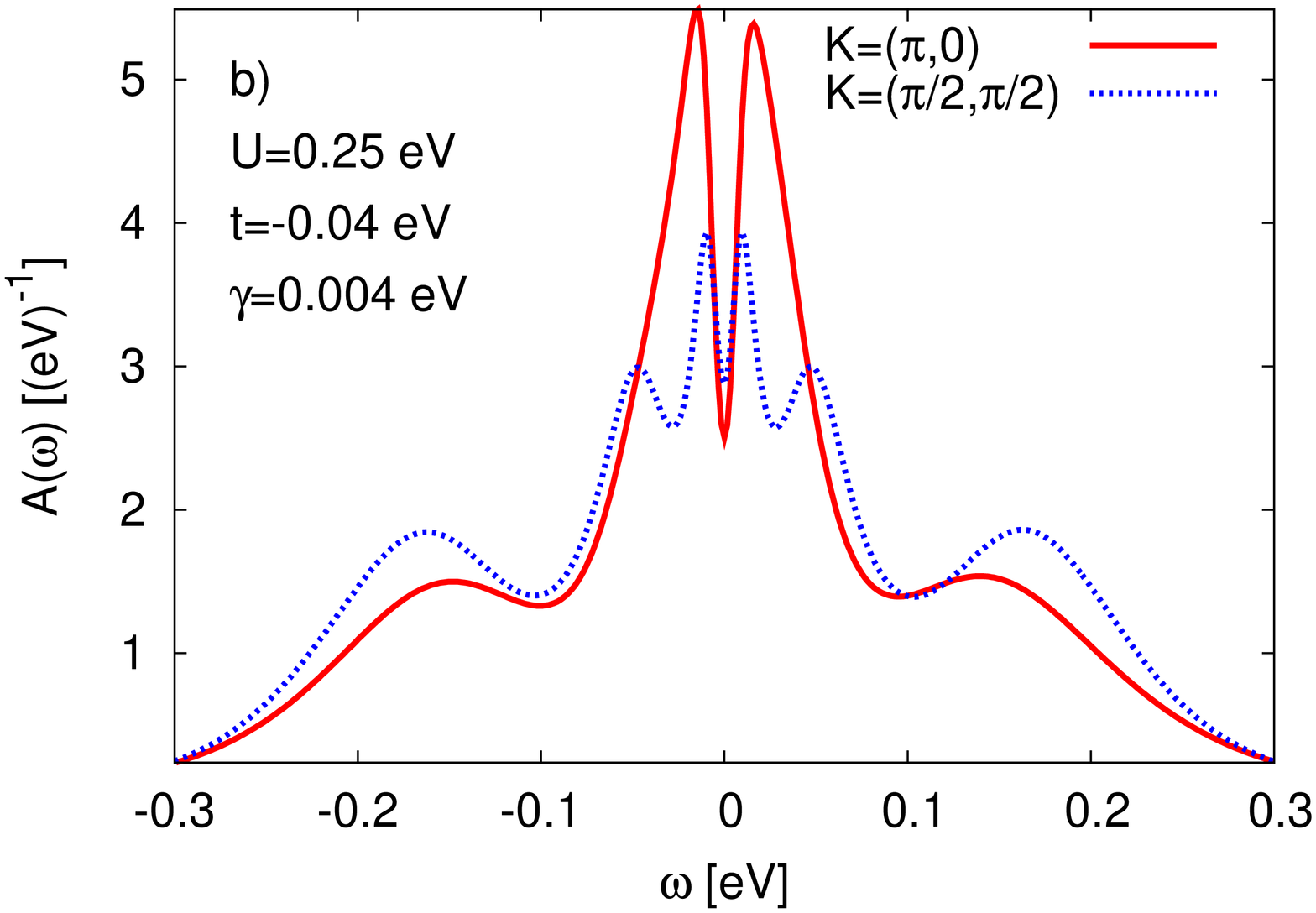}}}}
\caption{\label{fig:8.1}Spectra for a model including the antiferromagnetic potential [Eq.~(\ref{eq:8.1})]
on the cluster. A pseudogap is formed in the $(\pi,0)$ channel but not in the $(\pi/2,\pi/2)$ channel 
for $U=0.2$ eV, although the superconductivity fluctuations are strongly suppressed.
The parameters are $t=-0.04$ eV, $t'=0$, $\gamma=0.004$ eV and $T=38.4$ K.
}
\end{figure}

In Fig.~\ref{fig:8.1} we show spectra obtained for intermediate $U$.
Since the system is not homogeneous, the spectra have been calculated from 
the Fourier transform of the spatial cluster Green's function. This gives the
average spectrum over the patches around ${\bf K}=(\pi,0)$ and $(\pi/2,\pi/2)$. 
In the limit $U=0$ and $\gamma=0$, this approach 
would give the density of states over the patches around ${\bf K}=(\pi,0)$
and $(\pi/2,\pi/2)$. In particular the spectrum around ${\bf K}=(\pi/2,\pi/2)$
is therefore much broader than for other spectra in this paper. Those other spectra were 
calculated from Eq.~(\ref{eq:2.7}), and the spectra would just consist of a 
$\delta$-function in the limit $U=0$ and $\gamma=0$. It is then not surprising
that the $(\pi/2,\pi/2)$ spectrum is rather broad. The interesting aspect is that 
for $U=0.20$ eV the $(\pi,0)$ spectrum shows a pseudogap, while the $(\pi/2,\pi/2)$
spectrum does not. For $U=0.25$ eV both spectra show a pseudogap. For 
$U=0.20$ eV and $\gamma=0.004$ eV, the expectation value of the $d_{x^2-y^2}$ pairing   
operator is only 0.04 and for $U=0.25$ it is 0.05. For $\gamma=0$ a pseudogap forms 
for $U\sim0.30$ eV (in the first iteration). The corresponding expectation value 
of the pairing operator is then 0.18. Thus the pseudogap is formed for $\gamma=0.004$ 
eV although there are almost no superconductivity correlations in the system.
This is then consistent with the arguments in Sec.~\ref{sec:6} that a nondegenerate
lowest cluster state leads to a pseudogap.

\section{Conclusions}

Angular resolved photoemission spectra (ARPES) on cuprates show a pseudogap \cite{photoemission,Timusk}
in the antinodal direction around the antinodal [${\bf k}=(\pi,0)$] point and a peak 
around the nodal [${\bf k}=(\pi/2,\pi/2)$] point in a broad region of their phase diagram. The  
layered organic superconductors also show evidence for a pseudogap 
in susceptibility and $1/T_1T$-NMR experiments\cite{Kanoda06,Kawamoto97,Mayaffre94}.   
Although the Mott transition in the cuprates is driven by doping whereas in the half-filled 
organics $\kappa$-(BEDT-TTF)$_2$X it is driven by external pressure,   
the pseudogap phenomenon appears to be similar in both systems suggesting a common origin.\cite{McKenzie}   

A pseudogap in the many-body spectral function is found in the DCA for the Hubbard model 
consistent with ARPES experiments on cuprates and previous works.\cite{Civelli,Macridin,Kyung,Ferrero09,Millis,Liebscha,Sordi} 
In order to understand the origin of the  
pseudogap within DCA we have singled out the following key ingredients:
(i) Different bath-cluster couplings around antinodal   
and nodal points. (ii) The internal electronic structure of the cluster is crucial.
This gives rise to an important competition between Kondo-like states and
localization on the cluster.
(iii) The ground state of the isolated cluster should be non-degenerate. Degenerate cluster ground states 
can lead to a Kondo peak in the spectra instead of a pseudogap. 

The spectra of isolated clusters with $N_c=8$ or $16$ sites do not show any signs of the
pseudogap observed in DCA, {\it i. e.} the spectra at the $(\pi,0)$ and $(\pi/2,\pi/2)$ 
points obtained from ED are the same due to the special symmetries for these clusters. 
The absence of the pseudogap on small isolated clusters indicates that the pseudogap 
found in DCA is related to the much weaker coupling to the bath of the $(\pi,0)$ than  
$(\pi/2,\pi/2)$ sector. In fact, we have shown that if the baths
are switched the pseudogap occurs at $(\pi/2,\pi/2)$ instead of $(\pi,0)$.

While single-site DMFT calculations require self-consistency to produce a
gap, already first iteration DCA calculations with clusters can give a pseudogap. This shows that the
feedback effect on the bath is not necessary for the pseudogap opening in the DCA. 
This is due to the internal structure of the cluster and the formation of a 
nondegenerate localized state on the cluster, due to important correlations between the 
$(\pi,0)$ and $(0,\pi)$ sectors, missing in single-site DMFT calculations.
It is then important to understand the influence of the internal structure of the 
cluster on the pseudogap.  

For this purpose, we have analyzed the ground state wavefunctions of isolated clusters. 
The ground state of the smaller $N_c=$4, 8 isolated clusters 
is a short range NN-RVB state which     
can be expressed as linear combinations of configurations with doubly occupied 
$(\pi,0)/(0,\pi)$ and $(\pm \pi/2,\pm \pi/2)$ sectors. These correlations  
are found to dominate in the large-$U$ limit of DCA calculations, when 
the cluster is effectively decoupled from the bath. They are also present in the $(\pi,0)/(0,\pi)$ sector
at smaller $U$, in the pseudogap regime, indicating the importance of these characteristic
cluster correlations for pseudogap formation.

We have devised a four-level model which contains the essential ingredients
for describing the pseudogap observed in DCA for $N_c=4$. The model
includes $(\pi,0)$ and $(0,\pi)$ levels on the cluster, each coupling to one bath level. 
With appropriate parameters, it simulates the non-degenerate singlet ground state and the 
low energy spectra of the $N_c=4$ isolated cluster. We find that at small values of $U$
each cluster level forms a Kondo-like state with its corresponding bath level. As $U$ is increased, 
it becomes favorable to instead couple the $(\pi,0)$ and $(0,\pi)$ levels to each
other, forming a RVB singlet on an $N_c=4$ cluster, and opening 
up a pseudogap. This competition between Kondo and cluster RVB state formation is crucial for the physics.    
We show that the characteristic correlations found in the four-level model for the 
$(\pi,0)$ and $(0,\pi)$ levels also happen for DCA calculations with $N_c=4$ and 8 clusters.
For $N_c=8$ calculations, the $(\pm \pi/2,\pm \pi/2)$ form Kondo-like states
with the bath up to larger values of $U$, due to the stronger cluster-bath coupling in this channel.

For a particular choice of parameters of the four-level model, the isolated cluster has 
a triplet ground-state. Then the model has a Kondo state even for large values of $U$ 
and no proper pseudogap is formed.  This shows the importance of the lowest isolated
cluster state being nondegenerate. Similar work has been done having alkali-doped 
fullerides in mind.\cite{Ferrero07,Leo04,Capone}                                                

The suppression of density of states at the Fermi level leading to the pseudogap 
observed in DCA and in the four-level model can be understood in terms of destructive 
interference processes. In the weakly coupled cluster-bath limit, $U/|V|>>1$,
destructive interference can be captured for an infinite bath assuming that the
lowest cluster state is non-degenerate. In photoemission, intensity at the 
Fermi energy corresponds to transitions to final states where cluster neutral configurations
dominate. This can happen in essentially two ways. One way consists 
of removing an electron from the cluster when an extra electron has hopped from 
the bath into the cluster, while the other involves the removal of an electron directly from the 
neutral cluster, followed by the hopping of a bath electron into the cluster. 
In both ways the system can reach the same final state in which the cluster is in a 
neutral state and the bath is left with a hole at the Fermi energy. Due to the negative 
interference between these two processes the spectral weight of the photoemission 
process is suppressed at the Fermi energy. This is in contrast to the case in which the
the lowest state of the cluster is degenerate. As an example, we analyzed the degenerate impurity
Anderson model in the large degeneracy $N_f$ limit. In this case spin-flip terms
become important. These terms interfere constructively with one of the paths 
important for the nondegenerate case, while the other path is of lower order in $(1/N_f)$.
This leads to a peak at the Fermi energy.

There is a  strong gain in correlation energy when both the $(\pi,0)$ and $(\pm \pi/2,\pm \pi/2)$ 
sectors are allowed to localize simultaneously. This gives a strong tendency for all electrons in 
the cluster to localize simultaneously, leading to a common (pseudo)gap. This is counteracted 
by the large difference (factor 3-4) in the cluster-bath coupling for the $(\pi,0)$ and $(\pm \pi/2,\pm \pi/2)$ 
sectors. As a result the pseudogap opens up for a smaller $U$ for the $(\pi,0)$ sector. However, the 
difference in $U$ at which the pseudogap opens up in the two sectors is smaller than the
difference in cluster-bath coupling, indicating the importance of the correlation between
the $(\pi,0)$ and $(\pm \pi/2,\pm \pi/2)$ sectors.

The pseudogap occurs close to the Mott transition in a parameter region in which short 
range $d$-wave singlet formation is found to be substantial. When an electron is emitted 
in the photoemission process it is energetically favorable for an 
electron in the bath to hop into the cluster. Since the lowest ground state 
of the neutral cluster has negligible spectral weight due to destructive interference 
effects, most of the weight will concentrate on excited states  characterizing
the flanks of the pseudogap. These excited states have suppressed $d$-wave correlations
so that we can relate the pseudogap with the breaking of very short range $d$-wave pairs. 
However, we have also found that the observation of a pseudogap does not imply in general
the existence of preformed $d$-wave pairs.
We have constructed a model where such fluctuations are strongly suppressed, but still
find a pseudogap, provided that the lowest state of the isolated cluster is
nondegenerate.

Geometrical frustration has an important effect on the internal level structure of the
isolated cluster. At $t'/t \sim 0.7$ there is a crossing of many-electron states, leading
to a change of the cluster character. Interestingly, the $\kappa$-(BEDT-TTF)$_2$X family 
of organic superconductors displays a change from an antiferromagnetic (X=Cu[N(CN)$_2$]Cl) 
Mott insulator to a spin liquid Mott insulator (X=Cu$_2$(CN)$_3$) with\cite{yOrganics} 
the ratios $t'/t \sim 0.4-0.5$ and $t'/t \sim 0.8-0.9$, respectively, 
straddling  the ratio $t'/t \sim 0.7$ found above. In a model for the cuprates with, 
$t'=-0.3t$, on a frustrated square lattice, the pseudogap is sensitive to the doping, 
being destroyed for electron doping but not for hole doping consistent with experiments 
on cuprates. This behavior can be understood from the effective enhancement of the Kondo 
coupling to the $(\pi,0)$ bath  as the chemical potential is raised. An interesting 
prediction is that in the organics due to the opposite positive $t'/t$ ratio, the 
pseudogap would be enhanced for electron doping instead.

An important question arises in all DCA calculations. Are the
small cluster calculations representative of the thermodynamic limit behavior? 
Several works have shown\cite{Macridin,Maier,Millis} that the pseudogap is robust against 
increasing the cluster size up to $N_c=16$. We have found that the pseudogap 
can be associated with short range singlet and antiferromagnetic correlations
which characterize NN-RVB cluster states describing the ground state of small clusters ($N_c=4,8$). 
Cluster correlations on large clusters (up to $N_c=64$) also display significant 
very short range $d$-wave singlet correlations and slowly decaying antiferromagnetic spin correlations.
These correlations are consistent with RVB states including singlet bonds between sites 
further distant than nearest-neighbor sites. Hence, a mechanism  
in which Kondo and cluster singlet formation compete as in the smallest clusters discussed here 
can occur. Further work on larger clusters is needed to establish this important issue. 

\acknowledgments
 
JM acknowledges financial support from MINECO (MAT2012-37263-C02-01) and 
hospitality at Max-Planck-Institut f\"ur Festk\"orperforschung in Stuttgart during his stay there.

\appendix\label{sec:A.1}
\section{Coulomb integrals in four-level model}

Here we discuss the Coulomb integrals used in the four level model                 
for describing the $N_c=4$ cluster. Table~\ref{table:A.1} shows results for the 
four  lowest states of the isolated $N_c=4$ cluster. We fit the parameters
in the model to the three lowest levels, using the results in Eq. (\ref{eq:5.3}). 
These parameters are shown in Table~\ref{table:A.1}. We use the approximate
parameterization $\Delta U=0.03 U$ and $K=0.1 U$, appropriate for intermediate values of $U$.

Normally $U_{xx}>U_{xy}$. 
However, due to the short range of the Hubbard interaction $U_{xx}=U_{xy}$.
Furthermore, the interaction with the levels for ${\bf k}=(0,0)$ and
$(\pi,\pi)$  cannot be neglected.  This lowers the  energies $E_{1-}$ and $E_{1+}$ more
than $E_{2-}$ and $E_{2+}$. We describe this by putting $U_{xx}<U_{xy}$  and by adjusting $K$.

\begin{figure}
{\rotatebox{0}{\resizebox{1.8cm}{!}{\includegraphics {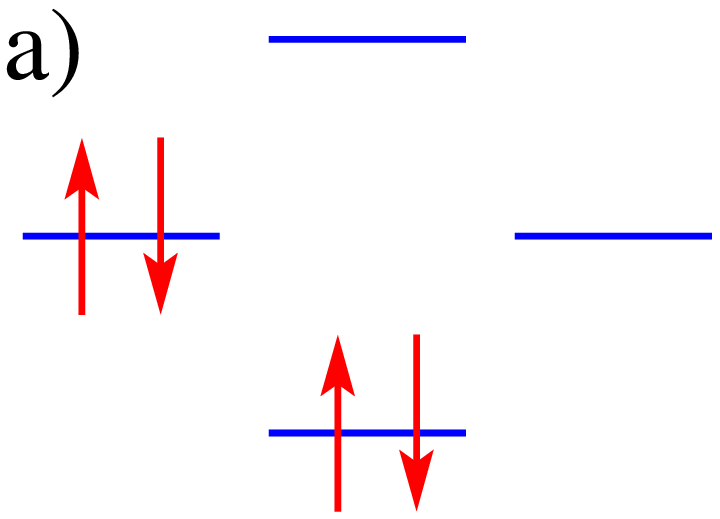}}}}\hskip0.4cm
{\rotatebox{0}{\resizebox{1.8cm}{!}{\includegraphics {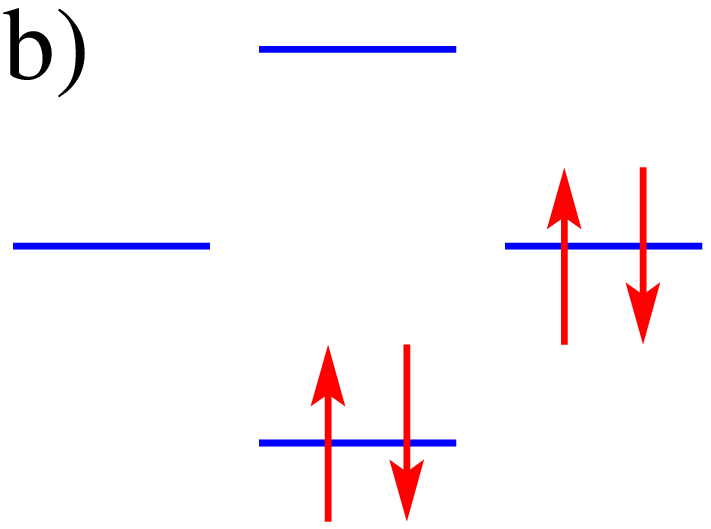}}}}\hskip0.4cm
{\rotatebox{0}{\resizebox{1.8cm}{!}{\includegraphics {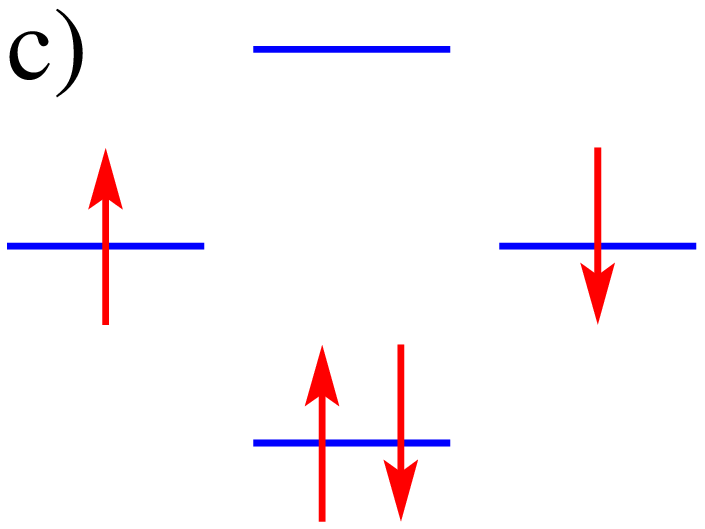}}}}\hskip0.4cm
{\rotatebox{0}{\resizebox{1.8cm}{!}{\includegraphics {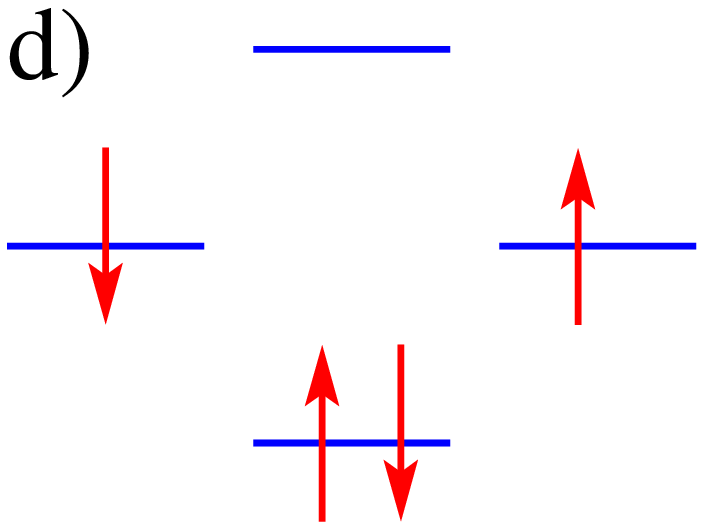}}}}\hskip0.0cm
\vskip0.3cm
{\rotatebox{0}{\resizebox{1.8cm}{!}{\includegraphics {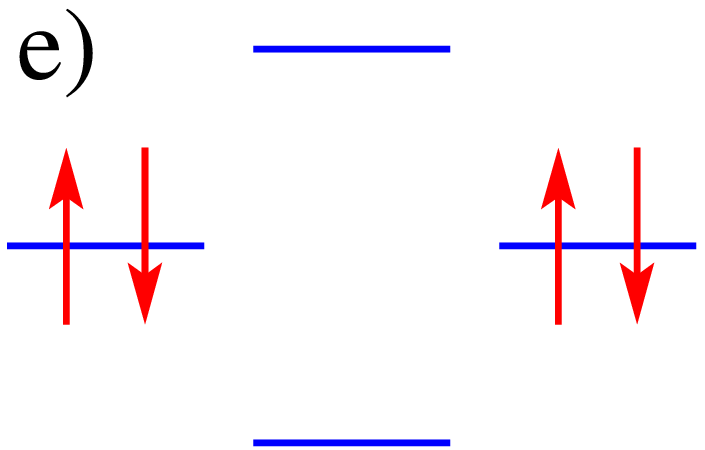}}}}\hskip0.4cm
{\rotatebox{0}{\resizebox{1.8cm}{!}{\includegraphics {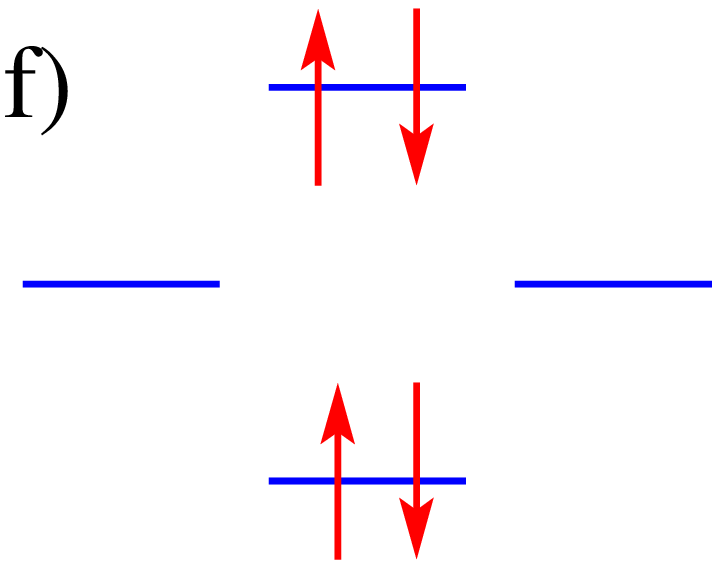}}}}\hskip0.4cm
{\rotatebox{0}{\resizebox{1.8cm}{!}{\includegraphics {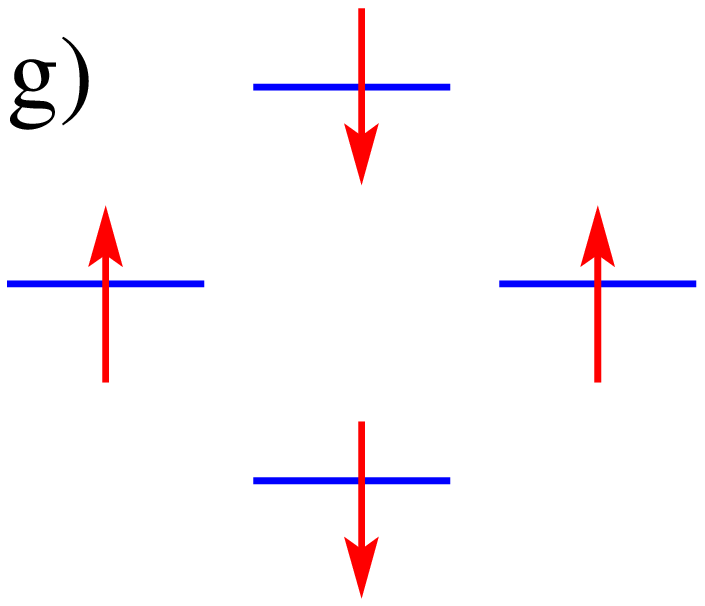}}}}\hskip0.4cm
{\rotatebox{0}{\resizebox{1.8cm}{!}{\includegraphics {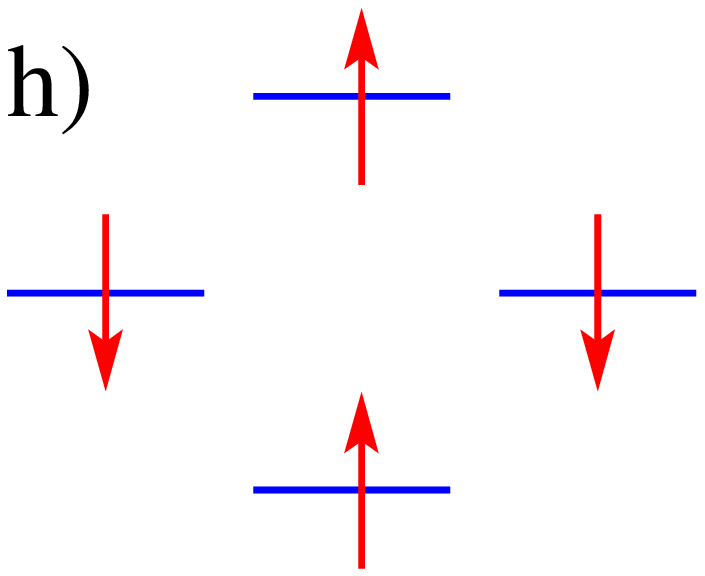}}}}\hskip0.0cm
\caption{\label{fig:A.1}Important configurations for the four-site cluster.
The linear combinations of states a) and b) form the lowest  and third lowest eigenstates,
$|1-\rangle$ and $|1+\rangle$, respectively.}
\end{figure}

The reason for these effects is illustrated in Fig~\ref{fig:A.1}. Due to the Coulomb
interaction, configurations e) and f) couple to a) and b) with the same sign and strength. 
These then lower $E_{1+}$ but not $E_{1-}$.  Instead, configurations g) and h) couple to 
a) and b) with opposite signs, due to the Fermion minus sign and an odd reordering of 
electrons. They then lower $E_{1-}$ but not $E_{1+}$. These couplings are
particularly efficient, since the same configuration couples to both a) and b).
The two couplings are then added before the sum is squared.
There are additional configurations coupling to a) and b), but then only to one or
the other and therefore less efficiently. In contrast, no configuration couples 
to both c) and d). This then makes the coupling to c) and d) less efficient.
Furthermore, fewer configurations couple to c) and d).
Both these facts lead to a larger lowering
of $E_{1-}$ and $E_{1+}$ than for $E_{2-}$ and $E_{2+}$. We describe this by making $U_{xx}<U_{xy}$.
\begin{table}
\caption{\label{table:A.1}Eigenvalues $E_i$ for a four-site Hubbard model with the
parameters $U$ and $t$. The parameters $U_{xx}=U-\Delta U$, $U_{xy}=U+\Delta U$
and $K$ are fitted to the three lowest levels. While the three lowest states
correspond to the three lowest states in Eq.~(\ref{eq:5.3}), the fourth state
is of different character than the fourth state in Eq.~(\ref{eq:5.3}) for these
values of $U$ and $t$.}
\begin{tabular}{cccccccc}
\hline
\hline
U & t& $E_1$ & E$_2$ & $E_3$ & $E_4$ & $\Delta U$ & $K$ \\
\hline
0.25 & -0.050 & -0.0922 & -0.0757 & -0.0422 & -0.0351& 0.0083 & 0.0250 \\
0.25 & -0.030  & -0.0383 & -0.0285 & -0.0148 & -0.0137& 0.0049 & 0.0118 \\
\hline
\end{tabular}
\end{table}

\end{document}